\documentclass[11pt]{article}
\pdfoutput=1
\usepackage[font=small,labelfont=small]{caption}

\usepackage{cite}
\usepackage{amsmath}
\usepackage{amsfonts}
\usepackage{amssymb}

\usepackage{subfig}
\usepackage{graphicx}
\usepackage{setspace}
\usepackage{geometry}
\usepackage{amssymb,epsfig}
\usepackage[hyperfootnotes=false]{hyperref}
\usepackage{color}
\usepackage{soul,xcolor}

\newcommand{\eq}[1]{(\ref{#1})}

\makeatletter
\renewcommand\section{\@startsection {section}{1}{\z@}%
                                 {-3.5ex \@plus -1ex \@minus -.2ex}
                                   {2.3ex \@plus.2ex}%
                                   {\normalfont\large\bfseries}}
\renewcommand\subsection{\@startsection{subsection}{2}{\z@}%
                                   {-3.25ex\@plus -1ex \@minus -.2ex}%
                                     {1.5ex \@plus .2ex}%
                                     {\normalfont\bfseries}}
\renewcommand\subsubsection{\@startsection{subsubsection}{3}{\z@}%
                                   {-3.25ex\@plus -1ex \@minus -.2ex}%
                                     {1.5ex \@plus .2ex}%
                                     {\normalfont\itshape}}
\makeatother

\def\pplogo{\vbox{\kern-\headheight\kern -29pt
\halign{##&##\hfil\cr&{\ppnumber}\cr\rule{0pt}{2.5ex}&\ppdate\cr}}}
\makeatletter
\def\ps@firstpage{\ps@empty \def\@oddhead{\hss\pplogo}%
  \let\@evenhead\@oddhead 
}
\def\maketitle{\par
 \begingroup
 \def\thefootnote{\fnsymbol{footnote}}
 \def\@makefnmark{\hbox{$^{\@thefnmark}$\hss}}
 \if@twocolumn
 \twocolumn[\@maketitle]
 \else \newpage
 \global\@topnum\z@ \@maketitle \fi\thispagestyle{firstpage}\@thanks
 \endgroup
 \setcounter{footnote}{0}
 \let\maketitle\relax
 \let\@maketitle\relax
 \gdef\@thanks{}\gdef\@author{}\gdef\@title{}\let\thanks\relax}
\makeatother

\numberwithin{equation}{section}

\newcommand{\be}{\begin{equation}}
\newcommand{\bea}{\begin{eqnarray}}
\newcommand{\ee}{\end{equation}}
\newcommand{\eea}{\end{eqnarray}}

\begin{document}
 
\setcounter{page}0
\def\ppnumber{\vbox{\baselineskip14pt
IFT-UAM/CSIC-15-045\\
CERN-PH-TH-2015-126\\
IPPP/15/27\\
DCPT/15/54
}}
\def\ppdate{\footnotesize{}} \date{}

\author{{\bf Guillermo Ballesteros}$^{1,2}$ and {\bf Carlos Tamarit}$^{3}$\\
[7mm]
{\normalsize $^1$ \it Instituto de F\'{\i}sica Te\'orica IFT-UAM/CSIC}\\[-2mm]
{\normalsize \it C/ Nicol\'as Cabrera 13-15, Cantoblanco, 28049 Madrid, Spain}\\[2mm]
{\normalsize \it $^2$ CERN, Theory Division, 1211 Geneva, Switzerland}\\[2mm]
{\normalsize \it $^3$ Institute for Particle Physics Phenomenology}\\[-2mm]
{\normalsize \it Durham University, South Road, DH1 3LE, United Kingdom}\\
[3mm]
{\tt \footnotesize guillermo.ballesteros$@$uam.es, carlos.tamarit$@$durham.ac.uk}}

\title{\bf \huge Higgs portal valleys, stability and inflation}
\maketitle

\begin{abstract} 

\noindent \normalsize The measured values of the Higgs and top quark masses imply that the Standard Model potential is very likely to be unstable at large Higgs values. This is particularly problematic during inflation, which sources large perturbations of the Higgs. The instability could be cured by a threshold effect induced by a scalar with a large vacuum expectation value and directly connected to the Standard Model through a Higgs portal coupling. However, we find that in a minimal model in which the scalar generates inflation, this mechanism does not stabilize the potential because the mass required for inflation is beyond the instability scale. This conclusion does not change if the Higgs has a direct weak coupling to the scalar curvature. On the other hand, if the potential is absolutely stable, successful inflation in agreement with current CMB data can occur along a valley of the potential with a Mexican hat profile.
We revisit the stability conditions, independently of inflation, and clarify that the threshold effect cannot work if the Higgs portal coupling is too small. We also show that inflation in a false Higgs vacuum appearing radiatively  for a tuned ratio of the Higgs and top masses leads to an amplitude of primordial gravitational waves that is far too high, ruling out this possibility.

\end{abstract}

\newpage

\begin{spacing}{0.01} 
\tableofcontents 
\end{spacing}

\section{Introduction} \label{introd}

While most models of primordial inflation involve one or more scalars to drive the accelerated expansion of the universe, the Higgs boson is the only known elementary scalar that has been found in nature so far. Its recent discovery at CERN by the ATLAS and CMS collaborations \cite{Aad:2012tfa,Chatrchyan:2012ufa} prompts the investigation of its possible connections, and those of the Standard Model of particle physics (SM) as a whole, with inflation. The Higgs is currently the part of the SM that is least known and its relation to the very early universe may open windows that could allow us to learn more about both.

Historically, it has been widely assumed that the Higgs field plays no role during inflation, being stabilized at its zero-temperature vacuum expectation value (VEV) of $246$ GeV. In standard slow-roll inflation, this can be understood geometrically as the assumption that inflation proceeds along a potential energy valley which extends in field directions other than the Higgs. In this traditional view, the Higgs is regarded as a mere spectator of the inflationary dynamics. There are however issues that complicate the picture of the connection between inflation and the SM which cannot be lightly neglected.

First, even in models in which the Higgs is not supposed to play an active role, it may still be affected by the process. A rapid expansion of the universe can induce large quantum perturbations on the Higgs field, which could potentially destabilize the universe \cite{Espinosa:2007qp,Degrassi:2012ry,Lebedev:2012sy,Kobakhidze:2013tn,Fairbairn:2014zia,Enqvist:2014bua,Hook:2014uia,Herranen:2014cua,Shkerin:2015exa,Kearney:2015vba,Espinosa:2015qea}. Any scalar whose mass is smaller than the Hubble expansion rate, $\mathcal H$, leads to an almost scale-invariant spectrum of perturbations on scales larger than $1/\mathcal H$. On a first approximation, the variance of the amplitude of these fluctuations is of order $\mathcal H^2$. If we assume the validity of the SM up to high energies and if the SM electroweak vacuum is metastable, which appears to be the case for the measured central values of the Higgs and top quark masses \cite{Degrassi:2012ry,Buttazzo:2013uya}, the instability region of the SM would likely be 
reached during inflation. This poses the disturbing question of how the Higgs ended on its right vacuum. Computing the likely fate of the Higgs during inflation assuming metastability of the SM is a subtle matter and the post-inflationary evolution needs to be accounted for as well \cite{Espinosa:2015qea}.  

Second, in models in which an inflaton coupled to the Higgs gets a large VEV, the interactions among the two may deform the potential energy valleys supporting inflation, causing them to reach into large values of the Higgs field and becoming sensitive to the destabilizing effect of the top quark. This effect could ruin the prospects for inflation itself, since the classical trajectories could be drawn towards the instability region, regardless of quantum fluctuations. 

And third, no model of inflation should  be considered fully complete if it does not deal with the question of generating the matter content of the universe when inflation ends. If inflation ever happened, we know positively that a (direct or ``hidden'') coupling between the inflationary sector and the SM must exist to reheat the universe. 

These issues show  that the consistency of the interplay between the SM and the inflationary sector should be clearly addressed. A possible approach to incorporate inflation into the SM consists in considering that the Higgs plays an active role. Unfortunately, the tree-level potential of the SM Higgs is unable to generate enough e-folds of inflation and primordial curvature perturbations in agreement with the measured spectrum. Since this failure is essentially due to the Higgs having a potential which grows too fast, some mechanisms that flatten the potential at large field values have been considered. These include the case of a plateau arising for tuned values of the top quark and Higgs masses and, also, a non-minimal coupling to gravity. As it is well-known, the first possibility again fails to fit the amplitude of scalar primordial perturbations while generating a sufficient amount of inflation, see e.g.\ \cite{Isidori:2007vm}. The second option requires very large values of the non-minimal coupling to 
the Ricci scalar \cite{Bezrukov:2007ep}, and there are ongoing discussions on whether the idea makes full sense within the context of effective theories, unitarity and viable ultraviolet completions \cite{Burgess:2009ea,Barbon:2009ya,Lerner:2010mq,Giudice:2010ka,Bezrukov:2010jz,Burgess:2014lza,Barbon:2015fla}. A further possibility is that the Higgs may be non-dynamical during inflation, but important in determining the energy scale. This is the case in  scenarios of Higgs false vacuum  inflation \cite{Masina:2011aa,Masina:2012yd}, which are based on adjusting the value of the top mass beyond the one giving rise to a plateau, in such a way that false vacuum with positive energy density appears. This vacuum can drive inflation, yet in order for it to end and generate curvature perturbations, new dynamics is needed. This can be done by introducing another field direction along which the false vacuum might be escaped, in such a way that the rolling in this direction generates the spectrum of primordial 
perturbations. However, recent calculations have found a strong tension in this model with the observed value of the Higgs mass \cite{Fairbairn:2014zia,Notari:2014noa}

Since a connection between the SM and the inflationary sector is necessary to reheat the universe, there are certain questions that any complete model of inflation should address. In particular, do the SM-inflaton interactions intervene in the inflationary dynamics? And, can the Higgs field be consistently ignored in models in which the dynamics is mostly driven by other fields?  

In principle, the SM-inflaton  interactions can be suppressed by assuming an anomalous shift symmetry for the inflaton field, which would make the inflationary background independent of the Higgs. Besides, concerning the Higgs perturbations, it should be noted that a positive coupling of the Higgs to the Ricci scalar can damp them, preventing it from falling into the instability region \cite{Espinosa:2007qp,Herranen:2014cua, Espinosa:2015qea}.  In this work, we will not impose a shift symmetry and we will mainly focus in negligible direct couplings to gravity.\footnote{See however, section \ref{subsec:nonminimal coupling}.} In this situation, the trouble with Higgs fluctuations should be solved by stabilizing the effective potential.

Assuming a $Z_2$ symmetry, we consider a simple extension of the SM in which an extra singlet acquires a large VEV and interacts with the SM through a quartic Higgs portal coupling. In general, the inflaton turns to be a combination of the singlet and the Higgs, though we will focus in the case in which it is mostly aligned with the former. This gives a minimal and appealing playground to start addressing the issues mentioned before in the absence of a shift symmetry and direct couplings to gravity. A Higgs portal coupling is strongly motivated by the requirement of reheating the universe at the end of inflation and the stability of the SM at large Higgs values. It is well known that the stability of the SM effective potential that can be improved  in the presence of couplings between the Higgs and other scalars \cite{Gonderinger:2009jp,Lebedev:2012zw,EliasMiro:2012ay,Cline:2013gha,Pruna:2013bma,Kadastik:2011aa}. In fact, a large VEV for the inflaton is needed for the stabilization mechanism of \cite{Lebedev:
2012zw,EliasMiro:2012ay}, which is based on a tree-level threshold effect in the presence of a heavy scalar. The Higgs portal has been included in some models of inflation, either with non-minimal couplings to gravity \cite{Masina:2011un,Lebedev:2011aq,EliasMiro:2012ay,Huang:2013oua,Khoze:2013uia} or for Higgs false-vacuum inflation \cite{Masina:2012yd,Fairbairn:2014nxa}. It has also been considered in inflation in relation to the stability of the SM \cite{Lebedev:2012sy,Bhattacharya:2014gva}, which we will explore here.  

We will see that a Higgs portal coupling to a heavy singlet with a large VEV is compatible with inflation in agreement with current CMB data, which can take place along potential energy valleys with Mexican hat profiles. While Planck data basically rule out monomial models of inflation with exponent higher than unity \cite{Ade:2015lrj}, Mexican hat inflation, which interpolates between negative and positive curvatures (i.e.\ between ``hilltop'' and quadratic potentials), provides a comfortable fit to the data. These models are characterized by a small tensor-to-scalar ratio $r\gtrsim 0.04$, which is expected to be testable with the precision of future CMB polarization experiments. 

In the (very particular) decoupling limit, where the Higgs portal coupling vanishes, the singlet alone is responsible for inflation, which thus is not directly connected to the SM. However, in a generic situation the inflationary valley mixes the singlet and the Higgs directions, and  it can reach values of $h$ larger than the SM instability scale $\Lambda_I\sim 10^{11}$ GeV.\footnote{The actual value of the instability scale depends on its precise definition and the values of the top quark and Higgs masses. If we define the instability scale as the value of the Higgs field at which the potential becomes negative,  choosing $m_t=173.15$ GeV and $m_h=125.09$ GeV, the instability scale is $5.0\cdot 10^{11}$ GeV.} As mentioned before, this is troublesome since the attractors of classical trajectories could then fall into the instability region. Moreover, even if inflationary trajectories overcoming this problem may be possible, quantum fluctuations of the Higgs during inflation could anyway send the Higgs to 
the instability region.  

It is therefore important to elucidate whether the threshold stabilization mechanism can work if the extra scalar provides inflation. We revisit the mechanism and show that if the mass scale of the inflaton is higher than the SM instability scale, so that the threshold effect induced by the inflaton is effectively decoupled in the instability region, the stabilization is not possible.\footnote{In  \cite{Bhattacharya:2014gva} a non-minimal setting with a complex scalar, additional fermions and a gauge field was considered. Stability in the Higgs direction was obtained with a large mass of the inflaton. However, in order to reproduce the current CMB constraints in this case, the potential in the inflaton direction needs to be unbounded from below due to radiative corrections.}

Even more, we show that regardless of a possible connection with inflation, the threshold stabilization mechanism does not work for a very small Higgs portal coupling. This is due to the existence of a scale which is inversely proportional to the square root of the portal coupling, and therefore grows unbounded in the decoupling limit. The appearance of this scale becomes apparent from the geometry of the valleys of the two-dimensional potential. 

If the extra scalar drives inflation, the CMB constraints imply that its mass has to be around $10^{13}$ GeV. Assuming a Higgs mass of 125.1 GeV, the instability scale of the SM  can be larger than $10^{13}$ GeV only if the top quark is lighter than 172.5 GeV.\footnote{We recall that the SM potential can in principle be stable for a sufficiently small top mass, which we estimate to be around $m_t\lesssim171.7$ GeV. However, taking into account the currently allowed range for the top quark mass (see \ \eqref{eq:mt}), and the most sophisticated calculations that are available \cite{Degrassi:2012ry,Espinosa:2015qea}  this possibility seems now unlikely in comparison with metastability of the SM electroweak vacuum.  In our calculations of the instability scale we take the strong coupling constant to be $\alpha_s(m_Z)=0.1885$, which is the current central value in \cite{Agashe:2014kda}. Variations of $\alpha_s(m_Z)$ within its $0.3\%$ error shift the instability scale by an irrelevant amount that does not change 
our results.} In spite of this, the appearance of the new scale in the stability conditions for the threshold mechanism precludes stabilization and, therefore, a different mechanism is required. This could be provided, for instance, by an additional heavy scalar that is stabilized at the origin.  We have checked that stabilizing the inflationary valleys in this way, the predictions for the cosmological parameters (including loop corrections) remain extremely close to the tree-level results with the SM plus the inflaton alone. 

We also revisit the idea of Higgs false-vacuum inflation \cite{Masina:2012yd,Fairbairn:2014nxa}, mentioned earlier, and conclude that it cannot produce successful inflation for the measured values of the SM couplings. A similar conclusion was already found in \cite{Notari:2014noa,Fairbairn:2014nxa}, and we confirm that the trouble is  related to the tensor-to-scalar ratio $r$, as argued in \cite{Notari:2014noa}. An accurate evaluation of the energy of the false vacua, compatible with the results of \cite{Degrassi:2012ry}, yields a lower bound of $r\gtrsim 2$ which excludes the result of \cite{Masina:2014yga}, according to which a value of $r$ compatible with the latest measurements of Planck could be generated inside the false-vacuum valley. 

The paper is organized as follows. In section~\ref{sec:SMS} we introduce the model: the SM coupled to a singlet through the Higgs portal with a $Z_2$ symmetry, to be referred to as SMS. The tree-level potential valleys are described in section~\ref{sec:valleystree}, and the corresponding inflationary dynamics is analyzed in detail in section~\ref{sec:inflation}, discussing the single-field and slow-roll approximations and the generation of curvature perturbations. We devote section~\ref{sec:quantum} to radiative effects. In section~\ref{subsec:RGimproved} we review the RG-improved potential, emphasizing the importance of the field-independent piece for cosmology. In section~\ref{subsec:stability} we analyze the issue of the stability of the effective potential, and in section~\ref{subsec:inflationloop} we study the implications for inflation. In section~\ref{subsec:nonminimal coupling} we study if a coupling of the Higgs to the Ricci scalar can affect the stability during inflation. In section~\ref{subsec:mtvalley}, we revisit the scenario of false-vacuum inflation. The conclusions are drawn in section~\ref{sec:Conclusions}. In addition, three appendices are provided: appendix~\ref{app:RGs} gives the two-loop RG equations that we use, appendix~\ref{app:SM_Pars} reviews the matching of the relevant SM parameters to experimental measurements, and appendix~\ref{sec:SMmatching} contains the details about the matching between the SM and SMS.

\section{\label{sec:SMS} Standard Model coupled to a real scalar}

 We consider the SM coupled to a real singlet $S$, with a tree-level scalar potential  given by
\begin{align} 
 \label{eq:V0}V^{\rm tree}(H,S;\delta_i)={m^2_H}H^\dagger H+\frac{m^2_S}{2} S^2+\frac{\lambda}{2}(H^\dagger H)^2+\frac{\lambda_S}{4!}S^4+\frac{\lambda_{SH}}{2}H^\dagger H S^2\,,
\end{align}
where $H$ is the Higgs $SU(2)$ doublet. The symbol $\delta_i$ is used to denote generically the couplings and squared masses  of the model. This encompasses not only the couplings ($\lambda$, $\lambda_S$, $\lambda_{SH}$) and the masses ($m_H^2$, $m^2_S$) of the scalar potential \eq{eq:V0}, but also the Yukawa and gauge couplings of the SM. If  a $Z_2$ symmetry is imposed, \eq{eq:V0} is the most general renormalizable potential for $S$ and $H$, excluding an allowed vacuum energy term $V_0$, which can be used to accommodate the measured value of the cosmological constant $\Lambda\sim (10^{-3} \text{eV})^4$.  For practical purposes, we can assume $\Lambda=0$, which does not change our results.\footnote{We assume that at the end of inflation the fields come to rest at a minimum of the potential corresponding to the cosmological constant that we measure today. Since this value is many orders of magnitude smaller than the energy scales involved during inflation we can safely 
take it to be zero.}  As we will later see, $V_0$ helps to compensate field independent contributions from radiative corrections, ensuring that the condition $\Lambda=0$ is met. 

Throughout this paper, we will consider the situation in which the singlet $S$ gets a large VEV and a large physical mass, which happens if $m^2_S$ is large (and negative) or if $\lambda_S$ is small. We will show that this scenario, and taking into account the observed properties of the Higgs boson, provides successful inflation compatible with current CMB data. Even though setting $\lambda_{SH}=0$ would decouple the SM from the singlet $S$, allowing a separate inflationary sector, generically the coupling $\lambda_{SH}$ plays a role by deforming the potential energy valley that supports inflation. This makes the inflationary valley reach into the Higgs direction and therefore become sensitive to quantum corrections from $H$. As we will see, the coupling $\lambda_{SH}$ is also important for the stability of the potential of the model at high energies, which we study in section~\ref{subsec:stability}.

The election of couplings in the model that contains the singlet $S$ (henceforth ``SMS'' for brevity) cannot be arbitrary, since at low energies one should recover the SM  Higgs field alone with its associated VEV and mass. The VEV $v\sim 246$ GeV is fixed by the measurements of the muon lifetime (see appendix~\ref{app:SM_Pars} for more details), while the physical Higgs mass is given by $m_h=125.09\pm 0.21(stat.)\pm 0.11(syst.)$ GeV from the latests combined ATLAS and CMS measurements \cite{Aad:2015zhl}. In order to ensure that the SMS is compatible with the most recent results coming from collider experiments, we will proceed by matching the theory at low energies with the SM, whose couplings are fixed by the experimental measurements. This two-step matching is appropriate because the singlet $S$ will be required to be very heavy (with a mass near $10^{13}$ GeV) as well as weakly coupled, so that it decouples at low energies, leaving the SM as an effective theory. In principle, the matching  to 
experimental particle physics data could be done in the high energy model by including the appropriate quantum corrections, but doing so numerical problems arise when demanding that the correct Higgs mass and VEV should be generated from large absolute values of $m^2_S$ and very small $\lambda,\lambda_{SH}$. In other words, the large range of scales between the physical masses $m_h\simeq 125$ GeV and $m_S\sim 10^{13}$ GeV motivates the  two-step approach. This also allows  to resum large logarithmic corrections involving the mass of the heavy singlet below the matching threshold
and, furthermore, this method will help to illustrate better the appearance of valleys in the potential.  

We will then consider the masses of all known elementary particles fixed at their central experimental values, except for the top quark, whose mass, $m_t$, will be allowed to vary in order to investigate the stability of the effective potential. The obtention of the SM parameters from the experimental measurements is reviewed in appendix~\ref{app:SM_Pars}. We  include one-loop strong and electroweak corrections in the determination of the SM Higgs parameters from the values of the Fermi constant $G_F$ and the Higgs mass. To relate a choice of $m_t$ with the corresponding top Yukawa we include one-loop electroweak corrections and up to three-loop strong corrections. 

Once the SM parameters are fixed, the parameters $m^2_S$, $\lambda_S$ and $\lambda_{SH}$ are regarded as inputs that are required to fit the constraints coming from inflation. After these parameters are fixed, the Yukawas, gauge couplings, and Higgs quartic and quadratic couplings in the SMS can be obtained from their SM counterparts. In short, the free parameters that we consider are $m_t$, $m^2_S$, $\lambda_S$ and $\lambda_{SH}$. Since we restrict the parameter space of the high energy model to the region which reproduces the measured Higgs mass, the Higgs quartic and quadratic couplings in the SMS are not independent of the parameters $m^2_S, \lambda_S$ and $\lambda_{SH}$.

Given the decoupling of the singlet at low energies, in order to  match the SM and the SMS  one should demand the equality of
the flat spacetime Green functions computed on both sides of the threshold at which the heavy singlet decouples. When it comes to  the parameters in the scalar potential $V$ (which encodes the Green functions at zero momentum) the decoupling of $S$ amounts to integrating it out using the zero-momentum equation of motion. This means that at sufficiently low scales, when quantum fluctuations of the singlet are suppressed, the field $S$ sits (on average) at the value which minimizes the potential energy for every $h\equiv \sqrt{2}H^0$, where $H^0$ is the neutral component of the Higgs doublet. As mentioned before, we are going to assume that $m_S^2<0$, which means that this minimum happens for $S^2\neq 0$. Notice that in principle the Higgs field could also be stabilized with $m_S^2>0$ (with $S=0$ at the minimum), but in this case the valley supporting inflation would not extend to large values of $h$. Inflation would then be exclusively driven by the field $S$ alone (as in a standard independent single-field 
model) with no role played by the Higgs.\footnote{If $m_S^2>0$, successful inflation in the SMS with $S$
playing the role of the inflation would then additionally require a
non-minimal coupling to gravity, in order to satisfy current CMB limits
\cite{Ade:2015lrj} on primordial gravitational waves.} We will later see that for $m_S^2<0$ successful inflation is actually mostly driven by $S$ as well, however the couplings of the effective potential that drives inflation in a single field approximation to the dynamics are affected by those of the Higgs in that case. 

After these considerations, the matching of the  parameters in the potential can be done in practice by considering the potential $V$  when the field $S$ is  set at the value $S_{min}$ that satisfies
\begin{align}
 \label{eq:ssol0}\left.\frac{d V}{dS}\right|_{S=S_{min}(h)}=0\,,
\end{align}
and demanding
\begin{align}
\label{eq:match0}  V^{SM}(h;\tilde \delta)=  V(h,S_{min}( h);\delta)+O(h^6/|m^2_S|).
\end{align}
 In other words, the (one-dimensional) potential in the SM should be understood as the value of the SMS  potential along a line which follows the minima with respect to the field $S$. This is the usual basic procedure for integrating out the heavy field at low-energies. Notice that we denote SM quantities (low-energy) with a tilde to distinguish them from the SMS ones (high-energy). As indicated, the equality in  \eqref{eq:match0} is valid up to terms that are suppressed by inverse powers of the heavy mass, corresponding to non-renormalizable terms in a polynomial expansion of the SMS potential. 
 
 At tree-level, and writing the SM potential as
\begin{align}
V^{SM}=\tilde m_H^2\, H^\dagger H + \frac{\tilde\lambda}{2}(H^\dagger H)^2\,,
\end{align} 
 taking derivatives of \eqref{eq:match0} with respect to the fields gives the following matching conditions, 
 \begin{align}
 \label{eq:treematching}\tilde m^2_H  =m^2_H - \frac{3\lambda_{SH}}{\lambda_S} m^2_S\,,\quad \tilde\lambda  =\lambda-\frac{3\lambda^2_{SH}}{\lambda_S},
\end{align}
which are the ones usually employed in the literature, see e.g.\ \cite{EliasMiro:2012ay,Fairbairn:2014nxa}. In our numerical calculations we  improve this matching by including higher order quantum corrections, making use of the one-loop effective potential with a two-loop RG improvement, whose construction is reviewed in section~\ref{subsec:RGimproved}. The matching with the quantum effective potential is done without relying on a polynomial expansion around the origin of field space; the technical details are provided in appendix \ref{sec:SMmatching}. However, to extract the features of the model for inflation and check its validity against current CMB data, the matching conditions \eq{eq:treematching} and a tree-level description of the dynamics are sufficient, as discussed in section \ref{subsec:inflationtree}.

Finally, the value of the field independent piece, $V_0$, mentioned in the beginning of this section, is fixed by demanding that the current Higgs vacuum should have zero energy. This can be imposed at the SM level, and then matching across the threshold  using \eqref{eq:match0} after $\lambda$ and $m^2_H$ are matched as in \eqref{eq:treematching}. At tree-level, this yields:
\begin{align}
\label{eq:treematching2}
 V_0=\frac{1}{2} \left(\frac{
 \tilde m^4_H}{\tilde\lambda}+3\, \frac{m^4_S}{\lambda_S}\right).
\end{align}

Before moving on, note that the matching equations \eqref{eq:treematching} can make the value of $m^2_H$ in the SMS substantially different from its SM counterpart, $\tilde m^2_H$, given the presence of corrections proportional to $m^2_S$, which in the models that we will consider here is large and negative ($|m^2_S|\gg m^2_H$). This large absolute value of $m^2_H$, which has an important effect in the shape of potential energy valleys (see next section), is perfectly compatible with the observed properties of the Higgs, as guaranteed by the matching procedure. From the point of view of the high energy model, the weak scale will arise from the large absolute values of $m^2_S$ and $m^2_H$  via an appropriate value of the dimensionless parameter $\lambda_{SH}$, as follows from \eqref{eq:treematching}. It is interesting to note that such a tuning of $\lambda_{SH}$ (a dimensionless coupling) is technically natural, as the beta function of $\lambda_{SH}$ goes to zero in the limit $\lambda_{SH}\rightarrow0$, 
making the choice of small $\lambda_{SH}$ radiatively stable, see \eqref{eq:betassinglet}.

\section{\label{sec:valleystree}Tree-level valleys}

This section describes the valleys that arise in the potential of the SM plus singlet model (SMS), for scenarios in which the singlet, $S$, acquires a large VEV, while the Higgs mass and VEV are compatible with the experimental measurements. Positivity of the potential energy for large field values  demands $\lambda>0$ and $\lambda_S>0$, while negative values of the Higgs portal coupling, $\lambda_{SH}$, are allowed as long as $\lambda_{SH}>-\sqrt{\lambda \lambda_S/3}$.\footnote{At large field values the potential is dominated by the quartic couplings. The  bound $\lambda_{SH}>-\sqrt{\lambda \lambda_S/3}$ can be easily obtained considering the effective quartic interactions along radial lines in field space, $h \propto S$, and demanding that they stay positive.} Then, the requirement of a large singlet VEV enforces a negative $m^2_S$, while $m^2_H$ can be positive or negative as long as the corresponding coupling in the SM, $\tilde m^2_H$, stays negative. This is needed, as usual, to have a nonzero 
electroweak symmetry breaking Higgs VEV, see \eqref{eq:treematching}.

We will proceed in section~\ref{sec:wiav} by introducing the notion of a valley in a multi-field potential, which defines in a mathematically precise way the intuitive idea of a physical valley running along a multi-dimensional surface. These valleys are important for inflation because they typically act as attractors capable of trapping the fields. The definition of valleys will be followed by an analysis of the curves described by the minima of the potential for constant $S$ (``$h$-line'') and constant $h$ (``$S$-line'').  We will see that in the limit $\lambda_{SH}=0$ at tree-level, each of these two lines corresponds to the bottom of a valley, which we call $h$- and $S$- valleys, and the potential along them reproduces, respectively, the potential in the singlet direction and the SM potential. The $h$-valley can support inflation in agreement with the data, and this is well described by a single-field model in which the resulting (Mexican hat) potential is a function of $S$. On the other hand, inflation 
compatible with the CMB measurements cannot happen within the $S$-valley, because the SM tree-level potential predicts curvature perturbations that are too large if enough e-folds are generated.

For small $\lambda_{SH}\neq0$, the $h$- and $S$- lines become distorted. The $h$-line remains a good approximation to the bottom of an actual valley far enough from the vacuum of the full potential. This distorted $h$-line can still support inflation. However, inflation will happen for larger values of the Higgs (with respect to the case $\lambda_{SH}=0$) and will be sensitive to quantum corrections involving $h$ through the Higgs portal coupling. In contrast, the $S$-line closely  describes the bottom of a valley in the vicinity to the vacuum of the model,\footnote{The reason why the $h$-line approaches an actual valley far enough from the vacuum and, conversely, the $S$-line does so close to the vacuum, is basically the large hierarchy between the mass parameters between the heavy singlet, $S$, and the field $h$:  $|m_S^2|\gg|m_H^2|$. See also the matching equation \eq{eq:treematching}.}  justifying the matching procedure of section~\ref{sec:SMS}. Near the vacuum and for small $\lambda_{SH}\neq0$, the $S$-line is 
nearly parallel to the $h$-direction, and gives a potential that reproduces the SM, which (as we just mentioned) cannot support successful inflation.  Far from the vacuum, the $S$-line curves away from the $h$-direction, and the rolling along the corresponding valley  starts to be dominated by the field $S$. This would seem to open the possibility that the dynamics of $S$ could generate successful inflation far from the vacuum; however, as the line curves it also tilts, due to the effect of the Higgs quartic, until it cannot trap the fields in the orthogonal direction and the line stops describing a valley. Therefore, for small non-zero portal coupling $\lambda_{SH}$ the $S$-line has a limited extension away from the vacuum, which makes it less likely to support inflation, since this could only happen far enough from the absolute minimum.

These features are illustrated schematically in figure~\ref{fig:valleysp}, corresponding to $\lambda_{SH}>0$,  and figure~\ref{fig:valleysm}, corresponding to $\lambda_{SH}<0$.  The $S$- and $h$- lines are represented by the dashed-red and dotted-blue curves, respectively. In the parameter range of interest, the valleys that can be seen in these figures have energies that monotonously decrease  from large values of $h$ towards the Higgs vacuum, located at $h=v_h$ and $S=v_S$, given by
\begin{align} \label{veVs}
v_{h}^2= -2\frac{\tilde m_H^2}{\tilde\lambda}\,,\quad \frac{\lambda_S}{6}v_S^2=\frac{\lambda_{SH}}{\tilde\lambda}\, \tilde m^2_H -m^2_S\,.
\end{align}
Using these quantities and the expressions \eq{eq:V0} and \eq{eq:treematching2}, we can write the tree-level potential (including the vacuum piece $V_0$) for the heavy singlet, $S$, and the neutral Higgs component, $h$, as follows\footnote{There is a factor of $1/\sqrt{2}$ between $H$ and $h$, see \eq{deco}.}
\begin{align} \label{Vvev}
V(h,S;\delta_i)=\frac{\lambda}{8}\left(h^2-v_h^2\right)^2+\frac{\lambda_S}{24}\left(S^2-v_S^2\right)^2+\frac{\lambda_{SH}}{4}\left(h^2-v_h^2\right)\left(S^2-v_S^2\right)\,.
\end{align}
This form of the potential helps to see clearly the structure of extrema at tree-level, which will be useful later on, in particular to study the stability of the model at large field values. If $\lambda_{SH}^2<\lambda_S\lambda$ (and focusing in the quadrant $h\geq 0$, $s\geq 0$) this potential has a minimum at $h=v_h$ and $S=v_s$. In addition, there is a local maximum at $h=S=0$ and two saddle points, each of them sitting on one of the axes $S=0$ and $h=0$. 

\subsection{\label{sec:wiav}What is a valley?}
\begin{figure}
\centering
\subfloat{\label{fig:valleys0}\includegraphics[width=3in]{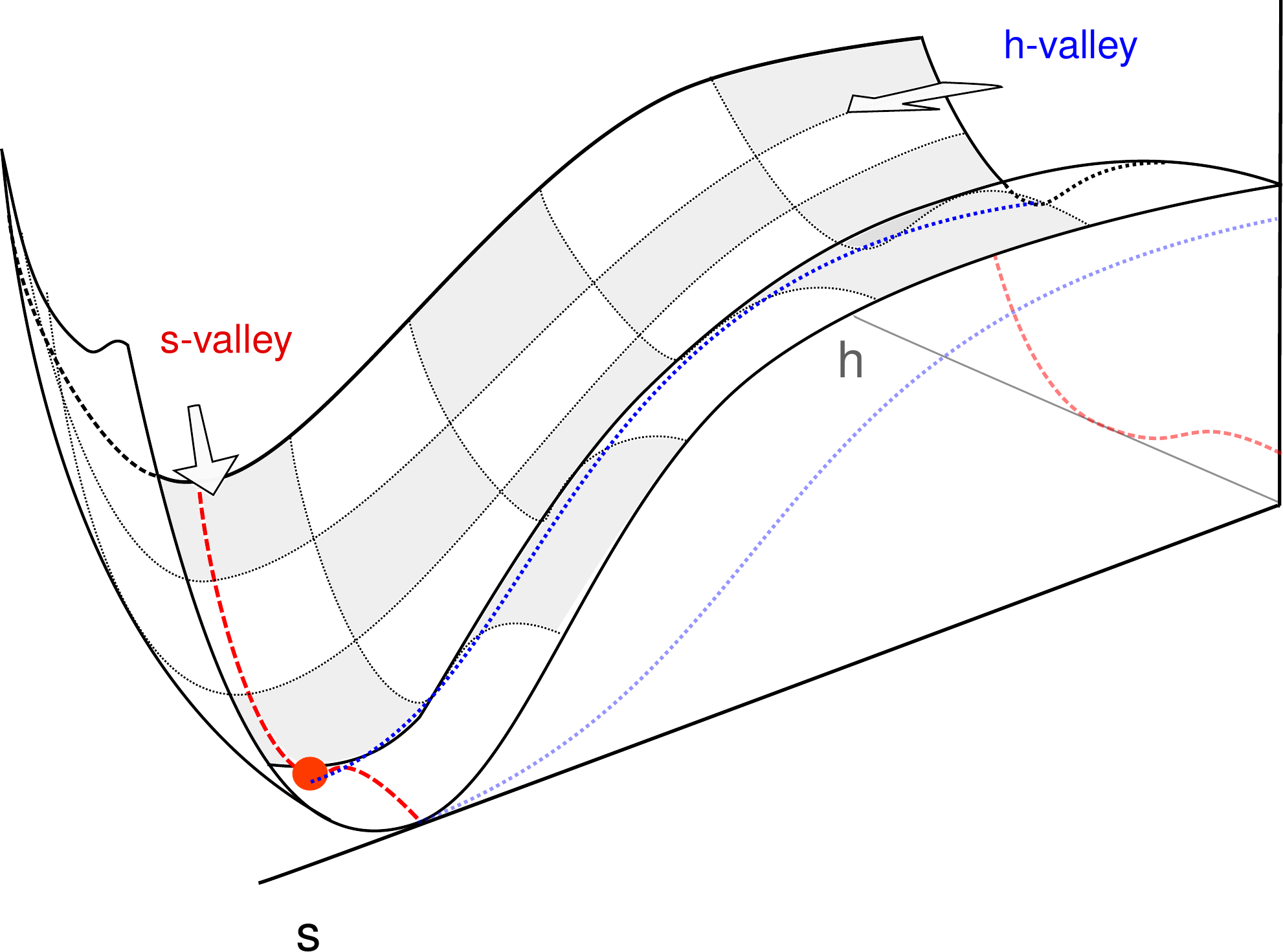}}\\
\vspace{-0.5in}
\subfloat{\label{fig:valleysp}\includegraphics[width=3in]{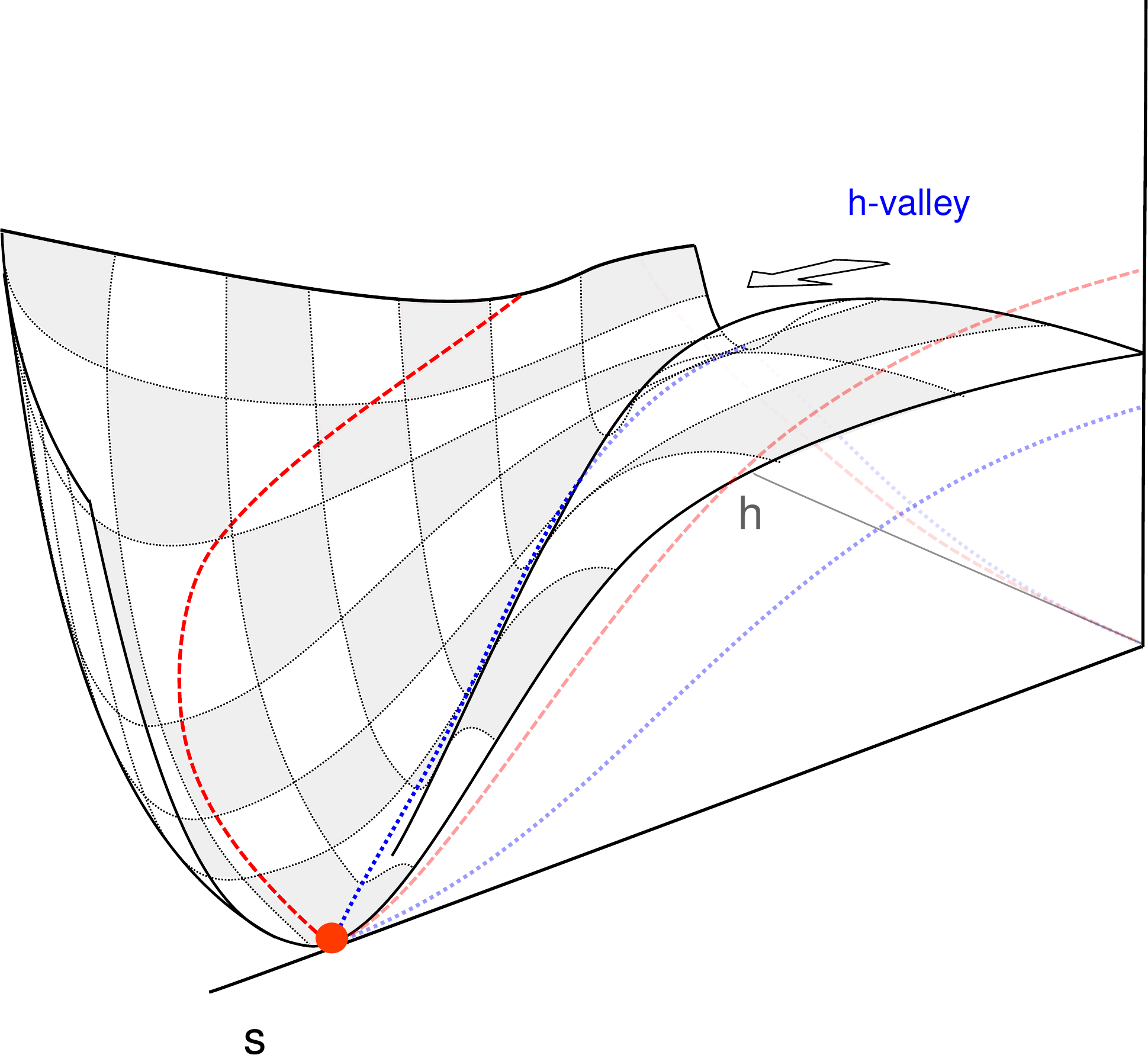}}\\
\vspace{0.03in}
\subfloat{\label{fig:valleysm}\includegraphics[width=3in]{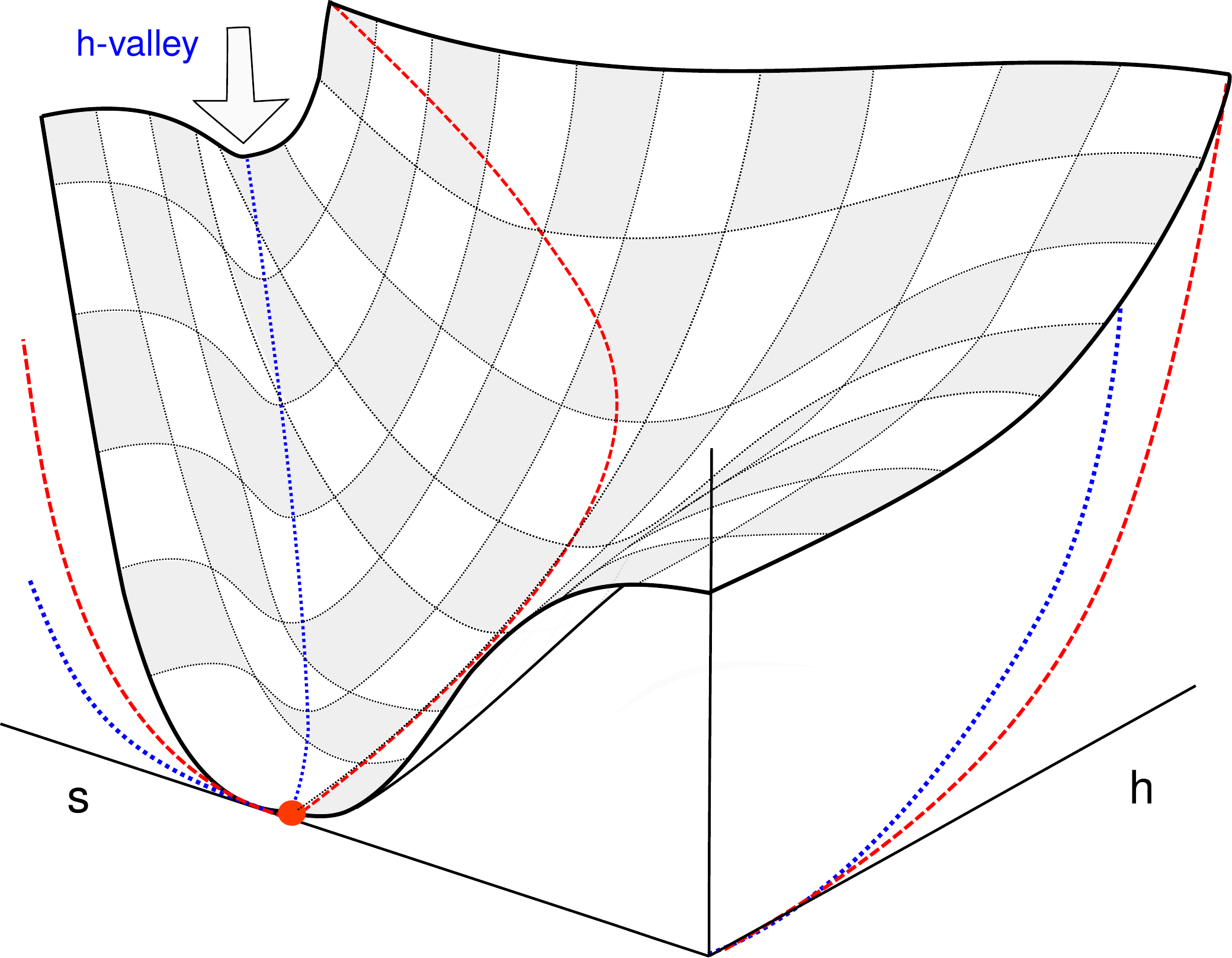}}
\caption{\small Schematic representation of the tree-level lines of minima (for derivatives with respect to $S$ and $h$) in the cases $\lambda_{SH}=0$ (top: \ref{fig:valleys0}),  $\lambda_{SH}>0$ (middle: \ref{fig:valleysp}) and $\lambda_{SH}<0$ (bottom: \ref{fig:valleysm}). The dashed red curves represent the potential along the $S$-line and its projections onto the coordinate planes, and the blue dotted curves represent the  potential along the $h$-line and its projections. The red dot represents in both cases the vacuum, which is the endpoint of inflation, lying at $S=v_{S}\gg h=v_{h}$, with $v_{h}\neq 0$ only visible in the top figure. The figures are merely illustrative and not to scale among them.}
\end{figure}

To understand the concept of a valley in primordial inflation driven by an arbitrary number $q$ of real scalar fields $\phi_i$, it is useful to think of the multi-field potential $V(\phi_1\,,\ldots\,,\phi_q)$ describing their (non-derivative) interactions as a $q$-dimensional hypersurface embedded in a space of $q+1$ dimensions.\footnote{We will assume that the fields have standard kinetic terms. For a more general treatment where kinetic mixing is allowed see \cite{Burgess:2012dz}. In that reference the term ``trough'' is used often instead of ``valley''. We prefer the second one, but both refer to the same concept.} Intuitively, inflation will preferably proceed along trajectories that fall in the valleys of the potential because the fields will tend to minimize their potential energy. Slow-roll inflation may occur along the floor of a valley if the floor is flat enough. In standard single-field inflation, the hypersurface is just a continuous curve, $V=V(\phi)$, living on a plane. In the SMS, the 
effective potential depends on two real scalar fields: the singlet $S$ and the real part of the neutral Higgs component $h$. In this case the potential energy surface is a two-dimensional region in $\mathbb{R}^3$, and can be visualized as we do with the figures \ref{fig:valleys0}--\ref{fig:valleysm}.

Following \cite{Burgess:2012dz}, a multi-field potential $V$ is said to have a valley if there exists a curve in field space along which the derivative of $V$ in the normal direction to the curve is always zero, and the second derivative in that direction is positive.\footnote{The condition \eq{vvsridge}  is needed to distinguish between valleys and ridges, which also satisfy the condition \eq{eq:valleydef}.} This means that the following two conditions have to be satisfied:
\begin{align}
\label{eq:valleydef}
 n^i V_{,i}=0\,,
\end{align}
and
\begin{align} \label{vvsridge}
m_\perp^2\equiv n^i n^j\, V_{,ij}>0\,,
\end{align} 
where we adopt the convention of summing over repeated indices and use commas to denote derivatives in field space, i.e.\  $V_{,i}=\partial V/\partial\phi_i$. In these expressions, and in the rest of the paper, we use the notation $n^i$ to indicate the components of the unitary normal vector to a curve in field space. 

We define the ``bottom of a valley'' (or ``valley's floor'') as  the curve on the hypersurface of the potential that is obtained by evaluating $V$ along a solution to \eqref{eq:valleydef}. We will normally use the word ``valley'' to refer to the region of the surface of the potential around the ``bottom of the  valley''. The physical motivation for requiring the vanishing of the normal derivative is that the potential is expected to trap the fields inside its valleys. However, notice that a solution to \eq{eq:valleydef} need not be as well a solution of the equations of motion. If the fields happen to follow precisely the valley's floor, a one-dimensional description to the background dynamics (as discussed in section~\ref{subsec:One-dimensionalapprox}) is possible. Similarly, if the fields move sufficiently close to the valley bottom (the region we call ``valley'') a one-dimensional approximation may also be accurate enough in practice. We will soon comment on the conditions that have to be met for such an 
approximation to be valid. 

It is convenient to describe the bottom of any valley parametrically, using the length, $\sigma$, of its projection in field space. This projection can then be written as a (continuous and sufficiently differentiable) curve $\phi_i=\chi_i(\sigma)$. The bottom of the valley is the map of  $\chi_i(\sigma)$ onto the surface described by the potential. The simplest situation corresponds to the case in which $\chi_i(\sigma)$ is an equipotential curve. However, in general the valley will be tilted, with the potential varying along its bottom. This is generically the case for slow-roll dynamics along the valley's floor. With this parametric description it is straightforward to define the tangent and normal at each point of $\phi_i=\chi_i(\sigma)$. The tangent is simply given by the unitary vector in field space whose components in the basis of $\phi_i$ fields are
\begin{align} \label{vlong}
t^i=\frac{d\chi_i}{d\sigma}\,.
\end{align}
Then, the equation $t^i\,d t^i /d \sigma =0$ gives the orthonormal vector to the tangent direction:
\begin{align} \label{vnorm}
n^i = \kappa\, \frac{d t^i}{d\sigma}\,,
\end{align}
where $\kappa$ is the normalizing factor that ensures $n^i n^i =1$ and characterizes the curvature of the projection in field space of the floor of the valley. With these definitions, the equation \eq{eq:valleydef} means
\begin{align}
V_{,i}{\big |}_{\phi_i=\chi_i(\sigma)}\,\frac{d t_i}{d\sigma}=0\,.
\end{align}
If the valley is parallel to a certain field direction, say $\phi_j$, we will have ${d t_i}/{d\sigma}=0$ for all $i \neq j$, and the valley will be defined by the solution to the equation
\begin{align}
V_{,j}=0
\end{align}
plus the condition \eq{vvsridge}. In the SMS, focusing in the quadrant $\{h\geq 0$, $S\geq 0\}$ and in the limit of $\lambda_{SH}=0$, the two curves (actually, straight lines in field space)
\begin{align} \label{lsh0h}
\lambda\, h^2 +2\,{m^2_H} & =  0\,,\\ \label{lsh0s}
\quad \lambda_S\, S^2+6\,{m^2_S} & =  0
\end{align}
do track the bottom of two different valleys,  satisfying $\partial V/\partial h=0$ and $\partial V/\partial S=0$, respectively. 

If the Higgs portal coupling, $\lambda_{SH}$, is not zero, there are no valleys that are strictly parallel to any of the two field directions. In order to study the geometry of the potential \eq{Vvev} in this general situation, it is useful to start from 
the analogous to the curves \eq{lsh0h} and \eq{lsh0s}, but now with $\lambda_{SH}\neq0$, and check if deforming them appropriately we can obtain the floors of two actual valleys. So, we first define two lines of minima, to be referred as  {\it $h$-} and {\it $S$-lines}, satisfying
\begin{align}
\label{eq:lineh}\frac{\partial V}{\partial h}=0\Rightarrow &\, h=f_h(S)\,,\quad \text{$h$-line},\\
\label{eq:lines}\frac{\partial V}{\partial s}=0\Rightarrow &\, S=f_S(h)\,,\quad \text{$S$-line}\,.
\end{align}
We want to search for small deformations to each of these curves, such that that the equation \eq{eq:valleydef} is satisfied in each case. Since $\lambda_{SH}$ will be a small number in the actual examples that will be relevant for us, we can expect the deformations that are needed to obtain valleys to be small as well. This implies that the $h$-line and $S$-line for $\lambda_{SH}\neq 0$ will generically be small modifications  of the lines \eq{lsh0h} and \eq{lsh0s}, respectively. Since \eq{lsh0h} and \eq{lsh0s} correspond to actual valleys when $\lambda_{SH}$ is zero, we also expect the $S$-line and $h$-line for $\lambda_{SH}\neq 0$ to lie close to the projections on the plane $\{h$, $S\}$ of the bottoms of true valleys, at least for some range of field values. Finally, since the valleys for $\lambda_{SH}=0$ are defined by straight lines in field space, see \eq{lsh0h} and \eq{lsh0s}, we expect that the deformations leading to valleys for $\lambda_{SH}\neq 0$ will be nearly perpendicular to those lines, 
wherever these approximations are valid.

Let us consider, for instance, the $h$-line for $\lambda_{SH}\neq 0$, as defined by  \eq{eq:lineh}. As we just explained, we can assume that this line lies close to the projection of an actual valley's bottom on the plane $\{h$, $S\}$, such that the deformation that distinguishes the two is small. Then, the directional derivative along the normal to the line, 
\begin{align} \label{exps}
n^iV_{,i}{\big |}_{h=f_h(S)}\,,
\end{align} 
which is non zero, can be expanded in a Taylor series in terms of quantities defined at the projection of the actual valley's floor, assuming that the normal to the line is approximately the same as the one to the projection of the bottom of the valley. In other words, we express the derivative of the potential along the line as $V_{,i}(line) \simeq V_{,i}(valley)-n^jV_{,ij}(valley)\delta_n$. Contracting this with $n^i$ and taking into account that, by definition, the sum $n^iV_{,i}$ vanishes on the valley, we get
\begin{align}  \label{eq:hdeviations}
n^iV_{,i}{\big |}_{h=f_h(S)}\simeq -m_\perp^2 \delta_h\,,
\end{align}
where $m_\perp^2$ was defined in \eqref{vvsridge}. The shift $\delta_n\simeq\delta_h$ denotes the normal deformation that brings the valley into the line. Since for $\lambda_{SH}=0$ the $h$-line is approximately parallel to the $S$-direction, the normal goes approximately along the field $h$ where the above expansion works. Proceeding analogously for the $S$-line we arrive to 
\begin{align}
\label{eq:sdeviations}
 n^i V_{,i}{\big |}_{S=f_S(h)}
 \simeq -m_\perp^2 \delta_S\,,
\end{align}
where now the normal is approximately parallel to the $S$-direction. Notice that the value of $m_\perp^2$ has to be evaluated in each case along the corresponding line.

The equations \eqref{eq:hdeviations} and \eqref{eq:sdeviations} make clear that the displacements $\delta_h$ and $\delta_S$  become small in the limit of large orthogonal mass $m_\perp^2$ and if the normal stays closely parallel to the $h$-axis, in the case of the $h$-line, or the $S$-axis in the case of the $S$-line. Knowing the curves $f_h$ and $f_S$, the previous equations allow estimations of the displacement with respect to the true valleys. When the Taylor expansion around the valley bottom would fail, one may still estimate the deviations by finding numerical solutions using the full tree-level potential. We will now study in turn the $S$- and $h$- lines of minima for $\lambda_{SH}\neq 0$, and their connection to actual valleys. These will be important to obtain the inflationary potentials and for the study of stability at high field values. 

\subsection{Line of \texorpdfstring{$S$}{S} minima}
In section~\ref{sec:SMS} it was explained that the SM potential can be viewed as the potential of the SMS along a line which follows the minima of the heavy field S. This $S$-line therefore solves the equation $\partial V/\partial S=0$ and is described at tree-level by the curve
\begin{align} \label{sline}
 {\lambda_S}\,S^2+3\lambda_{SH}\, h^2 +6\, m^2_S=0\,,
\end{align}
which defines a conic section in field space.\footnote{Clearly, in the limit $\lambda_{SH}=0$ the line \eq{sline} collapses to the valley \eq{lsh0s}.} Substituting \eq{sline} back into \eqref{eq:match0} and its derivatives with respect to $h$, and using the appropriate expressions for the potentials at each side of the threshold, yields the matching relations of \eqref{eq:treematching} and \eqref{eq:treematching2}.\footnote{See appendix \ref{sec:SMmatching} for more details on the matching procedure, including radiative corrections.}
We can identify the $S$-line \eq{sline} with the bottom of a valley (as defined in the previous section) for values of $h^2$ that are sufficiently small, because in that limit the $S$-direction in field space becomes orthogonal to the line, and  \eqref{eq:valleydef} becomes equivalent to the equation for the $S$-line, $\partial V/\partial S=0$. In other words, we can integrate out the field $S$ when it defines a direction that is orthogonal to the line along which we reconstruct the low-energy potential. For instance, if $|\lambda_{SH}|\sim 1$ the identification with a valley is valid for values of $h^2$ smaller than roughly $|m^2_S|$. Then,  performing the matching at scales smaller than $|m_S^2|^{1/2}$ is consistent. 

\begin{figure}
\centering
\includegraphics[scale=1]{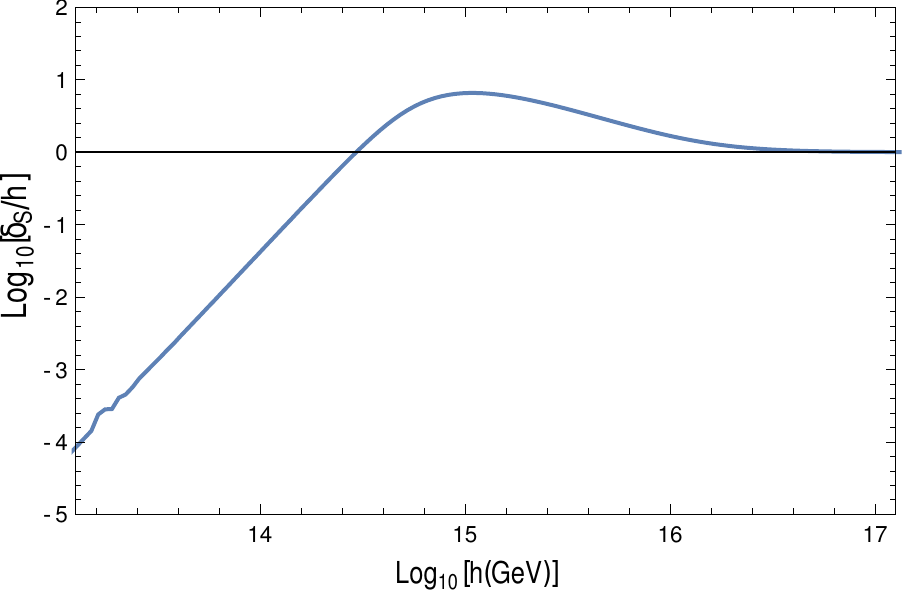}
\caption{\label{fig:sdeviations}\small Relative correction to the valley's location with respect to the $S$-line, parametrized in terms of $h$, for $m_t=171.7$ GeV, $\lambda_S=3.82\cdot10^{-13},\,\lambda_{SH}=3.67\cdot10^{-10},\, m^2_S=-1.06\cdot10^{26}\,{\rm GeV}^2$. The $S$-valley ends near $h=3\cdot10^{14}$ GeV.
  }
\end{figure}

In order to see this in more detail, and study how closely the $S$-line follows the bottom of an actual valley, we can resort to \eqref{eq:sdeviations}. From \eq{sline}, along the $S$-line we have
\begin{align}
 \frac{dS}{dh}=-\frac{3\lambda_{SH}}{\lambda_S}\frac{h}{S}\,.
\end{align}
It then follows that for very small $h$ the S-line is parallel to the $h$-axis, so that the l.h.s. of \eqref{eq:sdeviations} tends to zero and the $S$-line is a very good approximation to the projection of an actual valley floor. As $h$ grows, the $S$-line bends towards the $h$-axis, and its normal becomes increasingly parallel to it. This makes the curve get distorted with respect to the projection of the bottom of the true valley \eq{lsh0s}. For values of $h$ that are not too large, the deviation along the normal, $\delta_S$, can be estimated from \eqref{eq:sdeviations}. For very large values of $h$ the $S$-line will be nowhere near the projection of the bottom of the true valley. In other words, as $\delta_S$ increases the $S$-line stops being a good approximation to the actual valley. The reason why this happens is that the normal derivative becomes increasingly affected by the Higgs quartic coupling, so that the valley bends until it stops being a valley when the normal derivatives cannot become 
zero near the $S$-line; see the schematic representations of figures~\ref{fig:valleysp} and \ref{fig:valleysm}. The point at which the valley ends, and thus cannot trap fields any longer, can be estimated to be the one at which $\delta_S$ becomes of the order of the smallest of the fields, which for large VEVS of $S$ will typically be $h$. Solving for $\delta_S\sim h$ in \eqref{eq:sdeviations} yields, to lowest order in $\lambda_{SH}$,
\begin{align}
\label{eq:svalley_hmax}
h_{max}^{^6}\sim\frac{2^5\lambda_S}{3\tilde\lambda^2\lambda_{SH}^2}\left|m^2_S\right|^3.
\end{align}
This behavior is illustrated in figure~\ref{fig:sdeviations}, obtained by  solving \eqref{eq:valleydef} numerically, which clearly shows the growth of $\delta_S$ with $h$. 

Focusing in the region $h\geq0$, $s\geq0$, and assuming always $m^2_S<0$, we see that for positive $\lambda_{SH}$ the $S$-line of \eqref{sline} extends for $0\leq S^2\leq-{6m^2_S}/{\lambda_S}$ and $-{2m^2_S}/{\lambda_{SH}}\geq h^2 \geq 0$. Conversely, for $\lambda_{SH}<0$ the line runs for all $S^2\geq-{6m^2_S}/{\lambda_S}$, never touching the $h$ axis. Parametrizing the $S$-line in terms of $h$ yields the SM potential
\begin{align}
\label{eq:svalleyh}
 V_S(h)=\frac{\tilde\lambda}{8}\left(h	^2-v_{h}^2\right)^2\,,
\end{align}
where, $v_h$, the VEV of $h$, is given in \eq{veVs}. 
Using \eq{sline}, we can equivalently write \eq{eq:svalleyh}, along the line, as a function of the heavy singlet $S$ 
\begin{align}
\label{eq:svalley}
 V_S(S)=\frac{\tilde\lambda}{2}\,\left(\frac{\lambda_S}{6\, \lambda_{SH}}\right)^2\left(S^2-v_S^2\right)^2\,,
\end{align}
where $v_S$, the VEV of $S$, is also given in \eq{veVs}. The equations \eqref{eq:svalleyh} and \eqref{eq:svalley} correspond to the dashed-red lines in figures~\ref{fig:valleysp} and \ref{fig:valleysm} along the $S=0$ and $h=0$ planes, respectively. Note that the projection of the potential along the $S$-line onto the $h=0$ plane for $\lambda_{SH}>0$ gives the decreasing side of the Mexican hat potential of \eqref{eq:svalley}, while for $\lambda_{SH}<0$ it gives the (steeper) increasing part. 

In section \ref{sec:inflation} we will study the general inflationary dynamics of Mexican hat potentials, such as those of equations \eq{eq:svalleyh} and \eq{eq:svalley}. We will see that in order to produce successful inflation, the VEV has to be larger than the Planck scale, and inflation should proceed from smaller values of the field, towards the minimum. In addition, we will see that the quartic coupling of a generic inflationary Mexican hat (which is determined by the amplitude of the primordial perturbations) has to take a very small value, many orders of magnitude below unity. This prohibits successful inflation with the SM tree-level potential of \eq{eq:svalleyh}. 

We can consider two limiting dynamical regimes for inflation along the $S$-valley. In the first limit, for fields near the vacuum or when $\lambda_{SH}$ is  small enough, the valley is essentially parallel to the $h$-axis, so that $h$ is the relevant field for a one-dimensional description of the dynamics.\footnote{See section~\ref{subsec:One-dimensionalapprox} for more details about the one-dimensional approximation.} The potential along the $S$-line as a function of $h$ is approximately the SM potential \eqref{eq:svalleyh}, which as we just mentioned, cannot support successful inflation. The other limiting regime corresponds to the fields being far from the vacuum. Then, for $\lambda_{SH}\neq0$ the $S$-line becomes increasingly parallel to the $S$-axis. In this case the relevant dynamics is captured by $S$ instead of $h$, and the potential is given by \eqref{eq:svalley}. It would seem that inflation may work in this case, since a large VEV and small quartic coupling are possible. However, as explained 
before, far enough from the vacuum the $S$-line stops describing the bottom of a valley in which fields can be trapped, and thus the one-dimensional approximation of the dynamics is not justified and one cannot talk about inflation along the valley. When $|\lambda_{SH}|$ is small, so that  the extension of the $S$-valley increases, see \eqref{eq:svalley_hmax}, the first limiting case is recovered. Therefore, we conclude that there are important obstructions to achieve inflation within the $S$-valley, and so our attention will turn into the line of $h$-minima. As will be seen next, this closely describes a valley in a region of field space where successful inflation may be achieved.

\subsection{\label{subsec:hline}Line of \texorpdfstring{$h$}{h} minima}
We consider now the $h$-line, satisfying \eqref{eq:lineh}, i.e.\ $\partial V/\partial h =0$. 
The tree-level solution is
\begin{align} \label{htree}
\lambda\, h^2 +\lambda_{SH}\, S^2+2 m_H^2=0\,.
\end{align}
For $\lambda_{SH}>0$, and in the quadrant $h\geq0,S\geq0$, the $h$-line extends through the region limited by $0\leq S^2 \leq -2m_H^2/\lambda_{SH}$ and ${-2 m^2_H}/{\lambda}\geq h^2 \geq 0$. Note that in this case the vacuum in the $h$-direction for $S=0$ can correspond to a value of the field much larger than the Higgs VEV, since $|m^2_H|$ in the SMS will typically be of the order of $|m^2_S|\gg |\tilde  m^2_H|$, as follows from the SM matching condition \eqref{eq:treematching}. In the case $\lambda_{SH}<0$ the valley extends for $S\geq -2m_H^2/\lambda_{SH}$, never touching the $h$ axis, see figure\ \ref{fig:valleysm}.

It is straightforward to check that the resulting potential for the $h$-line can be written in terms of the one along the $S$-line, as a function of either $S$ or $h$:
\begin{align}
\label{eq:Vhvalley}V_h(h)=\left(1+\tilde\lambda\,\frac{\lambda_S}{3\lambda_{SH}^2}\right)V_S(h)\,,\quad
V_h(S)=\left(1+\tilde\lambda\,\frac{\lambda_S}{3\lambda_{SH}^2}\right)^{-1}V_S(S)\,,
\end{align}
where $V_S(h)$ and $V_S(S)$ are given by  \eq{eq:svalleyh} and \eq{eq:svalley}, respectively. 

As in the case of the $S$-line, the projections of the potential along the $h$-line onto the coordinate planes (see figures\ \ref{fig:valleysp} and \ref{fig:valleysm})  give Mexican hat potentials, which should be obvious from \eq{Vvev}. These projections are proportional to their $S$-line counterparts (at tree-level), as it is clear from the prefactors of \eqref{eq:Vhvalley}. They are represented as dotted blue lines in figures~\ref{fig:valleysp} and \ref{fig:valleysm}. The projection of the $h$-line on the vertical plane with $h=0$ is shallower than its $S$-line counterpart, whereas the projection on the plane with $S=0$ grows faster for large $h$, where the quartic coupling dominates. The figures also illustrate the fact that the potential along the $h$-line as a function of $S$ is given by the hilltop-like part of the Mexican hat for $\lambda_{SH}>0$, and by the steeper quartic-like (at sufficiently large $S$)  for $\lambda_{SH}<0$.

\begin{figure}
\centering
\includegraphics[scale=1]{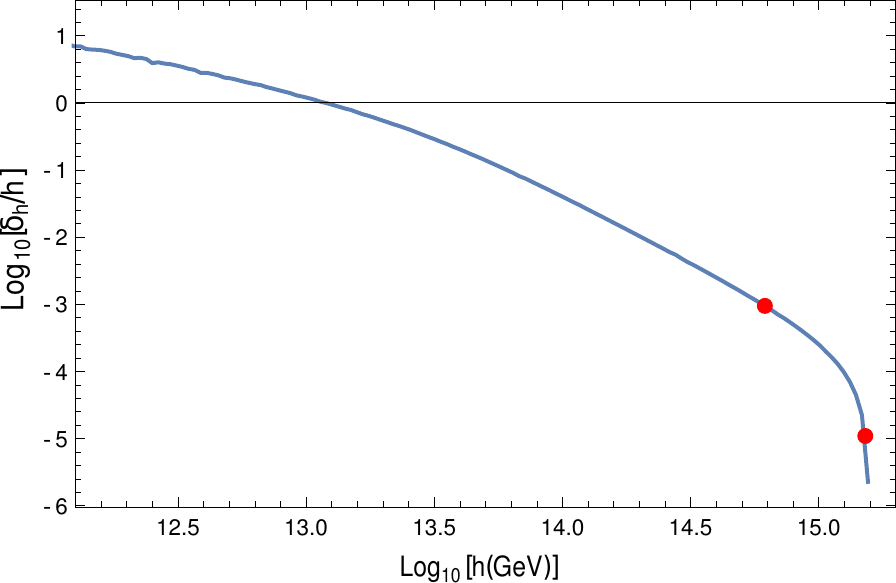} 
\caption{\label{fig:hdeviations} \small Relative correction to the valley's location with respect to the $h$-line, parametrized in terms of $h$, for $m_t=171.7$ GeV, $\lambda_S=3.82\cdot10^{-13},\,\lambda_{SH}=3.67\cdot10^{-10},\, m^2_S=-1.06\cdot10^{26}\,{\rm GeV}^2$. The rightmost red point marks where the observed values of $A_s$ and $n_s$ are obtained (which happens $59$ e-folds before the end of inflation, with $r=0.04$), while the left point marks the end of inflation. The corrections to the valley's location were estimated with the tree-level potential, while the cosmological parameters were calculated with the RG-improved effective potential.}
\end{figure}

In order to check whether the $h$-line corresponds to the bottom of an actual valley, we can use the equation \eqref{eq:hdeviations} giving the deviation of the line with respect to the projection in field space of the actual valley's floor. Along the $h$-line
\begin{align}
\frac{dS}{dh}=-\frac{\lambda}{\lambda_{SH}}\frac{h}{S}\,,
\end{align}
so that for large $h$ the normal to the trajectory becomes parallel to the $h$ axis. Therefore the l.h.s.\ of equation \eqref{eq:hdeviations} tends to zero, and so does the deviation $\delta_h$. Thus, the $h$-line is a very good approximation to the bottom of a valley for sufficiently large values of $h$, as shown in figure~\ref{fig:hdeviations}. For this reason we will often talk about the ``$h$-valley'' when we will describe inflation for large Higgs values. 

Since for large $h$ the line becomes increasingly parallel to the $S$-direction, the field relevant for the dynamics along the bottom of the valley is approximately given by $S$, with a Mexican hat potential given by \eqref{eq:Vhvalley} and \eqref{eq:svalley}. The potential  can have a small quartic and large VEV. Therefore, it can support inflation, which will be shown to satisfy all the CMB constraints, following from the analysis of inflation along Mexican hat potentials in section~\ref{subsec:inflationtree}.  In the example of figure~\ref{fig:hdeviations}, considering the one-dimensional rolling along the $h$-line (see section~\ref{subsec:One-dimensionalapprox} for details) successful inflation can be obtained starting around the red point on the right and ending at the red point on the left. In between these points the $h$-line remains a very good approximation to the bottom of a valley, which, as we just mentioned, we will call $h$-valley. Moreover, the corrections to the one-dimensional approximation of the 
rolling dynamics remain small, as will be seen in section~\ref{subsec:One-dimensionalapprox}.

To summarize this section, we have seen that the potential of the SMS has valleys which in certain regions are well approximated by the regions around the  lines of $h$- and $S$- minima. For $\lambda_{SH}=0$ these lines and the projections of the valley's floors coincide. In this (decoupling) limit, successful inflation can happen along the $h$-valley, whereas it is forbidden along the $S$-line (which gives the SM potential). If a non-zero Higgs portal coupling is present, the valleys are deformed. This causes the $S$-valley to have a limited range away from the vacuum, which does not ameliorate its prospects for supporting inflation. On the other hand, the  $h$-valley can still support inflation while reaching out to large values of $h$, being thus sensitive to Higgs quantum corrections, as we will later explore in detail.

\section{\label{sec:inflation}Tree-level inflationary dynamics}

We consider now the (tree-level) dynamics of the fields rolling down the valleys described earlier, with the aim of determining whether they can lead to successful inflation satisfying the current experimental and observational constraints. 

\subsection{\label{subsec:One-dimensionalapprox}One-dimensional approximation}

If a valley is sufficiently straight and if the derivatives of the potential along the projection of its floor in field space are small compared to the ones in the orthogonal direction, we can simplify the dynamics into a one-dimensional problem \cite{Burgess:2012dz}. 

Consider the dynamics of two real scalar fields\footnote{For the time being, these two fields are completely general, but we will later particularize them to the heavy singlet, $S$, and the real part of the neutral Higgs, $h$, that are relevant for the potential of the SMS.} with standard kinetic terms and coupled by a potential $V$
\begin{align} \label{L1}
 {\cal L}=\frac{1}{2}\partial_\mu \phi_1 \partial^\mu \phi_1+\frac{1}{2}\partial_\mu \phi_2 \partial^\mu \phi_2-V(\phi_1,\phi_2).
\end{align}
Let us assume that the potential has a valley, as defined in section~\ref{sec:wiav}. We recall that a valley is defined by the region around a curve in field space for which the first derivative of the potential along the orthogonal direction to the curve vanishes at every point, see \eq{eq:valleydef}. In addition, we require the mass along the orthogonal direction to be positive, see \eq{vvsridge}. In general, the mass matrix at any point of the projection of the bottom of the valley in field space can be written as
\begin{align} \label{massmax}
 V_{,ij}=\frac{d^2V}{d\sigma^2}\,t^i\,t^j+\kappa^{-1}\frac{dV}{d\sigma}\left(n^i\,t^j+n^j\,t^i\right)+m_\perp^2\, n^i\,n^j\,,
\end{align}
where the tangent and normal unit vectors were defined in \eq{vlong} and \eq{vnorm}. This matrix is symmetric and therefore has two eigenvalues along mutually orthogonal directions. In general, these two directions do not correspond to the directions defined by the tangent and normal field space vectors $t^i$ and $n^i$, as \eq{massmax} shows explicitly. In other words, the tangent and the normal are not mass eigenstates. The mass along the normal direction, $m^2_\perp$, was already defined in \eq{vvsridge}. The mass along the longitudinal direction can be read directly from \eq{massmax} and is simply given by the derivative with respect to length of the projection of the bottom of the valley:
\begin{align}
m^2_\parallel=\frac{d^2V}{d\sigma^2}\,.
\end{align}
The straightness of the projection of the  valley's floor in field space is controlled by a curvature parameter, $\kappa$, defined through the differential relation between \eq{vlong} and \eq{vnorm}. In the limit in which $d t^i/d\sigma$ vanishes,  the projection of the bottom of the valley in field space is a straight line and $\kappa$ goes to infinity. In this limit, there is no mixing between the normal and longitudinal directions in the mass matrix \eq{massmax}. This implies that for large enough $\kappa$, the mass eigenvalues correspond approximately to the masses $m_\parallel^2$ and $m_\perp^2$ in the directions parallel and orthogonal to the projection of the valley's floor. If, in addition, the bottom of the valley slopes down gently (thus, favouring a slow-roll trajectory) we will have that $m_\parallel^2 \ll m_\perp^2$\,, so we can identify the longitudinal and transverse directions with light and heavy degrees of freedom, respectively.  

If the conditions we have just described are met, it is convenient to make a change of basis in field space $(\phi_1,\phi_2)\rightarrow(\sigma,\phi_\perp)$, where $\phi_\perp$ parametrizes the direction perpendicular to the projection of the valley's bottom and $\sigma$ is the length travelled along it. This field redefinition generically induces a kinetic mixing between $\sigma$ and $\phi_\perp$. However, if the curvature of the projection of the valley is small and changes slowly, it is a good approximation to neglect the mixing, so that the Lagrangian of the system can be approximated by
\begin{align} \label{L2}
 {\cal L}=\frac{1}{2}\partial_\mu \sigma \partial^\mu \sigma+\frac{1}{2}\partial_\mu\phi_\perp \partial^\mu \phi_\perp-V(\phi_1(\sigma,\phi_\perp),\phi_2(\sigma,\phi_\perp))\,.
\end{align}
As we just discussed, if the valley is to allow for slow-roll, denoting with primes the derivatives with respect to the valley length, we have that $m_\parallel^2 \ll m_\perp^2$ and the field $\phi_\perp$ can be integrated out of the dynamics. This allows to reduce the problem to the rolling of a single field (that parametrizes the length travelled along the valley line) in an (effective) potential that measures the potential energy along the projection of the bottom of the valley in field space. This approximation is valid for large values of $m^2_\perp$ and $\kappa$, as it was shown in reference \cite{Burgess:2012dz}, where the first corrections to the one-dimensional effective theory were computed.  The masses $M_{-}^2$ and $M_{+}^2$ of the light and heavy eigenstates of \eq{massmax} can be  approximately written as $M_{-}^2 \simeq m^2_\parallel+\Delta\mu_\parallel^2$ and $M_{+}^2 \simeq m^2_\perp-\Delta\mu_\parallel^2$, where
\begin{align} \label{eq:valleycorrections1}
\Delta\mu^2_\parallel =-\frac{1}{\kappa^{2} m_\perp^{2}}\left(\frac{dV}{d\sigma}\right)^2\,.
\end{align}
Once the heavy field is integrated out, the effective potential for the (remaining) light field can be written as an expansion in $\sigma$, under the approximation that $\sigma$ describes the light eigenstate. This approximation is expected to work well provided that $\Delta\mu^2_\parallel$ is sufficiently small. Besides, the leading order correction in $1/\kappa$ and $1/m^2_\perp$ to quartic coupling of the potential of the light field is given by
\begin{align}  \label{eq:valleycorrections2}
\Delta\lambda^{}_\parallel\simeq-\frac{3}{\kappa^3}\frac{d \kappa}{d\sigma}\frac{d V}{d\sigma}\,.
\end{align}
Similarly, we could also compute the linear, cubic and subsequent corrections to the potential of the light field at higher orders in powers of $\sigma$. However, we focus here on the quadratic and quartic pieces because the tree-level potential \eq{Vvev} is an even function of the fields, leading to Mexican hat potentials along the lines of minima that we described in section~\ref{sec:valleystree}.

\begin{figure}
 \centering
\includegraphics[scale=.84]{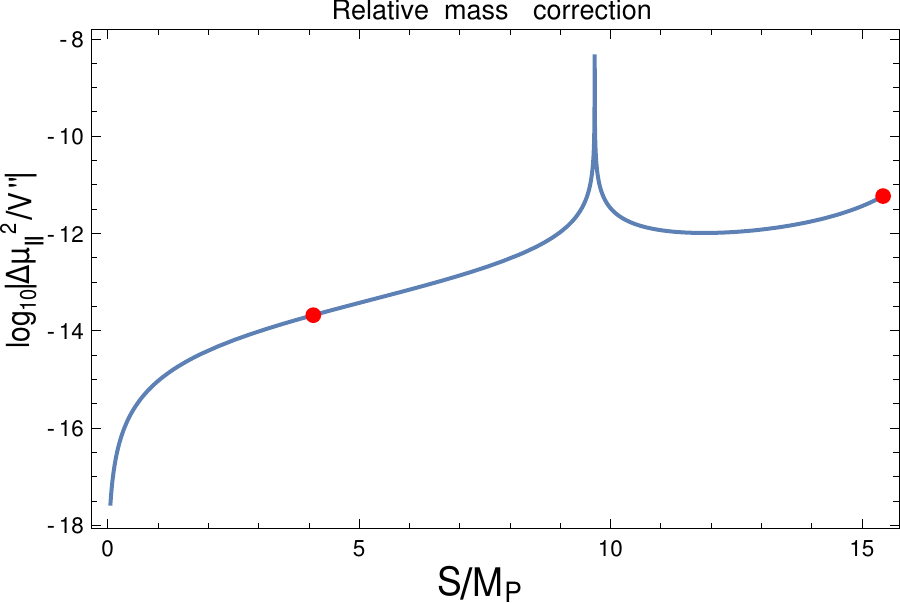}
\includegraphics[scale=.84]{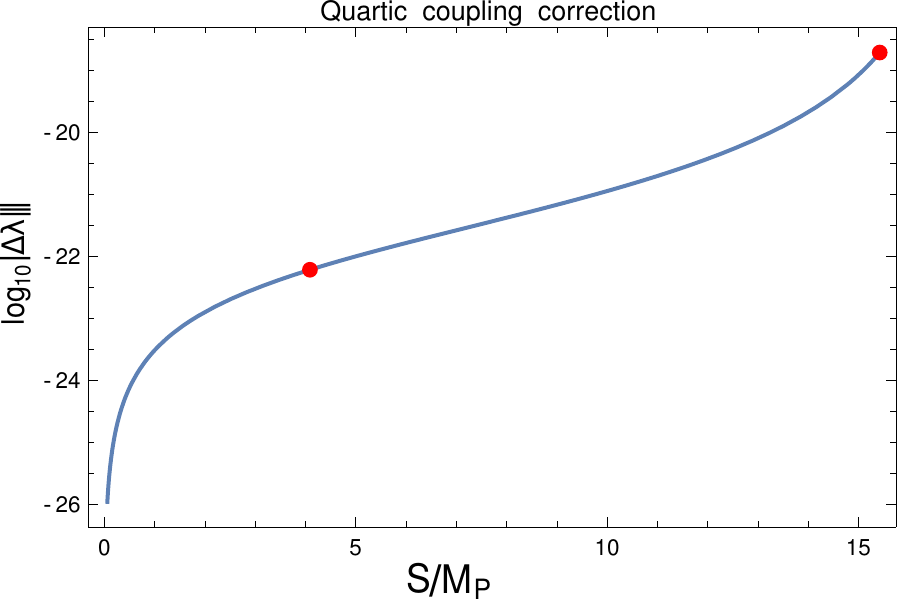}
 \caption{\label{fig:correctionsh} \small Corrections beyond the leading one-dimensional approximation to the mass parameter (left) and to the quartic coupling (right), in terms of the field $S$ along the $h$-line for the same scenario as in figure~\ref{fig:hdeviations}, with $m_t=171.7$ GeV, $\lambda_S=3.82\cdot10^{-13},\,\lambda_{SH}=3.67\cdot10^{-10}$, and $m^2_S=-1.06\cdot10^{26}\,{\rm GeV}^2$. The left red points mark the beginning of observable inflation, and the right points mark the end of inflation. The corrections to the mass and quartic parameters were estimated with the tree-level potential, while the cosmological parameters were calculated with the RG-improved effective potential. Notice that the corrections to the quartic coupling are much smaller than $\lambda_S$.}
\end{figure}

The equations \eqref{eq:valleycorrections1} and \eqref{eq:valleycorrections2} can be used, together with the tree-level formulae of the previous sections, to estimate the validity of the one-dimensional approximation for inflation along the $h$-valley in the SMS.  Doing so, we find that the approximation works with high accuracy, as the corrections to the couplings \eqref{eq:valleycorrections1} and \eqref{eq:valleycorrections2} are many orders of magnitude below the values obtained by simply considering the potential along the projection of the bottom of the valley as a function of the length $\sigma$. figure~\ref{fig:correctionsh}  shows the corrections evaluated along the $h$-line (which, as shown in section~\ref{subsec:hline}, is a good approximation to the projection of the valley's floor for large $h$) at tree-level, for a concrete choice of parameters which gives successful inflation. The peak in the size of the relative mass correction happens when the valley potential crosses an inflection point, so that $V''=0$.
 Away from this peak the relative corrections are very strongly suppressed.

\subsection{\label{subsec:slowapprox} Slow-roll approximation} 

In the one-dimensional and slow-roll approximation, we compute the primordial spectra produced during inflation in terms of the first three slow-roll (potential) parameters $\epsilon$,  $\eta$ and $\xi$, defined as
\begin{align}
\label{eq:one-dimensionalpars}
\epsilon =\frac{M_P^2}{2}\left(\frac{V'}{V}\right)^2\,,\quad \eta =M_P^2\frac{V''}{V}\,,\quad \xi = M_P^4\frac{V'V'''}{V^2}\,,
\end{align}
where $M_P=1/\sqrt{8\pi\,G}\simeq2.435 \cdot10^{18}$ GeV is the reduced Planck mass. In these expressions the potential is understood to be evaluated along the projection of the bottom of a valley in field space and, for simplicity, the primes denote derivatives with respect to the field $\sigma$, which parametrizes the valley's length. If the orthogonal corrections to the dynamics and the primordial spectra were not negligible, we would need a two-field description and similar parameters for the orthogonal direction as well, see e.g.\ \cite{Achucarro:2015bra}. However, as we explained in the previous section, the one-dimensional approximation turns out to be excellent along the $h$-valley.

The amplitude of scalar perturbations, $A_s$, the scalar spectral index, $n_s$, and its running, $\alpha$, are then given by 
\begin{align}
\label{eq:Asnsr}
 A_s\simeq \frac{V}{24\pi^2\,M_P^4\,\epsilon}\,,\quad n_s\simeq 1+2\eta-6\epsilon\,,\quad \alpha\simeq-2\xi+16\eta\epsilon-24\epsilon^2\,. 
\end{align}
Using them, the scalar primordial spectrum can be expressed, as usual, as 
\begin{align} \label{Ps}
P_s(k)=  A_s\left(\frac{k}{k_*}\right)^{n_s-1+\frac{\alpha}{2}\ln\frac{k}{k_*}+\cdots}\,,
\end{align}
where $k_*$ is an arbitrary reference scale that is often taken to be $k_*=0.05$ Mpc$^{-1}$, as in the most recent Planck study on inflation \cite{Ade:2015lrj}. In order to express the amplitude of the tensor power spectrum, $A_t$, this is conveniently related to the scalar one via the tensor-scalar ratio $r=A_t/A_s$, which in the slow-roll approximation is simply given by
\begin{align}
\label{eq:reps}
r\simeq 16 \epsilon\,. 
\end{align}
The scale dependence of the tensor spectrum in the slow-roll approximation in single-field inflation is essentially determined by $r$ through the so-called consistency relation, which says that the tensor index, $n_t$, is equal to  $-r/8$. We can then write the tensor spectrum at first order in slow-roll as
\begin{align} \label{Pt}
P_t(k) \simeq  r A_s \left(\frac{k}{k_*}\right)^{-r/8}\,.
\end{align} 
The primordial parameters whose values we are mostly interested in reproducing are  $A_s$, $n_s$ and $r$. We have also checked that the running $\alpha$ is typically very small in the numerical examples that we find, and well within current constraints, as we discuss later in more detail. Considering higher order slow-roll parameters is unnecessary to describe accurately the (Mexican hat) potentials that we deal with below. 

The number of e-folds $N_e(t)$ produced from some initial time $t_i$ during a lapse $t-t_i$ is defined as the integral over time of the Hubble function\footnote{We use the non-standard notation ${\mathcal H}$ to distinguish the Hubble function from the Higgs doublet field $H$.} $\mathcal{H}(t)$
\begin{align}
N_e(t)=\int_{t_i}^t \mathcal{H} d\hat t\,,
\end{align}
where $\hat t$ simply denotes the integration variable. During inflation, $\mathcal{H}(t)= da/dt\equiv \dot a$ is approximately constant and $N_e$ gives a measurement of the nearly exponential growth of the scale factor of the universe, denoted by $a$. In the slow-roll approximation, the number of e-folds can be rewritten as the integral 
\begin{align}
\label{eq:efoldsslow}
N_e\simeq\frac{1}{M_P}\,\int_{\sigma}^{\sigma^i}\frac{d\hat \sigma}{\sqrt{2\epsilon}}\,,
\end{align}
where $\sigma^i=\sigma(t_i)$ and we have assumed that $\sigma$ decreases with time. If instead the field grew as time flows, the integral \eq{eq:efoldsslow} picks a minus sign. Both possibilities will be analysed in the models that we study below. In order to solve the horizon problem, and depending on the specific details of the reheating process, approximately $50\sim60$ e-folds are needed to solve the horizon problem, see e.g.\ \cite{Liddle:2003as}. Achieving a sufficient amount of inflation is therefore an essential constraint that has to be fulfilled by a model in order  to be deemed successful. The end point of inflation is defined by the breaking of the condition $\ddot a>0$, which means that the accelerated expansion stops. This condition can be expressed in terms of the Hubble slow-roll parameter $\epsilon_{\mathcal H}$, defined as 
\begin{align}
\epsilon_{\mathcal H}=\frac{3\dot\sigma^2}{\dot\sigma^2+2V}\,,
\end{align}
by requiring $\epsilon_{\mathcal H}=1$. The dynamical equations in the one-dimensional approximation are
\begin{align}
\ddot\sigma+3 \mathcal{H}\dot\sigma+V' =0\,,\\
3 M_P^2 {\mathcal H}^2=  \frac{\dot\sigma^2}{2}+V\,,
\end{align}
and, indeed, it is straightforward to see from them that the condition $\ddot a >0$ is equivalent to $\epsilon_{\mathcal H}<1$. The point at which $\epsilon_{\mathcal H}=1$ can be determined by solving the dynamics of the field $\sigma$  as a function of the number of e-efolds $N_e$ . This is given by \cite{Ballesteros:2014yva}:
\begin{align} \label{eq:efolds}
\frac{d^2\sigma}{dN_e^2}+3\,\frac{d\sigma}{dN_e}-\frac{1}{2M_P^2}\left(\frac{d\sigma}{dN_e}\right)^3+\left(3M_P-\frac{1}{2M_P}\left(\frac{d\sigma}{d N_e}\right)^{2}\right)\sqrt{2\epsilon}=0\,,
\end{align}
where $\epsilon$ is defined in \eq{eq:one-dimensionalpars}. In the slow-roll approximation $V\simeq 3M_P^2 {\mathcal H}$ and $V' \simeq -2{\mathcal H} \dot\sigma$ and the end of inflation can be approximately identified by $\epsilon \sim 1$ or $|\eta| \sim 1$, whichever occurs first. Using the potential slow-roll parameters \eq{eq:one-dimensionalpars} to this end and for the computation of $N_e$ from \eq{eq:efoldsslow} has the advantage of avoiding the numerical resolution of the equation \eq{eq:efolds}, but comes at the price of a (typically small) inaccuracy in the determination of the number of e-folds. In our analysis we will use both methods and compare them. For more details on the slow-roll approximation we refer the reader to the appendix  of \cite{Ballesteros:2014yva} and to  \cite{Liddle:1994dx}.

\subsection{Isocurvature perturbations}

In the previous discussion we assumed that perturbations in the direction orthogonal to the projection of the bottom of the valley do not contribute to the power spectra of scalar and tensor perturbations. However, in multi-field models isocurvature modes sourced by these orthogonal fluctuations may be relevant, and can affect the evolution of the adiabatic modes. As we will show now, isocurvature perturbations are suppressed in our case  if the classical trajectory corresponds to the bottom of the $h$-valley, and they also have a negligible contribution to the scalar curvature perturbation, which validates the previous analysis.
 
Following the notation of \cite{Gordon:2000hv}, we consider scalar perturbations of a FLRW metric defined by the general line element
\begin{align}
 ds^2=-(1+2A)dt^2+2a(t)\, \partial_i B\, dx^i dt+[a(t)]^2[(1-2\psi)\delta_{ij}+2\partial_{ij}E]dx^idx^j.
\end{align}
Given two fields with canonical kinetic terms, $\phi_1$ and $\phi_2$, their perturbations $\delta\phi_i$ give rise to the following density, momentum and pressure perturbations of the common energy-momentum tensor
\begin{align}
 \delta\rho=[\dot\phi_i(\delta\dot\phi_i-\dot\phi_i A)+V_{,i}\delta\phi_i],\,\,\,\partial_J\delta q=-\dot\phi_i\,\partial_J\delta\phi_{i},\,\,\,
 \delta p=[\dot\phi_i(\delta\dot\phi_i-\dot\phi_i A)-V_{,i}\delta\phi_i],
\end{align}
where we sum over repeated indices $i$, and the subscript $J$ refers to spatial coordinates. Out of these perturbations, we can construct standard gauge-invariant quantities such as the  comoving curvature perturbation $\cal R$, the Bardeen potential $\Psi$ and the total isocurvature (or entropy) perturbation $\cal S$:
\begin{align}
{\cal R}=\psi-\frac{\mathcal{H}}{\rho+p}\delta q,\quad \Psi=\psi+\mathcal{H}(a^2\dot E-aB),\quad {\cal S}=\mathcal{H}\left(\frac{\delta p}{\dot p}-\frac{\delta\rho}{\dot \rho}\right)\,,
\end{align}
where we denote with dots the derivatives with respect to proper time and the Hubble function is $\mathcal{H}=\dot a/a$ to distinguish it from the Higgs doublet $H$. 

The power spectrum of the standard adiabatic curvature perturbations $P_s(k)$ is defined in terms of the correlation function:
\begin{align}
 \langle {\cal R}_{\rm \bf k} {\cal R}_{\rm \bf k'}\rangle=\frac{(2\pi)^3}{k^3}\delta^{(3)}({\rm \bf k+k'})\,P_s(k).
\end{align}
Similarly, we can define a power spectrum of isocurvature perturbations from $\cal S$. Taking the background trajectory as the projection of a valley's floor onto the space of fields, and changing variables to the coordinates $\sigma$ and $\phi_\perp$ defined earlier, the above perturbations can be written as follows, after making use of the energy-momentum constraints \cite{Gordon:2000hv}:
\begin{align}
 {\cal R}=\psi+\frac{\mathcal H}{\dot{\sigma}}\delta\sigma\equiv\frac{\mathcal H}{\dot\sigma} Q_\sigma\,,\quad {\cal S}=-\frac{4 M^2_P\,k^2\,V_\sigma}{3\dot{\sigma}^2(3\mathcal H\dot{\sigma}+2V_{\sigma})a^2}\Psi-\frac{2V_\perp}{3\dot\sigma^2}\delta\phi_\perp.
\end{align}
In the previous formulae $Q_\sigma$ represents the usual gauge-invariant Sasaki-Mukhanov perturbation of the field $\sigma$, and $V_\sigma$ and $V_\perp$ denote the parallel and orthogonal directional derivatives of the potential,
\begin{align}
 V_\sigma\equiv t^i V_{,i},\quad V_\perp\equiv n^i V_{,i}.
\end{align}
The first immediate consequence is that, independently of how large the perturbations $\delta\phi_\perp$ in the orthogonal direction may become, if the background trajectory corresponds to the bottom of a valley (with $V_\perp=0$), there is no entropy perturbation at super-Hubble scales (i.e.\ for $k\ll a\mathcal H$). Indeed, for $V_\perp=0$ the equation for ${\cal S}$ becomes identical to that in the single-field case, in terms of the field $\sigma$. Therefore, entropy perturbations at super-Hubble scales can only be generated if the fields roll away from the bottom of the valley. If the valley is sufficiently flat along its length, it will be a slow-roll attractor, which will suppress isocurvature perturbations. 

The vanishing of large-scale isocurvature perturbation applies in the usual picture of inflation, in which the Higgs sits classically at $v=246$ GeV. In this case $\phi_\perp=h$, and since $h$ sits at its minimum, $V_h=0$ and no entropy perturbations will be generated to leading order at large scales. 

The next question is whether $\cal R$ receives contributions from the orthogonal field $\phi_\perp$. To address this, we study the
equation for $Q_\sigma$, which is \cite{Gordon:2000hv}:
\begin{align}
 \ddot Q_\sigma+3 \mathcal{H}\dot Q_\sigma+\left[\frac{k^2}{a^2}+V_{\sigma\sigma}-\dot\theta^2-\frac{1}{a^3 M^2_P}\left(\frac{a^3\dot\sigma^2}{\mathcal{H}}\right)^{\cdot}\,\right]Q_\sigma=2(\dot\theta\delta s)^\cdot-2\left(
 \frac{V_\sigma}{\dot\sigma}+\frac{\dot{\mathcal{H}}}{\mathcal H}\right)\dot\theta\delta\phi_\perp.
\end{align}
In this equation, $V_{\sigma\sigma}=t^i t^i V_{,ij}$, while $\dot\theta$, which represents the rate of change of the angle between the tangent to the background trajectory and a reference direction,
is related to the trajectory's curvature $\kappa$ (defined in \eqref{vnorm}) as follows:
\begin{align}
 \dot\theta=\frac{\dot \sigma}{\kappa}.
\end{align}
In the limit $\dot\theta\rightarrow0$, the equation for  $Q_\sigma$ becomes identical to the corresponding one in the single-field case, which allows to conclude that perturbations in field directions orthogonal to the valley do not source curvature perturbations if the valley is straight. This is of course the case of inflation with the Higgs sitting at $h=246$ GeV, in which $\delta h$ perturbations do not feed $\mathcal{R}$. 

\begin{figure}
\centering
\includegraphics[scale=1]{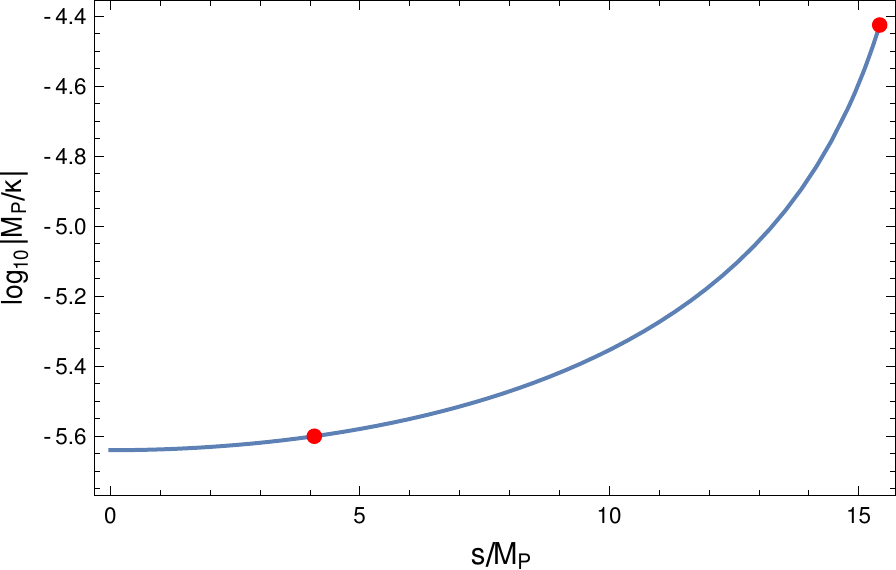}
\caption{\label{fig:Mp_kappa}$\dot\theta_{max}/\mathcal{H}=M_P/\kappa$ along the $h$-line, for the same scenario as in figures~\ref{fig:hdeviations} and \ref{fig:correctionsh}, with $m_t=171.7$ GeV, $\lambda_S=3.82\cdot10^{-13},\,\lambda_{SH}=3.67\cdot10^{-10},\, m^2_S=-1.06\cdot10^{26}\,{\rm GeV}^2$. The red points mark the beginning of observable inflation (left) and the end of inflation (right).
  }
\end{figure}

In \cite{Bartolo:2001rt} the equation for $Q_\sigma$ was solved in a slow-roll expansion, and it was shown that the solution deviates from the single-field case by factors proportional to $\dot\theta/\mathcal{H}$. We can estimate this ratio in the $h$-valley
by determining $\kappa$ from the equations of the $h$-line and using the slow-roll equations to evaluate the time derivative of $\sigma$. We get
\begin{align}
 \frac{\dot\theta^2}{\mathcal{H}^2}=\frac{2\epsilon V^2}{9 \mathcal{H}^2 \kappa^2 M^2_P}.
\end{align}
Since $\epsilon<1$ during inflation (at least in the slow-roll approximation), an upper bound is obtained by setting $\epsilon=1$. Applying Friedmann's equations this gives
\begin{align}
 \frac{\dot\theta^2}{\mathcal{H}^2}\leq \frac{M^2_P}{\kappa^2}.
\end{align}
To lowest order in $\lambda_{SH}$
\begin{align}
 \frac{M^2_P}{\kappa^2}\sim-\frac{\lambda_{SH}\lambda\lambda_S}{\tilde\lambda^2}\frac{M^2_P}{6m^2_S+\lambda_S S^2}\sim-\frac{\lambda_{SH}\lambda_S}{6\tilde\lambda}\frac{M^2_P}{m^2_S}\,.
\end{align}
This will be suppressed along the $h$-valley during inflation (with $S^2< -6 m^2_S/\lambda_S$) for $\lambda_{SH}\lambda\lambda_S\ll\tilde\lambda^2$. As will be seen in section~\ref{subsec:inflationtree}, inflation will typically require $\lambda_S\sim 10^{-13}, m^2_S\sim -10^{26}\,{\rm GeV}^2$. Taking $\tilde\lambda\sim 0.27$ as required by the Higgs mass and $\lambda_{SH}\sim 10^{-10}$, this gives a strongly suppressed $\dot\theta^2/\mathcal{H}^2\sim 10^{-13}$ for small values of $S$.  figure~\ref{fig:Mp_kappa} shows the value of $\dot\theta/M_P$ along a realistic inflationary valley, in terms of the field $S$, again showing a strong suppression. Finally, we note that when the large-scale isocurvature perturbations are negligible, and the inflationary trajectories sufficiently straight, as in the models analyzed here, non-Gaussianities will also be extremely small, see e.g.\ \cite{Rigopoulos:2005us}.

\subsection{\label{subsec:inflationtree}Tree-level dynamics}

In section~\ref{sec:valleystree} we concluded that in the presence of a non-zero Higgs portal coupling, $\lambda_{SH}$, the lines of $h$ and $S$ minima closely describe the bottom of valleys in the potential for large and small values of $h$, respectively. We recall that the potential along the $S$-line reproduces the SM potential, which predicts primordial curvature perturbations that are too large. In fact, we argued that the valley that runs close to the $S$-line cannot support inflation in the limit $\lambda_{SH}=0$ (in which line and valley match) and is also not likely to do so for $\lambda_{SH}\neq0$. The region around that valley where the dynamics would be dominated by the field $S$ (being then less sensitive to the Higgs' couplings, potentially allowing inflation) is precisely where the $S$-valley tends to disappear. For this reason, we will focus below on the possibility of inflation along the $h$-valley. 

For non zero $\lambda_{SH}$ the bottom of the $h$-valley is accurately described by the $h$-line for large values of $h$. This region of field space is also where we expect inflation to be possible, since the valley becomes increasingly parallel to the $S$-axis as $h$ grows and, contrary to the SM case, the rolling along the potential will not be constrained by the Higgs VEV and quartic coupling. In section~\ref{subsec:One-dimensionalapprox} we argued that the one-dimensional approximation works well for the $h$-valley. We saw that the relevant field for the dynamics near the valley's bottom is the length travelled along the valley, which in turn can be well approximated by the singlet $S$. In this section we use this approximation to prove the viability of inflation in the SMS and obtain its theoretical predictions. In section~\ref{subsec:inflationloop} we will compute the length travelled along the valley including loop corrections, in order to complement these results and cross-check the validity of the 
approximations used in this section.

The rolling  along the $h$-valley can be approximated by the dynamics of a field in a Mexican hat potential, which we proceed to study next using a parametrization that captures more general situations, not necessarily tied to the SMS and its valleys. The relevant  potential is then  given by \eqref{eq:Vhvalley} and \eqref{eq:svalley} and can be written as
\begin{align} \label{Mhat}
V(S)=\vartheta\left(S^2-v_S^2\right)^2\,,
\end{align} 
where $\vartheta>0$ is the dimensionless coupling
\begin{align} \label{effcoup}
\vartheta=\frac{\lambda_S}{4!}\frac{\tilde\lambda}{\lambda}\,.
\end{align}
This type of potential, with positive $\vartheta$ and $v_S^2$, has been studied in the context of inflation in various works and is known to be capable of providing a good fit to Planck data, see e.g.\ \cite{Martin:2014vha}. The potential \eq{Mhat} was probably first studied in \cite{Albrecht:1984qt} as a specific implementation of slow-roll inflation (back then called as well {\it new inflation}). It was pointed out there that a phase of accelerated expansion occurs if the symmetry breaking scale $v_S$ is of the order of the Planck mass (or larger), see also \cite{Moss:1985wn,Hu:1986exa}. Later, it was also considered in \cite{Vilenkin:1994pv,Cormier:1998nt,Cormier:1999ia} and more recently in \cite{Kallosh:2007wm,Smith:2008pf,Rehman:2010es,Okada:2014lxa}. Here we will give a detailed analysis, including some remarks about the slow-roll approximation and reheating, and discuss the implications for the Standard Model of particle physics, extended with the singlet $S$. 

Due to the $Z_2$ symmetry of the potential \eq{Mhat}, we can focus exclusively on the region $S\geq 0$ without loss of generality. The possible inflationary dynamics can be separated in two cases that turn to give rather different predictions. The first one corresponds to $S<v_S$, with the inflaton rolling from smaller to larger values, and corresponds to a sort of ``hilltop'' model \cite{Boubekeur:2005zm}. The second case is $S>v_S$, with $\dot S <0$ and may behave as a (displaced) quartic or quadratic potential depending on the concrete values of $\vartheta$ and $v_S$ and the field range. We will study in turn the two cases. 

Both possibilities share a property that is useful to highlight now. Of the two parameters on which the potential depends, only $v_S$ determines the amount of inflation that is produced. Since $V$ is proportional to $\vartheta$, the dependence on this parameter factors out from any expression involving the potential slow-roll parameters, which are homogeneous functions of $V$ of degree zero, see \eq{eq:one-dimensionalpars}. Therefore, the coupling $\vartheta$ does not affect the prediction for the number of e-folds, as the \eq{eq:efolds} shows. It does not intervene either in  any of the primordial parameters that we have defined, except $A_s$, see \eq{eq:Asnsr}, and thus it can be fixed solely from this number. 

We will denote by $S_*$ the value of $S$ for which a total of $N_e$ inflationary e-folds are produced. Then,  given $v_S^2$ and $S_*$ such that the slow-roll parameters and $N_e$ take appropriate values, $\vartheta$  is determined by the amplitude of the scalar perturbations through the expression
\begin{align} \label{xifromAs}
\vartheta = 192\pi^2 A_s \frac{S_*^2 M_P^6}{\left(S_*^2-v_S^2\right)^4}\,.
\end{align}
The primordial scalar amplitude at $k_*=0.05$ Mpc$^{-1}$ is approximately\footnote{The precise central value and range depend on the concrete data set and assumptions on parameters \cite{Planck:2015xua}.}
\begin{align} \label{measuredAs}
\log(10^{10} A_s) = 3.06 \pm 0.03
\end{align}
and implies that $\vartheta$ has to be considerably small. For instance, if $v_S-S_*\sim 10\, M_P$ (which is a good approximation for $v_S\lesssim 20\, M_P$) and assuming in particular $v_S=20\, M_P$, the expression \eq{xifromAs} gives $\vartheta\sim 5\times 10^{-14}$\,. Notice in passing that such a value of $\vartheta$ is very far away from the Higgs quartic coupling $\tilde\lambda\sim 1$, which confirms the well-known result that the Higgs potential cannot produce successful inflation at tree-level. Incidentally, we will also see that $v_S$ needs to be orders of magnitude larger than $v_h$ to produce successful inflation. 

The most recent constraints on the relevant primordial parameters besides $A_s$, i.e.\ $n_s$, $r$ and $\alpha$, can be found in \cite{Ade:2015lrj} and \cite{Planck:2015xua}. These parameters have quite simple expressions in the slow-roll approximation: 
\begin{align}  \label{eq:slowrollMex}
r=128M_P^2\frac{ S^2 }{\left(S^2-v_S^2\right){}^2}\,,\quad
n_s=1-8M_P^2 \frac{3 S^2+v_S^2}{\left(S^2-v_S^2\right){}^2}\,,\quad
\alpha=-64M_P^4\frac{3S^4+5S^2v_S^2}{\left(S^2-v_S^2\right)^4}\,.
\end{align}
Like for $A_s$, their concrete values depend on the datasets that are used in the analyses and the full set of parameters that are allowed to vary. Clearly, as more data are included the parameters become more constrained, but increasing the number of allowed parameters decreases the constraining power of the data. For instance, using only Planck CMB temperature and Planck polarization data at low multipoles, \cite{Ade:2015lrj} reports that at $95\%$ c.l.\ $n_s=0.9666 \pm 0.0062$ at $0.05$ Mpc$^{-1}$ and $r<0.103$ at $0.002$ Mpc$^{-1}$. If $\alpha$ is also included in the analysis, the same dataset makes those numbers become $n_s=0.9667\pm0.0066$ and $r<0.180$ at the same scales and c.l.. In that case, the value of $\alpha$ itself is constrained to be $-0.0126^{+0.0098}_{-0.0087}$ at $0.05$ Mpc$^{-1}$, which indicates some tendency towards small negative values.  Including B-modes from BICEP2/Keck affects primarily the constraint on the tensor-scalar ratio, which for the same Planck data as above, and in the 
case in which the running of the scalar spectral index is allowed to be non zero, becomes $r<0.10$ at $0.002$ Mpc$^{-1}$ and $95\%$ c.l. \cite{Ade:2015lrj}. Adding also lensing reconstruction and other datasets such as baryonic acoustic oscillations, supernovae data and measurements of the current Hubble parameter tighten somewhat the constrains on the scalar spectrum, but not too importantly for our purposes. For instance, the running becomes perfectly compatible with zero: $\alpha=-0.0065\pm 0.0076$ at $0.05$ Mpc$^{-1}$ and 95$\%$ c.l. \cite{Planck:2015xua}. The largest allowed values of $|\alpha|$ are associated to the values of $n_s$ that deviate the most from $n_s\simeq 0.965$. For example, the more negative is the running, the larger is $n_s-1$, which may become even $\sim 0.2$ at $95\%$ c.l., see \cite{Planck:2015xua}. Given these results, we can consider the following (approximate but rather conservative)  ranges for these parameters at $k_*=0.05$ Mpc$^{-1}$ as a goal:
\bea\label{eq:nsrange}
\begin{aligned}
n_s & = 0.967\pm 0.007\,,\\
\alpha &  = -0.006\pm 0.007\,,\\
r  & \leq 0.11\,.
\end{aligned}
\eea
 Notice that these values are all assumed at the same scale, and therefore at the same inflaton value in the single field approximation, since the inverse comoving distance scale, $k$, and the field are e.g.\ related through $d S/d \log k \simeq M_P\sqrt{2\epsilon}$, for $\dot S>0$. Although the constraints on $r$ are often given at $0.002$ Mpc$^{-1}$, since the tensor index $-r/8$ is small for small $r$ values, there is not a too large difference on $r$ between the scales $0.002$ Mpc$^{-1}$ and $0.05$ Mpc$^{-1}$, which typically changes by only $\sim 10\%$, see  \eq{Ps} and \eq{Pt}. For example, if $n_s=0.96$ and $r=0.11$ at $0.05$ Mpc$^{-1}$, the tensor-scalar ratio at $0.002$ Mpc$^{-1}$ is $r\simeq 0.10$ for values of $|\alpha|\sim 10^{-3}$. In what follows we analyse to which extent the potential \eq{Mhat} is compatible with the values \eq{measuredAs} and \eq{eq:nsrange} while producing sufficient inflation. 

\subsubsection{\texorpdfstring{$S < v_S$}{Smvs}} \label{infS}
Let us consider first the case in which the field $S$ rolls from $S<v_S$ towards its minimum at $S =v_S>0$\,, where, classically,  the field comes to rest after inflation ends and the universe reheats. As mentioned in section~\ref{subsec:hline} this captures the dynamics of inflation along the $h$-valley for $\lambda_{SH}>0$. The potential goes from being convex to concave as $S$ grows and, naively, we can expect results that interpolate between a pure Hilltop quadratic model and standard quadratic inflation. 

The derivative of $n_s$ with respect to $S$ is negative for all $S<v_S$ and therefore, for each $v_S>0$, the maximum of the scalar spectral index (as a function of $S$) is attained at $S=0$. This can be used to get a rather good estimate the range of $v_S$ that allows to obtain a reasonable set of values for $|1-n_s| \ll 1$. For instance, if we require $n_s(S=0)=1-8(M_P/v_S)^2$ to be within $0.94$ and $0.98$, we find that $v_S$ has to be approximately between 11.5 and 20 times larger than $M_P$. For $S\ll v_S$, the variation of $n_s$ is $d n_s/dS\simeq -80\, S\,M_P^2/v_S^4$, which, for the aforementioned range of $v_S$, gives $-d n_s/dS\sim 5\times 10^{-4} S/M_P$ -- $5\times10^{-3} S/M_P$. Since this variation is small, the previous estimate for the range of $v_S$ is expected to be roughly correct even if $n_s$ is not calculated exactly at $S=0$ but at some larger value. In other words, the running of the scalar spectral index is essentially negligible. Indeed, for $S\ll v_S$ we get $\alpha\simeq -5 (n_s-1)
^2(S/v_S)^2$. The limit $S_*\ll v_S$ turns out to be valid for $v_S$ up to $\sim 20\, M_P$, if $N_e\sim 60$. Then, $S_*/v_S \sim 0.1-0.3$ and hence $\alpha$ at $S_*$ will be about $10^{-4}$, at most $10^{-3}$.

In the slow-roll approximation, the number of e-folds produced between two field values $S_i$ and $S_f$ can be calculated with the expression \eq{eq:efoldsslow}, which gives: 
\begin{align} \label{efoldsap}
N_e=\frac{1}{4M^2_P}\left[\frac{S^2}{2}-v_S^2 \log \left(\frac{S}{M_P}\right)\right]^{S_i}_{S_f}.
\end{align}
 As discussed before, the endpoint of inflation, $S_e$, can be estimated assuming it is reached as soon as either $\epsilon$ or $\eta$ become of order 1. Since $2\eta/\epsilon=3-v_S^2/S^2$ and inflation ends when $S\sim v_S$ (i.e.\ when the field reaches the minimum), we see that $\epsilon\simeq |\eta|$ towards the end of inflation and we can use any of the two to estimate the endpoint. 
 
In the limit of large $v_S$, i.e.\ $v_S\gg M_P$,  the condition $\epsilon=1$ implies that the end of inflation occurs approximately for a value of the field equal to $S_e\simeq v_S-\sqrt{2} M_P$. 
 Correspondingly, using \eq{efoldsap}, the number of e-folds from a given value of $S$ until the end of inflation is approximately equal to
 \begin{align}
 \label{eq:Napprox}
  N_e\simeq\frac{v^2_S}{4M^2_P}\left(\frac{S}{v_S}-1\right)^2\,.
 \end{align}
 This implies that  the distance travelled along $S$ between the beginning of inflation at $S_*$ and the end of it is larger than the Planck Mass: $S_e-S_*\simeq (\sqrt{2}+2\sqrt{N_e})M_P$. In this regime of large $v_S$, the main primordial  parameters take approximately the following values in terms of the number of e-folds elapsed between the scale where they are measured and the end of inflation:
 \begin{align}
  \label{eq:cosmoapprox} A_s\simeq \frac{4\vartheta}{3\pi^2}N_e^2\frac{v^2_S}{M^2_P}\,,\quad n_s&\simeq 1- \frac{2}{N_e}\,,\quad r \simeq \frac{8}{N_e}.
 \end{align}
These formulae reproduce the behaviour of a quadratic potential, corresponding to the parabola that best fits the bottom of the Mexican hat. The product $A_s(n_s-1)^2$ in this limit is independent of $N_e$,
which gives a prediction for the mass scale related to $v_s$ and $\vartheta$. Recalling \eqref{effcoup} and the SMS relation $v_S^2\simeq -6{m^2_S}/{\lambda_S}$, valid in the large $m^2_S$ limit, see \eqref{veVs}, we get
\begin{align}
\label{eq:mapprox}
 m^2_S\simeq 4\vartheta\,v^2_S \simeq-\frac{3\pi^2}{4}M^2_P A_s (n_s-1)^2\sim  10^{26}{\rm \,GeV}\,,
\end{align}
where we have approximated $\lambda\sim \tilde\lambda$. Furthermore, from \eqref{eq:cosmoapprox}, we find that for $N_e\sim 60$
\begin{align}
\label{eq:rapprox}
 r\sim 0.13\,,
\end{align}
which is essentially ruled out by Planck \cite{Planck:2015xua}. 

\begin{figure}\centering
  \includegraphics[scale=1.2]{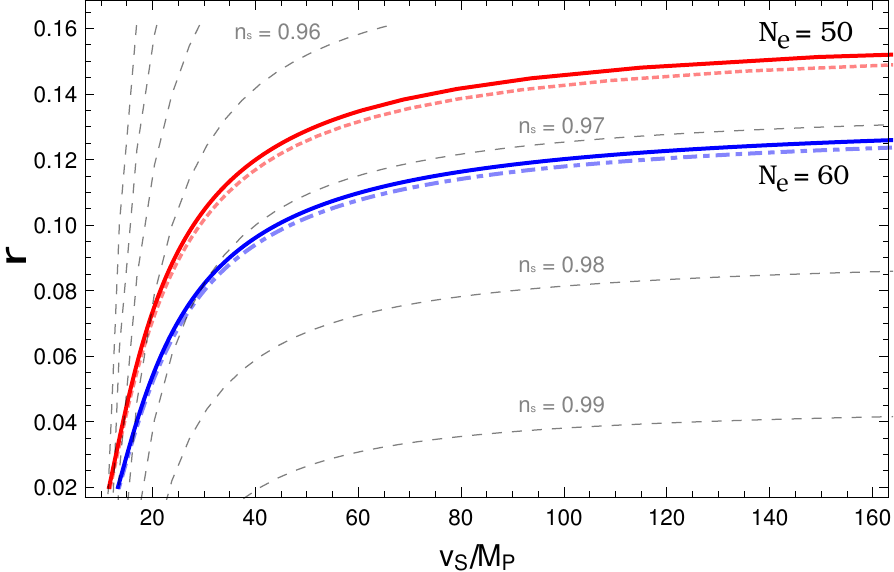}
\caption{\label{Frvs} \small  Tensor-scalar ratio as a function of $v_S$ in Planck units for the the potential \eq{Mhat} in the case $S<v_S$. The thicker (red and blue) lines give the curves of constant $N_e$ equal to 50 and 60, calculated with $\epsilon_{\mathcal H}$ and \eq{eq:efolds} (continuous lines), and with $\epsilon_V$ and \eq{eq:efoldsslow} (dashed). The thinner dashed lines represent the curves of constant scalar spectral index. Values of $r$ and $n_s$ compatible with current data are achieved for $v_S/M_P$ lower than $\sim 20$.}
  \end{figure}

However, in the opposite limit, where $v_s\ll M_P$, the cosmological parameters deviate from the quadratic regime. The end of inflation takes place at $S_e\simeq v^2_S/(2\sqrt{2}M_P)$, and the estimates for the primordial parameters become
\begin{align}
A_s\simeq \frac{v^4_S}{24\pi^2 M^4_P}e^{8N_e M^2_P/v^2_S}\,,\quad n_s\simeq 1-8\frac{M^2_P}{v^2_S}\,,\quad  r\simeq 16\,e^{-8N_e M^2_P/v^2_S}\,,
\end{align}
where
\begin{align}
 N_e\simeq \frac{v^2_S}{4 M^2_P}\log\frac{v^2_S}{2\sqrt{2}S M_P}\,.
\end{align}
Although the observed $n_s$ cannot be fitted in this limit (because it yields a too large value), these results suggest that small values of $r$ compatible with CMB data can be obtained away from $v_s\gg M_P$. A detailed numerical analysis away from these two limits confirms that this is indeed the case and that all the requirements can be simultaneously satisfied, see also \cite{Planck:2015xua}. Figure \ref{Frvs} confirms that \eqref{eq:rapprox} is recovered for large $v_S$ and $N_e \sim 50 - 60$. For lower values of $v_S$ it is possible to obtain $r\gtrsim 0.04$ while appropriately fitting $n_s$ and obtaining around 60 e-folds of inflation. This happens for $v_S\lesssim20\,M_P$, as was estimated at the beginning of this subsection when assuming that inflation started at small values of $S$.

\begin{figure}\centering
  \includegraphics[scale=1.2]{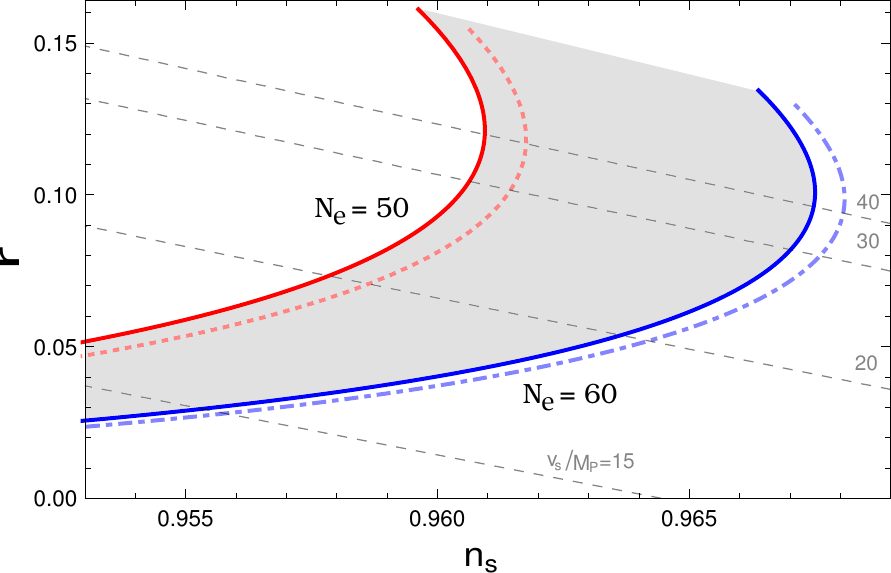}
\caption{\label{Frns} \small Tensor-scalar ratio as a function of the scalar spectral index for the the potential \eq{Mhat} and $S<v_S$. As in figure\ \ref{Frvs}, the thicker (red and blue) lines give the curves of constant $N_e$ equal to 50 and 60, calculated with $\epsilon_{\mathcal H}$ and \eq{eq:efolds} (continuous lines), and with $\epsilon_V$ and \eq{eq:efoldsslow} (dashed). The shaded area in between corresponds to an approximate region of plausible values for $N_e$. The thinner dashed lines represent the curves of constant $v_S$ in Planck units. Clearly, it is possible to have $r\lesssim 0.1$ and $n_s$ compatible with current measurements for sufficiently low $v_S$, while achieving $50$ -- $60$ e-folds. The upper cut of the shaded area is the limit of $v_S$ going to infinity, in which the quadratic relation $1-n_s=2/N_e$ applies.}
  \end{figure}

In figure \ref{Frvs}, the dashed lines for a fixed number of e-folds before the end of inflation equal to either 50 (red)  or 60 (blue) have been obtained assuming that inflation ends when $\epsilon=1$ and using the approximation \eq{efoldsap}. The continuous lines (on the same colours) correspond instead to the computation of the number of e-folds with the exact condition $\epsilon_{\mathcal H}=1$ and using the solution of \eq{eq:efolds} to determine the field value at which the primordial parameters are obtained. The small difference between the two methods that can be observed in figure\ \ref{Frvs} becomes relevant in the plane $r$ -- $n_s$. In figure \ref{Frns} we plot the tensor-scalar ratio, $r$, versus the scalar spectral index, $n_s$, in the slow-roll approximation \eq{eq:slowrollMex}.  For a given value of $r$ the error committed on $n_s$ by using the dashed curves instead of the continuous ones can be at most of the order of $\sim 0.1 \%$. This number is however comparable to the error with 
which Planck can measure the scalar spectral index; see the discussion at the beginning of this section. If instead we fix the scalar spectral index, the (bending) shapes imply that for certain values of $n_s$ the error on $r$ introduced by the the dashed lines can be as large as $\sim 30\%$. With the current upper limits on $r$, this error is sufficiently high to turn a point from being excluded by the data to be allowed, or vice versa. The dashed lines (i.e.\ the computation of the endpoint of inflation from $\epsilon$, or $\eta$, and of the number of e-folds from the approximation \eq{eq:efoldsslow}) have often been used in the literature to test inflationary models. However, these results show that a more accurate treatment of the inflationary predictions is necessary as the quality of the data improves, specially taking into account that future probes may be able to measure $r$ with a precision of $10^{-2}$ or even $10^{-3}$, see e.g.\ \cite{Matsumura:2013aja}.

It has been pointed out in several works, e.g.\ in \cite{Liddle:2003as,Martin:2010kz, Martin:2014rqa,Martin:2014nya,Munoz:2014eqa,Cook:2015vqa,Domcke:2015iaa}, that a good a understanding of the reheating process is needed for testing inflationary models properly. This is basically because the number of e-folds required to solve the horizon problem depends on how the reheating of the universe takes place \cite{Liddle:1993fq,Liddle:2000cg,Liddle:2003as}. The number of e-folds happening since a scale $k_*$ exited the horizon during inflation until the end of it depends on the ratio $a_0 {\mathcal H}_0/k_*$ to the size of the current Hubble scale, but also on the details of reheating. An approximate expression is given by \cite{Liddle:2003as} (see also e.g.\ \cite{Domcke:2015iaa})
\begin{align}
N_e^*\simeq 67+\log\frac{a_0 \mathcal{H}_0}{k_*}+\frac{1}{4} \log \frac{V_*}{M_P^4}+\frac{1}{4}\log\frac{V_*}{V_e}+\frac{1-3w}{12(1+w)}\log\frac{\rho_{r}}{V_e}\,,
\end{align}
where the parameter $w$ represents the equation of state of the universe during the reheating phase and typically varies between 0 (for matter domination) and 1/3 (for radiation). The energy density of the universe at the end of reheating is represented by $\rho_r$; and $V_*$ and $V_e$ denote the inflaton potential corresponding, respectively, to the scale $k_*$ and the end of inflation. An uncertainty on $w$, $\rho_r$ or the relevant values of the potential can easily change $N_e^*$ by a few e-folds. Looking at figure \ref{Frns}, and recalling the discussion of the previous paragraph, we see that these uncertainties can be comparable to the error introduced by an imprecise use of the slow-roll approximation, specially on the determination of the endpoint of inflation. We therefore advocate the use of the equation \eq{eq:efolds} and the condition $\epsilon_{\mathcal H}=1$, instead of the approximation \eq{eq:efoldsslow}. Notice also that since the SMS gives a complete picture of the connection between the 
inflationary sector and the Standard Model of particle physics, the details of the reheating process are, in principle, calculable. 

\begin{figure}\centering
  \includegraphics[scale=1.2]{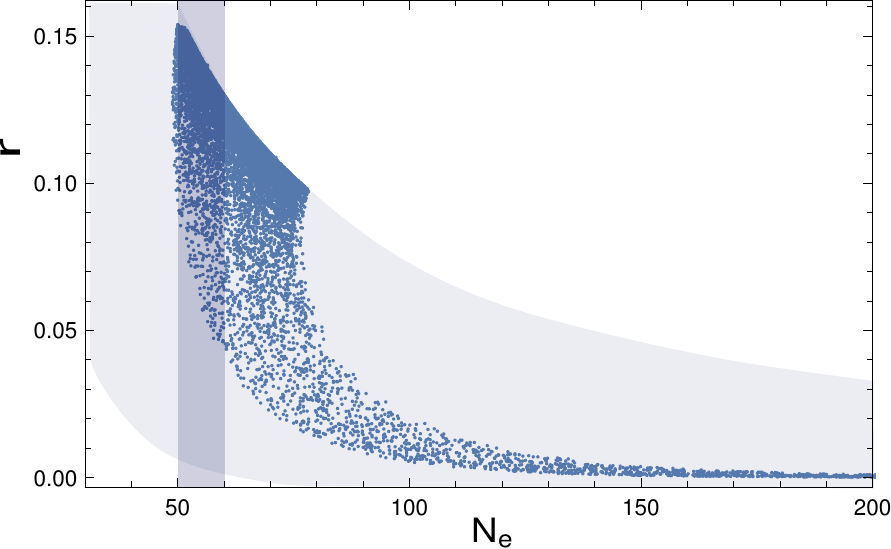}
\caption{\label{FrNe} \small Scatter plot for $v_S>S$ in the plane $N_e$--$r$ of points that fit current measurements of $A_s$ (we assume $A_s=(2.142\pm 0.049)\times 10^{-9}$)  and have a scalar spectral index in the range given in \eq{eq:nsrange}. For all blue points, the running of $n_s$ is compatible with the data. The location at which the primordial spectrum is evaluated has been computed using the condition $\epsilon_{\mathcal H}=1$ for the end of inflation and the equation \eq{eq:efolds} to track the dynamics of the inflaton. The dashed grey area that contains the blue points marks the boundaries of the region of parameters explored for the plot.}
  \end{figure}

\begin{figure}\centering
  \includegraphics[scale=1.2]{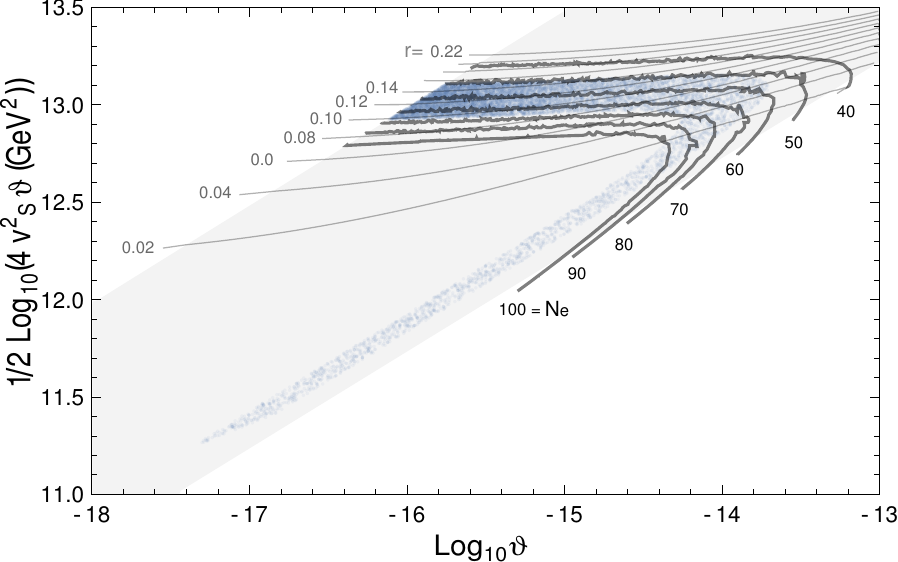}
\caption{\label{Fmasstheta} \small Mexican hat inflation for $S<v_S$. The figure shows the mass scale $4 \vartheta v_S^2\simeq m_S^2$, introduced in \eq{eq:mapprox}, as a function of the effective quartic coupling $\vartheta$. The blue points match those shown in figure \ref{FrNe}, while the grey ones lay outside the 68$\%$ c.l.\ interval $A_s=(2.142\pm 0.049)\times 10^{-9}$. Curves of equal $r$ and $N_e$ are also displayed. The dashed diagonal band marks the boundaries of the region of parameters explored for the plot and maps to the corresponding region of figure\ \ref{FrNe}. The present figure shows that the data selects a small coupling $\theta\sim 10^{-13}$ and a mass scale of the order of $10^{13}$ GeV.}
  \end{figure}

As discussed earlier, the value of the effective quartic coupling $\vartheta$ is determined by the amplitude of the scalar primordial perturbations through \eq{xifromAs}. Figures \ref{FrNe} and \ref{Fmasstheta} show scatter plots of points (in blue) for which the right amplitude of perturbations at $68 \%$ c.l.\ is obtained. In particular, for these figures we have imposed $A_s=(2.142\pm 0.049)\times 10^{-9}$, see \cite{Planck:2015xua} and footnote 5. The blue points have $n_s$ in the range given in \eq{eq:nsrange}. The dashed region of figure\ \ref{Frns} maps into a portion of the vertical grey band of figure\ \ref{FrNe}, where we represent $r$ vs $N_e$. The seemingly higher density of points towards high values of $r$ is simply a result of the sampling method and does not correspond to a tendency of the model. For all the points, the running of the spectral index is well within the allowed limits, in agreement with our previous estimates. In addition, figure \ref{Fmasstheta} shows that 
the absolute value of the mass scale of \eqref{eq:mapprox} is indeed of the order of $10^{13}$ GeV, not just in the limit of very large $v_S$, but for all points having around 60 e-folds of inflation and values of $n_s$ compatible with observations. This plot also confirms that, as anticipated before, very small values of $\vartheta<10^{-13}$ are required for successful inflation. 

In summary, Mexican hat inflation with $S<v_S$ is compatible with current constraints, requiring a small quartic coupling $\vartheta<10^{-13}$ and a VEV $v_S\gtrsim  15 M_P$. This implies an associated mass scale in the SMS of the order of $|m_S^2|^{1/2}\simeq10^{13}$ GeV. From the point of view of CMB measurements, the value of $m_S^2$ is approximately determined by $A_s$ and $n_s$, through \eq{eq:mapprox}. The value of $\vartheta$ is fixed by the amplitude $A_s$ of scalar primordial perturbations, see \eq{xifromAs}. We point out that a measurement of the tensor-scalar ratio, $r$, would allow to determine the value of $v_S$, as illustrated by figure\ \ref{Frvs}. As it can be seen in figure\ \ref{Frvs} the prediction of the model is that the tensor-scalar ratio should be larger than $\sim 0.04$ and smaller than $r\sim 0.15$, for a number of e-folds before the end of inflation between $50$ and $60$. Whereas the largest values of $r$ are ruled out by Planck,  the range $0.04\lesssim r\lesssim 0.1$ is allowed 
by the data and will be reached by the precision of near-future probes. Besides, given that $v_S^2\simeq -6{m^2_S}/{\lambda_S}$ for large $m_S^2$, we would then be able to have an estimate of the SMS quartic coupling $\lambda_S$.  Unfortunately, constraining in this model the Higgs portal coupling, $\lambda_{SH}$, is not feasible, since the effective inflationary potential along the longitudinal direction depends only on two parameters: $\vartheta$ and $v_S^2$. What we have just seen is that we can estimate $m_S^2$ and $\lambda_S$ from CMB measurements under the assumption of small $\lambda_{SH}$. Although, strictly speaking, the possibility of $\lambda_{SH}=0$ cannot be excluded from the data, the Higgs portal coupling is unavoidable for reheating if we assume there are no other fields. This could allow to put theoretical bounds to the value of the coupling. In addition, the corrections in the normal direction to the trajectory, see e.g.\ equations \eq{eq:valleycorrections1} and \eq{eq:valleycorrections2}, are 
sensitive to $\lambda_{SH}$ through the curvature $\kappa$, which might allow an estimate of $\lambda_{SH}$, but only for relatively large values of it where the deformation of the valleys described here would be significant. Needless to say, the prospects for measuring $\lambda_{SH}$ from a particle physics experiment are bleak, since the effects of the singlet at low energies will go like $\tilde m_H^2/m_S^2$\,, which is an insignificant ratio.
  
  \begin{figure}\centering
  \includegraphics[scale=1.2]{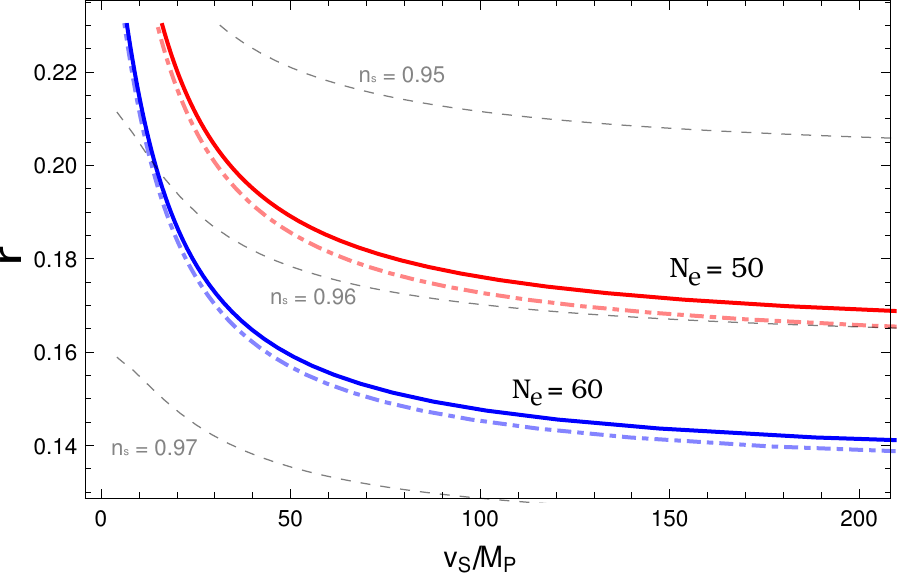}
\caption{\label{Frvs2} \small  Tensor-scalar ratio as a function of $v_S$ in Planck units for the the potential \eq{Mhat} in the case $S>v_S$. The thicker (red and blue) lines give the curves of constant $N_e$ equal to 50 and 60, calculated with $\epsilon_{\mathcal H}$ and \eq{eq:efolds} (continuous lines), and with $\epsilon_V$ and \eq{eq:efoldsslow} (dashed). The thinner dashed lines represent the curves of constant scalar spectral index. For $n_s$ close to 0.96 -- 0.97 and $N_e$ around 50 -- 60, the value of $r$ turns to be too high to be allowed by CMB data from Planck. Compare to figure \ref{Frvs}.}
  \end{figure}
  
  \begin{figure}\centering
  \includegraphics[scale=1.2]{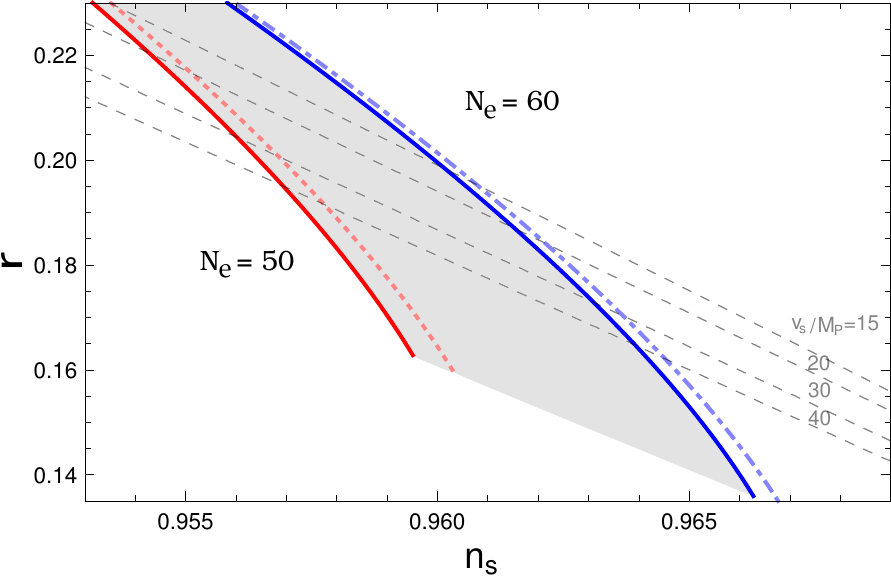}
\caption{\label{Frns2} \small Tensor-scalar ratio as a function of the scalar spectral index for the the potential \eq{Mhat} and $S>v_S$. As in figure\ \ref{Frvs2}, the thicker (red and blue) lines give the curves of constant $N_e$ equal to 50 and 60, calculated with $\epsilon_{\mathcal H}$ and \eq{eq:efolds} (continuous lines), and with $\epsilon_V$ and \eq{eq:efoldsslow} (dashed). The shaded area in between corresponds to an approximate region of plausible values for $N_e$. The thinner dashed lines represent the curves of constant $v_S$ in Planck units. As in figure\ \ref{Frns}, the lower cut in the shaded area corresponds to the limit of $v_S$ going to infinity, which reproduces the predictions of a quadratic potential.}
  \end{figure}
  
      \begin{figure}\centering
  \includegraphics[scale=1.2]{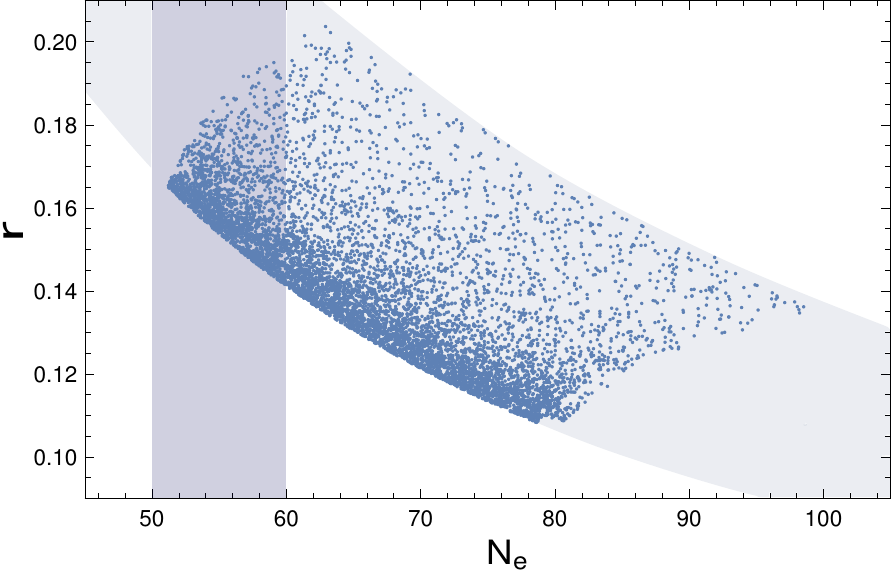}
\caption{\label{FrNe2} \small Scatter plot for $v_S<S$ in the plane $N_e$--$r$ of points that fit current measurements of $A_s$ (we assume $A_s=(2.142\pm 0.049)\times 10^{-9}$)  and have a scalar spectral index in the range given in \eq{eq:nsrange}. The location at which the primordial spectrum is evaluated has been computed using the condition $\epsilon_{\mathcal H}=1$ for the end of inflation and the equation \eq{eq:efolds} to track the dynamics of the inflaton. The dashed grey area that contains the blue points marks the boundaries of the region of parameters explored for the plot.}
  \end{figure}
  
      \begin{figure}\centering
  \includegraphics[scale=1.2]{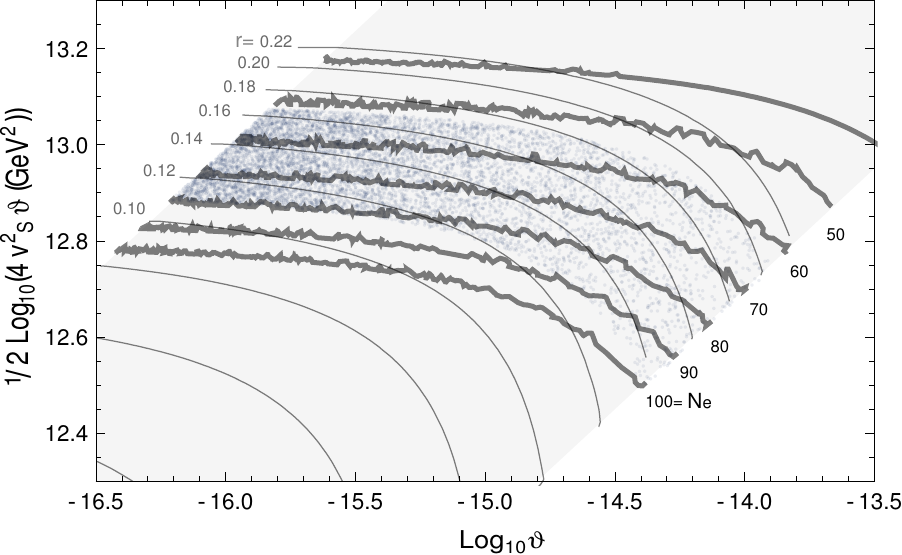}
\caption{\label{Fmasstheta2} \small Mass scale $4 \vartheta v_S^2\simeq m_S^2$, introduced in \eq{eq:mapprox}, as a function of the effective quartic coupling $\vartheta$. The blue points match those shown in figure \ref{FrNe2}, while the grey ones lay outside the 68$\%$ c.l.\ interval $A_s=(2.142\pm 0.049)\times 10^{-9}$. Curves of equal $r$ and $N_e$ are also shown. The dashed band marks the region of parameters explored for the plot and maps to the corresponding area of figure\ \ref{FrNe2}.}
  \end{figure}

\subsubsection{\texorpdfstring{$S > v_S$}{SMvs}}

We consider now the fields rolling down the steeper part of the Mexican hat potential towards the vacuum. As mentioned in section~\ref{subsec:hline}, this scenario will capture the inflationary dynamics along the $h$-valley for $\lambda_{SH}<0$. In this region, the potential is concave and dominated by $(S-v_S)^4$  where $S \gg v_S$; and then by $(S-v_S)^2$ for $S\sim v_S$. We expect the results to be a mixture between these two behaviours. The primordial parameters will be basically dominated by the quartic behaviour, while the number of e-folds gets an important contribution from the quadratic one. In particular, we can already guess that imposing $50$ -- $60$ e-folds, the resulting values of $r$ will be large (as it happens in an standard chaotic quartic model) and hence ruled out by Planck. Proceeding as before, we can consider the limits of large and small $v_S$. In the first case, one obtains the same results as in \eqref{eq:Napprox}--\eqref{eq:rapprox}, reproducing the behaviour of a quadratic 
potential. In the opposite limit, i.e.\ for small $v_S$, we obtain that the number of e-folds can be approximated by
\begin{align}
 N_e\simeq \frac{S^2}{8 M^2_P}-1
\end{align}
and then, the primordial parameters can be expressed as
\begin{align}
 A_s\simeq \frac{8\vartheta(1+N_e)^3}{3\pi^2}\,,\quad n_s\simeq 1-\frac{3}{N_e+1}\,,\quad r\simeq \frac{16}{N_e+1},
\end{align}
which reproduce the results of a quartic potential. Noting that $r\simeq-16/3(n_s-1)$ and substituting the central value for $n_s$ of \eq{eq:nsrange} gives
\begin{align}
\label{eq:rapprox2}
 r\sim 0.18\,.
\end{align}
Therefore, in these scenarios, for which $S>v_S$, we expect values of $r$ in between those given by \eqref{eq:rapprox} and \eqref{eq:rapprox2}, which are ruled out by experimental constraints. A detailed numerical analysis beyond these approximations gives the results displayed in figures~\ref{Frvs2} -- \ref{Fmasstheta2}, which are analogous to figures\ \ref{Frvs} -- \ref{Fmasstheta} and confirm the qualitative features just discussed.

In figure\ \ref{Frvs2} we see that the value of the tensor-scalar ratio is too large to fit CMB data comfortably. In the limit of large $v_S$ the relation to the spectral index is given by $1-n_s=r/4$. This means that $r\lesssim 0.1$ requires $0.975\lesssim n_s$, which is just above the upper value of the range for $n_s$ that we have indicatively taken in \eq{eq:nsrange}. In the limit of small $v_S$ the corresponding relation reads $n_s=1-3r/16$, as we have just seen. This implies that $r\simeq 0.1$ is associated to $n_s\simeq 0.981$, which is even larger than the large $v_S$ value. In figure\ \ref{Frns2} we can see clearly that a large number of e-folds and a large $v_S$ tend to lower $r$, therefore enhancing the compatibility with the data in the case $S>v_S$. However, this comes at the expense of raising $n_s$ to values that are outside the allowed range. The figure \ref{FrNe2} illustrates the same idea from a different point of view. The blue dots have $n_s$ and $A_s$ within the current limits, but they 
correspond to values of $r$ that are above the upper bound. This can also be appreciated in figure\ \ref{Fmasstheta2}. In conclusion, the Mexican hat potential can easily fit current CMB data but only for $S<v_S$, i.e.\ for $\dot S>0$, in which case the tensor-scalar ratio turns out to be $r \gtrsim 0.04$.

\section{\label{sec:quantum}Stability and quantum effects}

In this section we study the role of quantum effects on the inflationary scenarios analyzed previously and the stability of the effective potential at high energies. We do so by introducing the renormalization group (RG) improved effective potential in section~\ref{subsec:RGimproved}, followed by a discussion on stability in section~\ref{subsec:stability}. Then, we study the implications for inflation in section~\ref{subsec:inflationloop}, including radiative corrections to the $h$-valley inflation scenario, which was studied at tree-level in section~\ref{subsec:inflationtree}. We also study the viability of the Higgs false-vacuum inflation in section~\ref{subsec:mtvalley}. 

\subsection{\label{subsec:RGimproved}RG-improved effective potential}

Here we review the construction of the two-loop improvement of the one-loop effective potential, which allows an accurate treatment of the quantum corrections and their effects on the inflationary dynamics. These corrections can be specially important at high field values. Concretely, we compute the radiative corrections in a background of the the neutral Higgs component, $h$, and the singlet $S$. Decomposing the Higgs doublet into 
\begin{align} \label{deco}
H=\left(h_i+i\, h_r\,, h+i\,\chi\right)/\sqrt{2}\,,
\end{align}
and writing the loop expansion of the effective potential as
\begin{align} \label{sp}
 \bar V=V^{\rm tree}+V^{(1)}+\cdots\,,
\end{align}
the one-loop correction is:  
\begin{align} \nonumber
 \label{eq:V1}V^{(1)}(h,S,\mu;\delta_i)=&\frac{1}{16\pi^2}\left(\frac{3}{4}\sum_{a}(m^2_a(h,S))^2\left[\log\frac{m^2_a(h,S)}{\mu^2}-\frac{5}{6}\right]\right.\\
+\frac{1}{4}\sum_{i} (m^2_i & (h,S))^2 \left.\left[\log\frac{m^2_i(h,S)}{\mu^2}-\frac{3}{2}\right]-\frac{1}{2}\sum_{I}(m^2_I(h,S))^2\left[\log\frac{m^2_I(h,S)}{\mu^2}-\frac{3}{2}\right]\right).
 \end{align} 
In this expression the subscripts $a,i,I$ refer to vectors, scalars and (Weyl) fermions, respectively, and $m^2_k(h,S)$ stand for the mass eigenvalues in the background of the fields $h$ and $S$. 

Only the effective masses in the scalar sector differ from those in the SM, which can be found e.g.\ in \cite{Quiros:1999jp}.  The new masses (with respective multiplicities 3, 1 and 1) are
\begin{align} \label{m1}
 m^2_1& =\frac{1}{2} \left({h}^2 {\lambda}+{\lambda_{SH}} S^2+2 {{m^2_H}}\right)\,,\\\label{m2}
 m^2_2 & =\frac{1}{4} \left({h}^2 (3 {\lambda}+{\lambda_{SH}})+S^2 ({\lambda_S}+{\lambda_{SH}})+2 {{m^2_H}}+2 {{m^2_S}}-\sqrt{\Delta}\right)\,,\\ \label{m3}
 m^2_3 & =\frac{1}{4} \left({h}^2 (3 {\lambda}+{\lambda_{SH}})+S^2 ({\lambda_S}+{\lambda_{SH}})+2 {{m^2_H}}+2 {{m^2_S}}+\sqrt{\Delta}\right)\,,
\end{align}
where $\Delta  =\tilde\Delta^2+2\lambda_{SH}\left(h^2-S^2\right)\tilde\Delta +\lambda_{SH}^2\left(h^4+S^4+14 h^2 S^2\right)$ and
 $\tilde\Delta  = \lambda_S S^2-3 \lambda h^2+2\left(m_S^2-m_H^2\right)$\,.

 Clearly, the radiative corrections involve parameters of the SM (e.g.\ quark Yukawa couplings) on which the tree-level potential does not depend explicitly. Naively, it might seem that once the loop corrections are included, the potential would depend as well on the parameter  $\mu$, which represents the renormalization scale. However, as it is well known, physical observables do not depend on the renormalization scale, thanks to the properties of the renormalization group. Indeed, after an appropriate redefinition that we discuss below, the potential is exactly scale invariant if the loop corrections are implemented at all orders in perturbation theory. In practice, only a residual (and controlled) dependence on $\mu$ occurs in actual calculations, due to the need of truncating the perturbative expansion at a finite order. 
 
The scale independence of the potential happens as a cancellation of the explicit $\mu$-dependence coming from the radiative corrections with an implicit dependence through the couplings (including masses) and fields, which get renormalized under changes of $\mu$. This is determined by the RG equations for the (scale-dependent) couplings $\delta_i(t)$ and fields $H(t)\equiv Z_H(t)H, S(t)\equiv Z_S(t)S$ as follows:
\begin{align}
  \nonumber \frac{d \log  Z_H(t)}{dt} &=-\gamma_H( \delta_i(t)) Z_H(t),\quad Z_H(0)=1,\\
  \label{eq:betadef}  \frac{d \log  Z_S(t)}{dt} & =-\gamma_S(\delta_i(t))Z_S(t),\quad  Z_S(0)=1,\\
  \nonumber  \frac{d \delta_j(t)}{dt} & =\beta_{\delta_j}( \delta_i(t)), \quad  \delta_i(0)=\delta_i(\mu_0).
\end{align}
In these equations, 
\begin{align}
\label{eq:t}
t = \log\frac{\mu(t)}{\mu_0}
\end{align} 
is a convenient rescaling parameter, where $\mu_0$ is a reference energy that can be chosen at will. For example, $\mu_0$ can be set as a scale at which couplings are matched to experimental collider results.  The renormalization of the fields is determined by $\gamma_H$ and $\gamma_S$, whereas the beta functions $\beta_{\delta_i}$ control the running of masses and couplings.
With these functions we define the Callan-Symanzik operator
\begin{align}
\label{CS}
\mathcal{D}=\mu\frac{\partial}{\partial \mu}+\sum_i\beta_{\delta_i}\frac{\partial}{\partial\delta_i}-\gamma_S S\frac{\partial}{\partial S}-\gamma_H h\frac{\partial}{\partial h}\,.
\end{align}
Applied to the effective potential $\bar V$, this operator gives
\begin{align} \label{CSe}
\mathcal{D}(\bar V-\Omega)=0\,,
\end{align}
where $\Omega$ is a field independent function of the scale  whose form depends on the specific renormalization scheme that is employed. In our case (and working in the $\overline{\rm MS}$ renormalization scheme) the function $\Omega$ can be approximated  at one-loop order by the expression \eq{Omega1} below.  Then, using \eq{CSe} we can define a new potential
\begin{align} \label{newpot}
{\cal V}=\bar V - \Omega\,,
\end{align}
which is scale independent, since it satisfies $\mathcal{D}{\cal V}=0$, see \cite{Ford:1992mv}. In standard particle physics calculations (e.g.\ cross sections and decay rates in flat spacetime) this redefinition can typically be omitted since the vacuum energy is irrelevant. The subtraction of $\Omega$ affects the overall height of the potential, but not its shape, as long as one makes choices of $\mu$ that do not depend on the fields. However, when considering gravitational effects, as we will do here, it becomes important to render the full effective potential (including its vacuum piece) scale-invariant. This is generally the case in cosmology, and specially for inflation, where the value of the vacuum energy plays a central role.

Let us then discuss the relevance of the field independent piece $V_0$ that can be added to the potential $V$. As we discussed earlier, when the minimum of the potential is reached at the end of inflation (i.e.\ $h=h_{min}$, $S=S_{min}$), the value of the cosmological constant $\Lambda$ must be zero (by assumption). Using the previous expressions, it can be easily checked that $V=\bar V - \Omega$ vanishes at $h=S=0$ by construction (and this holds too at any order in the loop expansion). Therefore, defining
\begin{align}
V = {\cal V}+V_0= \bar V - \Omega+V_0\,,
\end{align} 
we can set the value of the potential to be $\Lambda$ at its minimum in $h=h_{min}$, $S=S_{min}$. It is important to remark that $V_0$ can be calculated in such a way that $V$ satisfies $\mathcal{D}V=0$, i.e.\ maintaining scale invariance. Since $\bar V-\Omega$ is scale-invariant, as follows from \eqref{CSe}, its value ${\cal V}_{min}$ at the minimum $h=h_{min}$, $S=S_{min}$ will also be scale-invariant. Therefore $V_0=\Lambda-{\cal V}_{min}$  is guaranteed to be independent of the choice of $\mu$. 

As we have mentioned earlier, it is usually not feasible to compute  the loop corrections to the potential at all orders and we have to truncate the series \eq{sp} at some order, introducing a residual scale dependence. For instance, if we include only radiative corrections at one loop and work in the $\overline{\rm MS}$ scheme, we get
\begin{align} \label{Omega1}
\Omega^{(1)}=&\frac{1}{64\pi^2}\left[4 m^4_H\left[\log\frac{m^2_H}{\mu^2}-\frac{3}{2}\right]+m^4_S\left[\log\frac{m^2_S}{\mu^2}-\frac{3}{2}\right]\right]\,.
\end{align}
Then, the equation $\mathcal{D}(V^{\rm tree}+V^{(1)}-\Omega^{(1)})=0$ only holds up to two-loop effects, which are suppressed by a multiplicative factor $1/({16\pi^2})^2$. 

A more precise treatment of the radiative corrections can be achieved by inserting into the potential the running of the fields and couplings with the renormalization scale. In particular, we will consider the two-loop RG improvement of the one-loop effective potential, which is given by
\begin{align}
\label{eq:Veff}
\hat V(h,S,t)=V^{\rm tree}( h(t),S(t),\mu(t); \delta_i(t))+V^{(1)}( h(t),  S(t), \mu(t); \delta_i(t))-\Omega^{(1)}+V_0^{(1)}\,,
\end{align}
where the masses, couplings and fields run with the RG at two loops. As it was proven in \cite{Bando:1992np} for a simpler $m^2\phi^2+\lambda \phi^4$ model, the L-loop effective potential and (L + 1)-loop RG give an effective potential which is exact up to L-th-to-leading log order. This means that by taking the one-loop effective potential and the two-loop RG we are in practice resumming all the  log terms up to NLO appearing at each order in the loop expansion. This is the form of the potential that we will use in this paper for the numerical computations. The relevant two-loop beta functions are provided in appendix \ref{app:RGs}. 

An important point concerning scale-dependence is that the subtraction of $\Omega$ in \eq{newpot} (and the consistent addition of $V_0$) allows to choose field dependent values for the the renormalization scale, $\mu=\mu(S,h)$, without spoiling the shape of the potential. This is often applied in models with a single scalar to ameliorate truncation errors in the loop expansion by minimizing logs, since the effective masses in that case are typically proportional to the field for large values of it. In this work we will also make field dependent choices of $\mu$, as described in the next section.  

Once radiative corrections are included, we can proceed as it was done (at tree-level) in section~\ref{sec:SMS}. We can match the SMS to the SM by integrating out the heavy singlet $S$ using the same method described there. A detailed discussion of the matching procedure used in the calculations of section~\ref{subsec:inflationloop} is given in appendix  \ref{sec:SMmatching}.

\subsection{\label{subsec:stability}Stability}

It is well known that the negative contribution of the top Yukawa to the beta function of the Higgs quartic coupling can destabilize 
the SM effective potential by driving it towards negative values. Indeed, with the recent measurements at CMS and ATLAS measurements, the SM electroweak vacuum appears to be metastable for the vast majority of the allowed range of Higgs and top quark masses \cite{Degrassi:2012ry,Espinosa:2015qea}.  This effect may also change qualitatively the potential around the lines of minima in the SMS, the $h$- and $S$-lines that were described in section~\ref{sec:valleystree}. In particular, it may destabilize the $h$-valley, that was shown to be able support inflation at tree-level. Naively, this could ruin the possibility of obtaining inflation, as the energy density could become negative inside the valley were the field should roll. Since the valley acts as an attractor for the dynamics of fields rolling in its vicinity, inflation would then have to be discarded for initial conditions in a wide region around the valley. In addition to this geometrical effect, there is also the crucial issue of large quantum 
fluctuations of the Higgs field induced by inflation, which can displace it directly into the instability region. It is therefore important to know under which conditions a potential SM instability can be cured in the SMS, which we analyze now. 

For values of $S$ below the VEV of $S$ in the Higgs vacuum, and  in the limit in which $h$ is larger than the other mass scales, a well motivated choice for the renormalization scale is $\mu\sim h$ \cite{EliasMiro:2012ay}. Then, using the tree-level potential, neglecting terms other than the quartic Higgs coupling and ignoring the field-renormalization factor, we have that for $S=0$
\begin{align}
 \frac{\partial V}{\partial h}\simeq \frac{1}{2}\left(\lambda(h)+\frac{1}{4}\beta_\lambda(h)\right)h^3\,.
\end{align}
For $\beta_\lambda(h)<0$, which causes $\lambda(h)$ to be a decreasing function, the derivative of the potential can become negative at high enough values of $h$, triggering an instability. In the SM, for $m_t=173.15$ GeV and $m_h=125.09$ GeV, after matching the experimental measurements to the SM parameters as detailed in appendix~\ref{app:SM_Pars}, the scale at which the potential becomes negative is around $\Lambda_I\sim 5\cdot10^{11}$ GeV. This effect is absent in the other two quartic couplings of the SMS, since their  beta functions lack the top-Yukawa driven contributions present in $\lambda$. Indeed, the one-loop beta functions in the SMS are the following:
\begin{align} \label{betal1}
 \beta_\lambda & =\frac{1}{16\pi^2}\left[-12 y_t^4+\lambda  \left(-\frac{9}{5}  g_1^2-9 g_2^2+12 y_t^2\right)+\frac{27}{100}g_1^4+\frac{9}{10} g_2^2 g_1^2+\frac{9}{4}g_2^4+12 \lambda ^2+ \lambda_{SH}^2\right],\\
 \beta_{\lambda_S} & =\frac{1}{16\pi^2}\left[3 \lambda _S^2+12 \lambda_{SH}^2\right],\\
 \beta_{\lambda_{SH}} & =\frac{1}{16\pi^2}\left[ \lambda_{SH} \left(-\frac{9}{10}  g_1^2-\frac{9}{2}g_2^2+6 \lambda +\lambda _S+6 y_t^2\right)+4 \lambda_{SH}^2\right].
\end{align}
Notice that the negative contribution to $\beta_\lambda$ coming from $y_t^4$ may in principle be compensated by $\lambda_{SH}^2$ (and this possibility is of course absent in the SM). However, this will typically require rather large values of $\lambda_{SH}$.

If the $S$- and $h$-lines of minima extend to values of $h$ that are large enough to sense the instability, there will be a value of the top quark mass, $m_t$, above which the potential along them will end up developing a runaway behavior. It will be seen in the next section that for the large values of $v_S$ needed for successful tree-level inflation along the $h$-valley, see section~\ref{subsec:inflationtree}, small values of $\lambda_{SH}$ suffice to make the $h$-valley reach values of $h$ larger than the instability scale.  The lower bound of $\lambda_{SH}$ for which this happen is given by \eqref{boundsh}. An example is provided by the choice of parameters shown in figures \ref{fig:hdeviations} and \ref{fig:correctionsh}, for which inflation takes place at tree-level for $h>10^{14}$ GeV with $\lambda_{SH}\sim 10^{-10}$.

Stability bounds are usually obtained by demanding absolute stability, i.e.\ that the potential does not become  smaller than the Higgs vacuum anywhere.  Therefore, for the discussion that follows, we will say that the potential is stable if the minimum corresponding to the Higgs vacuum is the one of lowest potential energy and if the potential does not develop a runaway behaviour in any direction in field space. If there are other vacua (different from the Higgs one) with higher energy, those minima will be unstable with respect to the Higgs vacuum (since tunneling to lower energies is always possible, in principle) but the potential as a whole is deemed stable. By Higgs vacuum we normally refer to the minimum of the potential which corresponds to the standard electroweak symmetry breaking vacuum with $v=246$ GeV in the low-energy model. Since we arranged for the Higgs vacuum to have zero cosmological constant (because for our purposes this makes no practical difference), the condition of absolute stability 
is equivalent to requiring that the potential should be positive (or zero) at all points. With this criterion, we recall that the instability scale, $\Lambda_I$, can be defined in the SM with the value of $\mu=h$ at which the potential of the Higgs crosses zero towards negative values, i.e. $\Lambda_I\simeq 5\cdot10^{11}$ GeV as mentioned before. 

The stability in the SMS was already discussed in \cite{EliasMiro:2012ay,Lebedev:2012zw}. It was found there that in addition to the possible stabilizing effect of $\lambda_{SH}$ via the RG running that we mentioned above, there is a tree-level effect which may be sufficient on its own to guarantee stability at large $h$ values. The threshold correction 
\begin{align}
\label{eq:delta}
\delta_{th}= 3\,\frac{\lambda^2_{SH}}{\lambda_S}
\end{align}
appearing in the matching of the Higgs quartic coupling, see \eqref{eq:treematching} and \eq{eq:Vhvalley}, plays a key role in this mechanism.

Concretely, it was argued in \cite{Lebedev:2012zw,EliasMiro:2012ay} that any potential with $\lambda_{SH}>0$ (which is the case of interest for us) can be stabilized by a sufficiently large $\delta_{th}$, provided that $|m^2_S|$ is smaller than (roughly) the instability scale (squared) at which the (low-energy) potential becomes negative. Being careful with factors involving dimensionless couplings, the threshold effect would stabilize the potential if the scale
\begin{align} \label{lambdath}
\Lambda_{th} ^2\sim6\frac{\lambda_{SH}}{\lambda_S \lambda}|m^2_S|
\end{align}
is smaller than $\Lambda_\lambda$, which is the scale at which the quartic coupling $\tilde\lambda$ becomes negative.\footnote{The SM instability scale, $\Lambda_I$, is larger than $\Lambda_\lambda$. At one-loop order in perturbation theory and taking the physical Higgs and top masses to be $m_h=125.09$ GeV and $m_t=173.15$ GeV we find $\Lambda_I=5\cdot 10^{11}$ GeV, whereas $\Lambda_\lambda=8\cdot 10^{10}$ GeV. Clearly, the physically meaningful scale for the stability is $\Lambda_I$, and the scale $\Lambda_\lambda$ appears as a consequence of using the approximation of the (RG-improved) tree-level potential.} It was concluded in \cite{EliasMiro:2012ay}  that in order to have absolute stability from this mechanism, the quartic coupling $\lambda(\mu)$ should satisfy the following condition\footnote{In \cite{EliasMiro:2012ay} the condition was actually formulated assuming $\Lambda_{th}\sim |m_S^2|^{1/2}$. Note that this only holds if $6 \lambda_{SH}\sim \lambda_S \lambda$, but the meaning of the two scales is 
different in general. While $|m_S^2|^{1/2}$ gives an estimate of the regime of validity of the SM (as a low-energy theory), the scale $\Lambda_{th}$ puts a bound to the range where $\lambda>\delta_{th}$ is needed for stability.}
\begin{equation}\label{condst}
\lambda(\mu)>\left\{ 
  \begin{aligned}
  \delta_{th} &\\ 
  0 &
  \end{aligned}
\right.\quad\text{for}\quad
\begin{aligned}
 & \mu\lesssim \Lambda_{th}\\ 
 & \mu\gg \Lambda_{th}
  \end{aligned}\,.
\end{equation}
Notice that equation \eq{eq:treematching} tells us that the SM quartic coupling is $\tilde\lambda= \lambda-\delta_{th}$. Thus, the upper condition in \eqref{condst} is equivalent to the SM stability condition. If the scale $\Lambda_\lambda$ at which $\tilde\lambda$ would become negative   is sufficiently larger than $\Lambda_{th}$, the relevant stability condition would be the less restrictive (lower) condition in \eqref{condst}. Given that $\lambda=\tilde\lambda+\delta_{th}$, the instability could then be avoided by a large enough $\delta_{th}$. The condition \eq{condst} was inferred in \cite{EliasMiro:2012ay} by minimizing the tree-level potential at $S=0$, a choice that is motivated because for small $S$ and $v_h$ (in comparison to $h$) the potential is susceptible to becoming negative due to the combined effects of the quadratic and quartic $h$ terms, see \eq{Vvev}. We recall that the SMS potential will be positive at large field values provided that $\lambda>0$, $\lambda_S>0$ and $\lambda_{SH}>-\
sqrt{\lambda \lambda_S/3}$. 

We will now see how the condition \eq{condst} should be completed by including another relevant scale. The SM instability as an RG effect is due to the beta function of $\lambda$ becoming negative due to a large $y_t$ contribution, see \eq{betal1}. Let us then suppose that the SM effective potential appears to be unstable due to a heavy top quark. According to \eq{condst}, it would then seem possible to cure this instability by coupling the SM to a singlet $S$, even very weakly, by introducing a sufficiently big threshold $\delta_{th}$. And this would only work provided that the instability occurs at a scale beyond $\Lambda_{th}$.  However, it is clear that we can send $\lambda_{SH}$ and $\lambda_S$ to very small positive numbers while keeping the value of $\delta_{th}$ unchanged. In such a limit, we are effectively decoupling the singlet from the SM and it would  be counterintuitive if stability could still be achieved for very small values of $\lambda_{SH}$. In fact, rewriting \eq{lambdath} as:
\begin{align} \label{contrast}
\lambda_{SH}\,\Lambda_{th}^2\sim 2\,\delta_{th}\frac{|m_S^2|}{\lambda}\,,
\end{align}
we see that if we reduce $\lambda_{SH}$, the value of $\Lambda_{th}$ has to increase for fixed $m_S^2$ (to keep constant the right-hand side). At some small $\lambda_{SH}$, the value of $\Lambda_{th}$ will then become larger than $\Lambda_I$, preventing altogether the possibility of curing the instability with $\delta_{th}$ for fixed $m^2_S$.   This suggests that the coupling $\lambda_{SH}$ may also play an important role in the mechanism of tree-level stabilization, which cannot depend only on the threshold $\delta_{th}$.

Another puzzle appears if the SM instability scale, $\Lambda_I$, happens to be below $|m_S^2|^{1/2}$ but above $\Lambda_{th}$, as can happen for small $\lambda_{SH}$, see \eqref{lambdath}. According to the stability conditions described above by \eq{lambdath} and \eq{condst} the stabilization with a threshold should be possible in this case. On the other hand, since the RG of the SM is to be trusted up to scales of the order of $|m_S^2|^{1/2}>\Lambda_I$, the stabilization does not appear to be feasible because the instability is reached before the threshold can have an effect. This issue could be resolved if another scale, higher than $\Lambda_I$, would forbid the stabilization. If there is such a scale, the argument of the previous paragraph tells that it should be related to $|m_S|$ and determined by $\lambda_{SH}$. We will now see how such an scale can actually become relevant.

After the discussion in section~\ref{sec:valleystree} about the lines of minima of the potential, we can gain a more intuitive understanding of the stability conditions. These must ensure that the potential along the $h$- and $S$-lines of minima is always positive, simply because absolute stability demands that the potential must be positive everywhere. Since these lines are good approximations to the actual valleys of the potential for small $\lambda_{SH}$, and the potential grows in the directions orthogonal to the bottom of the valleys, the potential along the lines will provide stringent stability conditions. 

The condition \eq{condst} can be obtained following the potential along the $h$-line as a function of the field $h$. Effectively, this means that we identify the renormalization scale, $\mu$, with $h$. This is an appropriate choice to study the potential at sufficiently high energies (where the instability region is). Assuming that $\lambda_S>0$ and $\lambda_{SH}>0$, the condition $\tilde\lambda = \lambda-\delta_{th}>0$ is immediately implied by \eqref{eq:Vhvalley}. The region of applicability of $\lambda>\delta_{th}$ follows from for the range of scales for which the $h$-line exists, which is given by \eq{htree}, i.e. $h^2\lesssim {2|m^2_H|}/{\lambda}\sim\Lambda_{th}^2$, where we have used \eq{eq:treematching} and $\tilde m_H^2\ll m_H^2$.  For scales much larger than $\Lambda_{th}$, which are not reached by the $h$-line, the potential near $S=0$ is dominated by the Higgs quartic and therefore the stability condition reduces to $\lambda>0$, in agreement with \eq{condst}. 

Similarly, we can follow the potential along the $S$-line expressed as a function of $h$, given by \eq{eq:svalleyh}, for the same reason as for the $h$-line. Since the potential along the $S$-line reproduces to lowest order the tree-level SM potential,\footnote{Notice that the potential along the $S$-line can be identified with the result of integrating out the heavy field $S$, which is the usual procedure to study the model for $h$ at low energies, i.e.\ much smaller than $|m_S|$. Indeed the $S$-line first appeared in section \ref{sec:SMS} for matching the SMS at low energies to the SM.} absolute stability requires  that the SM quartic coupling must be above zero, i.e.\ $\tilde\lambda>0$.  As before, this should occur for the whole range of scales for which the $S$-line exists, that is: $ h^2\lesssim 2|m^2_S|/\lambda_{SH}$. And again, for scales much larger than this one, the potential is dominated by the quartic couplings and the stability condition is simply $\lambda>0$. 

Therefore, we see that once we consider the $S$-line, a new scale enters into the game:
\begin{align} \label{lambdath2}
\hat\Lambda_{th}^2 \sim \frac{2|m^2_S|}{\lambda_{SH}}\,,
\end{align}
which makes explicit the relevance of $\lambda_{SH}$ for the tree-level stabilization mechanism. The scale $\Lambda_\lambda$ must be larger than both $\Lambda_{th}$ and $\hat\Lambda_{th}$ for the threshold effect to be able to cure the instability. The stability conditions can then be phrased as follows: the quartic coupling $\lambda$ must satisfy 
\begin{equation}\label{condstfull}
\lambda(\mu)>\left\{ 
  \begin{aligned}
  \delta_{th} &\\ 
  0 &
  \end{aligned}
\right.\quad\text{for}\quad
\begin{aligned}
 & \mu\lesssim \Lambda\\ 
 & \mu\gg \Lambda
  \end{aligned}\,,
\end{equation}
where the scale $\Lambda_\lambda$ at which the SM quartic Higgs coupling becomes negative must be such that
\begin{align} \label{cd1}
\Lambda_\lambda \gtrsim \Lambda
\end{align}
and we define
\begin{align} \label{cd2}
\Lambda \sim \text{Max}\left\{\Lambda_{th}\,,\hat\Lambda_{th}\right\}\,. 
\end{align}

We can now re-interpret the relation \eq{contrast} and use it to connect the two scales:
\begin{align}
\label{eq:lambdasrel}
\Lambda_{th}\sim\frac{\delta_{th}}{\lambda}\hat\Lambda_{th}\,.
\end{align}
As we argued above, in the decoupling limit (i.e.\ $\lambda_{SH}\rightarrow 0$) the scale $\Lambda_{th}$ will surpass $\Lambda_\lambda$ if $\delta_{th}$  and $m^2_S$ are kept fixed, violating the condition $\Lambda_\lambda \gtrsim \Lambda_{th}$ (which is necessary to implement the mechanism). By definition, this limit sends $\hat\Lambda_{th}$ to infinity, which clearly forbids the possibility of stabilization, according to \eq{cd1} and \eq{cd2}. The new scale $\hat\Lambda_{th}$ also explains the apparent puzzle that we discussed before for $\Lambda_{th}<\Lambda_I < |m_S|^2$. The definition of the scale \eq{cd2} tells us that in this situation the hierarchy of scales will be such that $\Lambda\sim\hat\Lambda_{th}>\Lambda_I$, preventing stabilization. In general, given the dependence of the scales $\Lambda_{th}$ and $\hat\Lambda_{th}$ on $\lambda_{SH}$, at least one of the scales will always be larger than $\lambda_\lambda$ (and even $\Lambda_I$) for sufficiently small $\lambda_{SH}$, precluding stabilization. 
 This result is in contrast to the analyses of \cite{Lebedev:2012zw} and \cite{EliasMiro:2012ay}, which left open the possibility of stabilization for very small $\lambda_{SH}$. Also, we note that stabilization is impossible (for perturbative values of the couplings) whenever 
\begin{align} \label{massscl}
|m^2_S|>\Lambda_\lambda^2\,.
\end{align}

\subsection{\label{subsec:inflationloop}Implications for inflation} 

As we have just seen, the threshold mechanism cannot stabilize the potential if $|m^2_S|>\Lambda_\lambda^2$. Notice that if the inequality $|m^2_S|>\Lambda_I^2$ holds, it also implies $|m^2_S|>\Lambda_\lambda^2$, since $\Lambda_I>\Lambda_\lambda$. As we have shown before, the required mass scale (squared) for inflation is $|m^2_S|\sim10^{26}\,{\rm GeV}^2$, which is larger than $\Lambda_\lambda^2\simeq 6.4\cdot 10^{21}$ (and  the instability scale of the SM: $\Lambda_I^2\simeq 2.5\cdot 10^{23}$ GeV$^2$) for the central values $m_t=173.15$ GeV and $m_h=125.09$ GeV. Therefore,  the threshold mechanism cannot stabilize the potential in this case.

The natural question that arises at this point is whether inflation in the SMS is doomed if the potential is unstable for large values of $h$. In order to answer this question, and according to the arguments we gave in section~\ref{introd}, we should consider if quantum fluctuations generated during inflation would put the Higgs beyond the instability scale. However, a simpler way of approaching the issue in this case consists in checking if inflation itself classically probes the region where the potential becomes unstable. More concretely: whether the values of $h$ reached during inflation fall in the instability region. 

Recalling the results of section~\ref{subsec:inflationtree}, inflation compatible with current CMB constraints can take place along the bottom of the $h$-valley, whose projection on field space can be very well approximated (for large $h$ values) by the line of equation \eq{htree}. As we just mentioned, in these scenarios the required mass scale is $|m^2_S|\sim10^{26}\,{\rm GeV}^2$. Assuming the central values for the Higgs and top masses, the possible instability (induced by the top Yukawa coupling) will affect inflation if the value of $h^2$ during the process needs to be of the order of $\Lambda_I^2\sim 10^{23}$ GeV$^2$ or bigger. We can easily estimate the maximum $h$ reachable in inflation along the $h$-valley by setting $S=0$ in the equation \eq{htree}. This gives $h^2=-2 m_H^2/\lambda$. If we now use the matching expression \eq{eq:treematching} for the Higgs mass parameter and the expression \eq{veVs} for the VEV of the singlet $S$, we get $ h^2 \simeq \lambda_{SH}\,v_S^2/\tilde\lambda$, where we have 
neglected the contribution of $\tilde m_H^2$ and we have approximated $\lambda$ by $\tilde\lambda$. Then, the condition for inflation to be safe from the instability, i.e.\ $h\ll \Lambda_I$, translates into
\begin{align} \label{boundsh}
\lambda_{SH}\ll \tilde\lambda\, \frac{\Lambda_I^2}{v_S^2}\,.
\end{align}
The value of $\tilde\lambda$ at the electroweak scale is about $0.27$, and roughly one order of magnitude smaller than this number when evaluated at $\Lambda_I$. Besides, as explained in section~\ref{subsec:inflationtree}, the value of $v_S$ needed for successful inflation is approximately $10$ to $20$ times larger than $M_P$. Inserting these numbers in \eq{boundsh} we obtain $\lambda_{SH}\ll 10^{-17}$, approximately.

This result already tells us that if the SM parameters are such that the effective potential of the SM is unstable, as the most up-to-date measurements and calculations point to, and if it cannot be stabilized, a necessary condition for inflation from the SMS requires that the Higgs portal coupling, $\lambda_{SH}$, has to be extremely weak. With the central values of the Higgs and top quark masses, for Higgs portal couplings larger than $10^{-17}$ the inflationary valley can sense the instability and hence acquire negative energies, which would forbid inflation with $h$ finishing on the right (electroweak) vacuum. Therefore, for values of $\lambda_{SH}$ larger than $10^{-17}$, the potential would have to be stabilized. In addition, we stress that if we take into the account large Higgs fluctuations, even a very small coupling between the Higgs and the inflaton may not be sufficient to make inflation safe \cite{Espinosa:2015qea}. We recall that the stabilization cannot be achieved for the central values of 
the Higgs and top masses via the threshold effect, since $|m^2_S|>\Lambda_I^2$, as explained in the previous section.

\begin{figure}\centering
\includegraphics[scale=1.2]{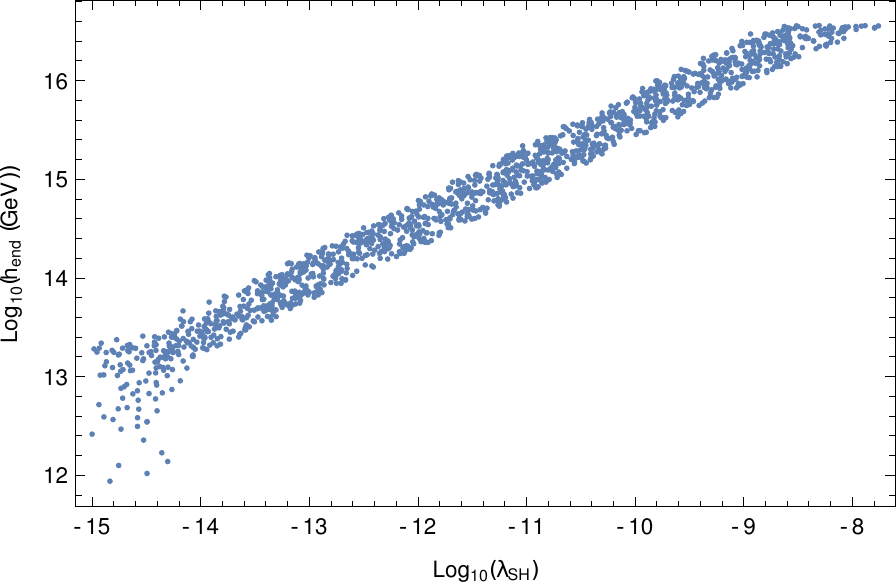}
\caption{\label{boundhfig} Values of $h$ at the end of inflation along valleys stabilized with $m_t=171.7$ GeV and yielding values of $A_s,n_s,r$ compatible with observations, as a function of $\lambda_{SH}$.}
\end{figure}

To illustrate further the depth in the direction of $h$ that is probed by inflation we have performed a parameter scan for models stabilized by a small enough value of $m_t$, as discussed  below in detail, using the two-loop RG-improved one-loop effective potential. In  figure\ \ref{boundhfig} we show the value of $h$ at which inflation ends, as a function of the coupling $\lambda_{SH}$. This is done for several choices of the other parameters, leading to successful inflation. The figure shows that for values of $\lambda_{SH}$ that are clearly above the limit \eq{boundsh} beyond which the possible instability starts to be worrisome (from a classical point of view), the value of $h$ at the end of inflation is indeed larger than $\Lambda_I$. Therefore, we conclude that if there is an instability in the SM (coming from the top Yukawa) it must be cured for successful inflation in the SMS (for not too small values of $\lambda_{SH}$), regardless of any consideration about quantum fluctuations of the Higgs during 
inflation. 

We have seen that if the actual Higgs and top masses correspond to the currently measured central values, the healing of the instability cannot come from the threshold effect discussed in the previous section because $|m^2_S|>\Lambda_I^2$. However, other values for $m_h$ and $m_t$ can raise the instability above $|m^2_S|^{1/2}$, potentially making the threshold stabilization viable. Assuming e.g.\ a Higgs mass of $125.09$ GeV, it turns out that $m_t<172.25$ GeV  is sufficient (at one-loop) to make $\Lambda_\lambda$ become larger than $|m^2_S|^{1/2}$. This value of the top mass  is above the current LHC bound of $m_t\gtrsim 171.6$ GeV \cite{CMS:2014hta,Aad:2015nba} and could in principle suffice to make the mechanism viable, provided that the SMS parameters satisfy  the condition \eq{condstfull}. Unfortunately, it is easily checked that \eq{condstfull}  cannot actually be fulfilled in this situation, due to the scale $\tilde\Lambda_{th}$, which remains above $|m^2_S|^{1/2}$ for perturbative values of $\lambda_
{SH}$. 

Smaller values of $m_t$ would raise the instability to even higher values of $h$. For instance, taking $m_t\simeq 172.0$ GeV, we obtain $\Lambda_I \simeq 1.2\cdot 10^{15}$ GeV and $\Lambda_\lambda \simeq 1.1\cdot 10^{14}$ GeV at one-loop. Using \eq{lambdath2}, and approximating $\Lambda_\lambda \sim 10^{14}$ GeV, we get that $\lambda_{SH}$ would have to be larger than $\sim 10^{-2}$, which implies $\lambda_S>10^{-3}$ from the definition \eq{lambdath}. However, these numbers are in tension with the values coming from inflation, because in section~\ref{infS} we obtained $\lambda_S\sim\vartheta\sim 10^{-13}$ and we were assuming a very small $\lambda_{SH}$. Furthermore, the value of $m_t$ for which the SM potential becomes stable in our calculations  (for the same Higgs mass of $125.09$ GeV)  is approximately  $171.75$ GeV, and hence the window of top masses where the threshold stabilization could be potentially playing an interesting role is necessarily extremely narrow, if it exists at all. We can actually 
estimate the width of this region imposing that the couplings are of the adequate orders of magnitude for successful inflation. Choosing $|m_S^2|\sim 10^{26}$ GeV and $\lambda_S\sim 10^{-13}$ GeV, the scale $\Lambda$ of \eq{cd2} becomes a function of $\lambda_{SH}$ alone. Then, the minimum of the function $\Lambda(\lambda_{SH})$, which is $\sim 5\cdot10^{16}$ GeV,  is a rough estimate of the minimum value of $\Lambda_I$ that would be needed for the threshold mechanism to work. Assuming $m_h=125.09$ GeV, we find that this occurs for $m_t \simeq 171.82$ GeV, which is very close to the value  ($171.75$ GeV with the one-loop effective potential, $171.76$ GeV with the  two-loop potential, both with a two-loop RG improvement) for which we find that the SM potential becomes absolutely stable. Note that this value of the top mass is marginally compatible with the ones measured at the LHC, but not so with the latest Tevatron combined result. Indeed, the currently allowed (and still relatively large) width of values 
of the top mass is constrained by the experiments at the LHC and Tevatron as follows:
\begin{equation}
\begin{aligned}
\label{eq:mt}
 m_t=& 172.38\pm0.10 (stat.)\pm0.65 (syst.) {\rm\,\,GeV},\quad \text{CMS \cite{CMS:2014hta}},\\
 m_t=& 172.99\pm0.48(stat.)\pm0.78(syst.) {\rm\,\,GeV},\quad \text{ATLAS \cite{Aad:2015nba}},\\
 m_t=& 174.34\pm 0.64 {\rm\,\,GeV},\quad \text{CDF + D0 \cite{Tevatron:2014cka}}\,.
\end{aligned}
\end{equation}
Our result for the limiting value of $m_t$ for $m_h=125.09$ GeV differs from the analyses of \cite{Degrassi:2012ry} and \cite{Espinosa:2015qea} by around +0.7 GeV. This difference could be attributed to our lower precision in the RG, since we did not include three-loop effects, or differences in the renormalization conditions and/or the matching of couplings to experimental measurements.  Checking whether this very narrow region may actually be of any relevance for the threshold effect would require a very accurate numerical analysis of the running of the couplings in the SM and the SMS and the stability conditions. We then conclude that if the top and Higgs masses are such that the SM potential is unstable when extrapolated to high field values, and if the singlet $S$ responsible for inflation couples to the SM, the stabilization through the threshold effect is not viable for most (and quite possibly all) of the parameter space. 

This conclusion about the compatibility of the threshold stabilization mechanism with inflation, although obtained in a minimal model, is expected to hold in more complicated examples involving additional fields coupling to the inflaton, provided that radiative corrections in the inflaton direction do not significantly alter  the Mexican hat profile. An exception is the scenario of \cite{Bhattacharya:2014gva}, which studied a non-minimal model in which the inflaton is coupled to fermions and a gauge field. In that case the Higgs direction can be stabilized by choosing $|m^2_S|<\Lambda^2_I$, but then the potential along the $S$ direction is dominated by $\lambda_S$, and the only way to make inflation compatible with CMB constraints is by means of radiative corrections coming from the additional fermions. These make the potential in the singlet direction unstable, and thus Higgs stability comes in that scenario at the price of an inflaton instability.

 We remark that the instability problem with inflation in the SMS simply would not arise if the actual top mass value would be such that the SM potential does not become negative. However, the current data indicate that this is just a restricted possibility. This exemplifies the relevance that the measurement of the top mass may have for primordial cosmology. Indeed, we can expect a similar issue for other models of inflation that we could think of coupling to the SM. 
 
For the values of $m_S^2$ singled out by the tree-level analysis, i.e.\ $|m^2_S|\sim 10^{26}$ GeV, taking $m_t\lesssim171.7$ GeV ensures absolute stability. This limiting value of $m_t$ corresponds to the situation in which the SM potential develops a false vacuum degenerate with the Higgs vacuum  for a Higgs mass of $m_h=125.09$ GeV. For greater values of $m_t$ the potential would no longer be absolutely stable. We have obtained these numbers matching the SM parameters to the experimental measurements as detailed in appendix \ref{app:SM_Pars}, which yields results for the Higgs couplings in terms of the physical masses that agree with those of \cite{Degrassi:2012ry} with relative deviations smaller than 0.24\% at the scale $m_t$, well within the theoretical error of 1\% reported there.\footnote{Using two-loop thresholds for the determination of the Higgs quartic from experimental data we get a value of $\lambda(m_t)$ which only differs by 0.20\% from \cite{Degrassi:2012ry}.}  Varying $m_h$ within its 
experimental error yields variations of this limiting value of $m_t$ by $\pm0.16$ GeV, while varying the strong coupling constant $\alpha_s(m_Z)$ changes it by $\pm0.24$ GeV. The theoretical uncertainty from varying the RG scale $\mu$ between $\mu={h}/{10}$ to $\mu=10 h$ in the RG-improved effective potential is even smaller, of the order of $\pm0.05$ GeV. From these numbers we can conclude that within experimental and theoretical uncertainties, the Higgs and top quark masses can be such that the SM and SMS remain absolutely stable. 

Figure \ref{boundhfig} corresponds to $m_t=171.7$ GeV (ensuring absolute stability at the level of approximation that we work) and has been obtained using the one-dimensional approximation of section \ref{subsec:One-dimensionalapprox}, but this time using the full one-loop potential, improved with the two-loop RG equations, and computing numerically the length travelled along the valley instead of approximating it by the value of the singlet field. The potential was calculated by starting with the SM potential, matched to experimental measurements as reviewed in appendix \ref{app:SM_Pars} --including one-loop electroweak thresholds and up to three-loop strong effects in the top Yukawa-- and subsequently matching at one-loop the SM to the high-energy model with the singlet (the SMS) as detailed in section \ref{sec:SMmatching}. The optimal renormalization scale at a given region of field-values is of the order of the dominant field-dependent mass, since this will minimize the logarithms of the form $\log (m^2_i(h,S)/\mu^2)$ 
appearing in perturbation theory. Given this, on the SM side the renormalization scale $\mu$ is chosen to interpolate between $v=246$ GeV for small values of the fields and $h$ for large ones. Similarly, on the high-energy side the scale is chosen to interpolate between $|m^2_S|^{1/2}$ and $h$. This is because in the region $S\lesssim v_S$, the largest effective masses are determined by $m^2_S$ for small $h$, while for large $h$ the Higgs interactions dominate and the largest masses are set by $h^2$. At the value of $h$ at which the potentials are matched, the renormalization scales differ across the threshold. Regarding the values of the cosmological parameters $A_s, n_s, r$, they were obtained from the slow-roll formulae of section~\ref{subsec:One-dimensionalapprox}. The number of e-folds was determined from the differential equation \ref{eq:efolds} rather than from the slow-roll approximation of \eqref{eq:efoldsslow}, and the end of inflation was determined imposing $\epsilon_{\mathcal H}=1$ instead of the 
approximate criterion $\epsilon=1$. 

\begin{figure}[t]\centering
\includegraphics[scale=.89]{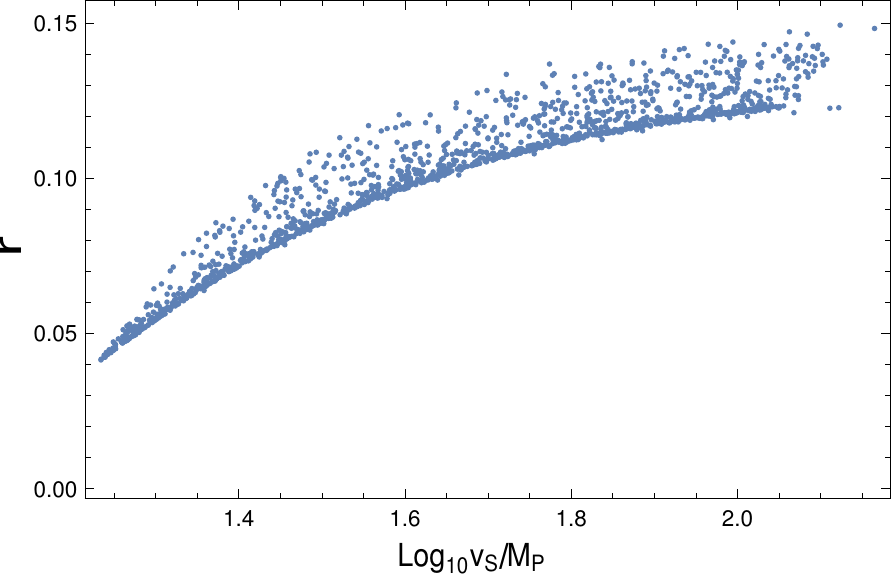}
\includegraphics[scale=.89]{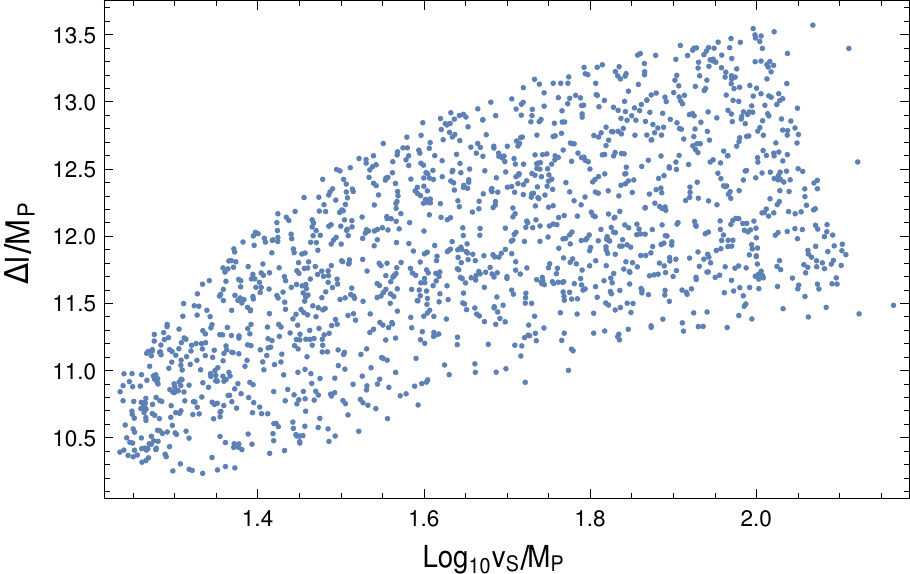}
\caption{\label{fig:loopscan}For $m_t=171.7$ GeV, scatter plots of points in which the two-loop improved one-loop potential along the $h$-valley producing successful inflation. Left: $r$ vs $v_S$. Right: distance travelled in the last 50 to 60 e-folds of inflation vs $v_S$.}
  \end{figure}

After a preliminary scan confirming the tree-level results of section~\ref{subsec:inflationtree}, we performed a scan of $10^4$ points focusing on the following region of parameter space: $\lambda_S\in\{10^{-10},10^{-14}\}$, $\lambda_{SH}\in\{10^{-6},10^{-15}\}$ and $|m^2_S|\in\{10^{25.6},10^{26.4}\}$, which selects the smallest values of $r$. In these scans we required that the lines of $h$-minima, which is reconstructed numerically, have to be connected to the Higgs vacuum without any intermediate barrier. In addition we enforced the values of $A_s$ and $n_s$ can be reproduced along the line, between 50 and 60 e-folds before the end of inflation. We chose a window of $A_s$ given by  Planck's more constraining 68\% $\Lambda{\rm CDM}$ confidence limit, $A_s=2.142\pm0.049$ \cite{Planck:2015xua}, plus an additional error coming from the theoretical uncertainty in the value of the potential at the matching scale, which we estimate  to be of the order of 5\% from varying the RG scale within a factor of 10. For $n_s$ 
we choose the window of \eqref{eq:nsrange}, that we already employed in the analysis at tree-level. The results are summarized in figures~\ref{boundhfig} and \ref{fig:loopscan}. Note that the minimum value of the tensor to scalar ratio sits around $r=0.04$, in perfect agreement with the tree-level results of section~\ref{subsec:inflationtree}, as it is clear from looking at figure~\ref{FrNe}. As was anticipated before, the length travelled along the valley is of the order of 10 $M_P$, and the minimum value of $v_S$ lies near 17 $M_P$. In the allowed points, the stabilized $h$-valleys can reach values of $h\sim10^{17}$ GeV. In fact, for these stabilized valleys we find a striking agreement between the tree-level results and those obtained with the more elaborate one-loop matching and RG-improved effective potential. In particular, the potentials along the bottom of the valleys obtained with the latter method are essentially identical to the ones obtained using the RG-improved tree-level expressions resulting from 
substituting the couplings and mass parameters in \eqref{eq:Vhvalley} and \eqref{eq:svalley} with their scale-dependent values, using the same field-dependent choice as the one described in section~\ref{subsec:mtvalley}.

If a future (and more precise) determination of the top and Higgs masses, together with a highly accurate calculation of the potential such as the one of \cite{Degrassi:2012ry} and \cite{Espinosa:2015qea}, definitively confirm an instability in the SM, the instability (and the scenario of inflation we have discussed) could be cured reverting the runaway behaviour of the potential at large $h$ by coupling the SMS to another singlet, $\overline{S}$, with a large enough Higgs portal coupling that stabilizes the potential without altering the properties of the $h$- and $S$-lines of minima at tree-level. This allows  to construct successful inflation along the $h$-valley, even with $m_t\gtrsim 173$ GeV, if the new singlet has a positive tree-level mass which stabilizes it at the origin. The potential of this extended model (the SMSS) would be:
\begin{align}
V^{\rm tree}(H,S,\overline S;\delta_i)=&V_0+V^{\rm tree}(H,S,\delta_i)+\frac{\overline m^2_S}{2} \overline S^2+\frac{\overline\lambda_S}{4!}\overline S^4+\frac{\overline\lambda_{SH}}{2}H^\dagger H \overline S^2+\frac{\overline\lambda_{SS}}{4}\overline S^2 S^2,
\end{align}
where $V_0$ is given in \eq{eq:treematching} and $V^{\rm tree}(H,S,\delta_i)$ is our starting potential of \eq{eq:V0}, which defines the SMS. If both $\overline S$ and $S$ are very massive, they can be integrated out and the SM potential will be recovered along the line in field space following their minima (as a function of $h$). For $\overline m^2_S>0$ (and positive $\overline\lambda_S$), the minima with respect to $\overline S$ are always at $\overline S=0$, and then the location of the minima with respect to $S$  have exactly the same tree-level dependence on $h$ as in the SMS, given by \eqref{sline}. This implies that the tree-level matching conditions with the SM will be the same as in the SMS: equation~\eqref{eq:treematching}, with no additional threshold contributions. Since the potential always increases for growing $|\overline S|$ and reduces to the SMS potential for $\overline S=0$, we will have the same valleys as before, with identical valley floors sitting at $\overline S=0$. Thus, all the 
conclusions regarding inflation reached at tree-level for the SMS carry over to the extended (SMSS) model. Quantum effects are different, though, and in particular stability
may be achieved thanks to the additional contributions of the couplings of $\overline S$ to the beta function of the Higgs quartic, which at two-loop order is now given  by\footnote{See appendix \ref{eq:betas2} for additional beta functions in the SMSS.}
\begin{align}
 \beta_{\lambda}=&\tilde\beta_{\lambda}-\frac{1}{16\pi^2}\left(\bar\lambda_{SH}^2+ \lambda_{SH}^2\right)+\frac{1}{(16\pi^2)^2}\left(4 \bar\lambda_{SH}^3+5\lambda  \left(\bar\lambda_{SH}^2+\lambda_{SH}^2\right)+4 \lambda_{SH}^3\right)\,,
\end{align}
where $\tilde\beta_{\lambda}$ is the beta function in the SM. We have checked, using the two-loop RG improvement of the one-loop effective potential in the extended SMSS, with appropriate one-loop matching conditions to the SM (adapting to this model the methods detailed in section~\ref{subsec:RGimproved} and appendix \ref{sec:SMmatching}) that the example of figures \ref{fig:sdeviations}--\ref{fig:correctionsh} can be stabilized choosing $m_t=173.15$ GeV and  $\overline\lambda_{SH}=0.7$. Moreover, the results for the cosmological parameters change by less than 3\% using either the stabilization by a low enough value of $m_t$ or by adding the extra singlet $\overline S$.

\begin{figure}[t]\centering
 \includegraphics[scale=.95]{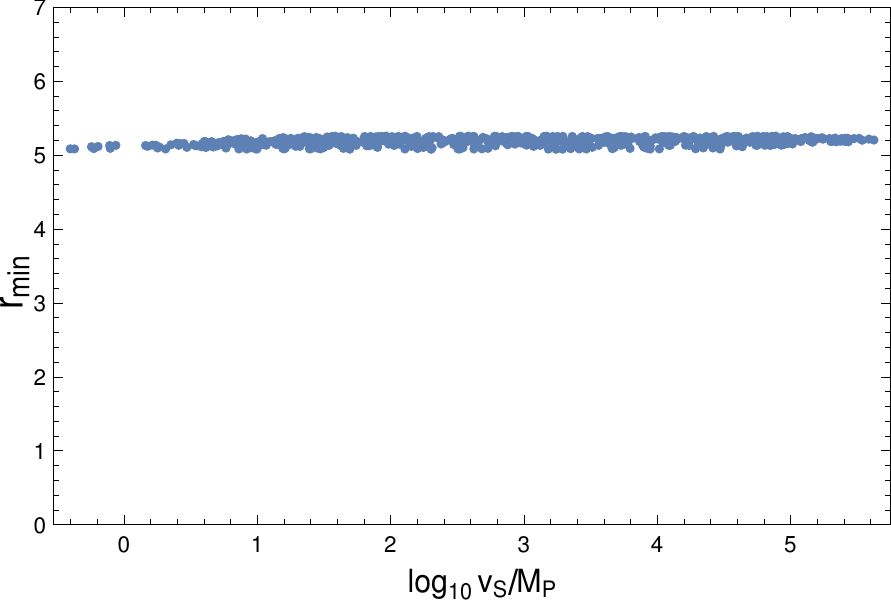}%
 \caption{\label{fig:mtscan}Lower bound for the tensor-to-scalar-ratio inside the top-valley, obtained with the RG-improved effective potential by looking for the minimum value of the potential along the valley whenever the latter ends before intersecting the line of $S$-minima.
 }
 \end{figure}

\subsection{\label{subsec:nonminimal coupling}Stability in the presence of a non-minimal gravitational coupling}

So far, we have have neglected direct couplings of the scalar fields to the curvature, $R$. This type of coupling is the basis for the simplest model of inflation in which the Higgs is the inflaton \cite{Bezrukov:2007ep}. There, the Higgs is coupled to the metric through a term in the Lagrangian of the form
\begin{equation} \label{coup}
{\cal L}\supset\sqrt{-g}\, \xi H^\dagger H R\,, 
\end{equation}
where $\xi$ is a large ($\sim 10^{3}$--$10^{4}$) positive number. This large value of the coupling can flatten the SM potential sufficiently at large Higgs values, if it is stable, allowing for inflation. 

In the SMS, there is no need for such a coupling to produce inflation, since the potential along the $h$-valley can easily be flat enough without it. However, it is nonetheless interesting to consider a coupling, such as \eq{coup}, of the Higgs to $R$ in the SMS. The first reason to do it is that this kind of couplings are always generated radiatively through the RG (even if they are set to zero at some scale).\footnote{See e.g.\ \cite{Lerner:2009xg} for the relevant one-loop RG equations.} Although we can always assume that they are negligible at the scales of interest for inflation (as we have been doing up until now), they will be present in the most general case. Besides, a positive value of $\xi$ suppresses the quantum fluctuations of the Higgs during inflation, because for positive $R$ (as in a de Sitter background) the interaction  \eq{coup} rises the effective Higgs mass, which is shifted by an amount
\begin{align} \label{mshift}
 \delta m^2_H \simeq 12\,\xi \mathcal{H}^2\,,
\end{align} 
since in a FLRW metric $R=6/a^2(\ddot a  a+\dot a^2)$ and $\mathcal{H}=\dot a/a$ is approximately constant during inflation. This effect, which has been studied  e.g.\ in \cite{Espinosa:2007qp,Herranen:2014cua, Espinosa:2015qea}, can prevent the Higgs from falling into the instability region of the SM potential (through quantum fluctuations during inflation), provided that the coupling $\xi$ satisfies $\xi(\mu=m_t)\gtrsim 10^{-2}$, depending on the concrete value of $\mathcal{H}$. 

In spite of the effect of $\xi$ for suppressing those dangerous quantum fluctuations, the classical trajectories of the fields might still fall in the instability regions of the potential if its shape allows them to do so. We recall that the threshold mechanism is most likely unable to stabilize the SM potential with the values of the SMS parameters that are needed for inflation in the case $\xi=0$. In section~\ref{subsec:stability} and section~\ref{subsec:inflationloop}, we found that the region of parameter space where the threshold stabilization mechanism might work is very small and possibly empty. This is basically due the large value of the singlet mass: $|m_S|\sim10^{13}$ GeV that is needed to reproduce the amplitude of primordial perturbations. In addition, we saw that if the SM potential is unstable, the Higgs field reaches during inflation values that probe the instability region whenever $\lambda_{SH}\gtrsim 10^{-17}$.  To understand what happens if the coupling \eq{coup} is present, we can study the dynamics 
of inflation using the modified Friedmann equation:
\begin{align} \label{friedxi}
3\left(M_P^2-\xi h^2\right)\mathcal{H}^2=\left(\frac{\dot h^2}{2}+\frac{\dot S^2}{2}+V\right)
\end{align}
and the dynamical equations for the fields $S$ and $h$
\begin{align}
\label{eq:heq}\ddot h +3\mathcal{H}\dot h+\frac{\partial V}{\partial h}+\xi h R=&0\,,\\
\ddot S +3\mathcal{H}\dot S+\frac{\partial V}{\partial S}=&0\,,
\end{align}
where we are assuming for simplicity that the singlet $S$ does not couple directly to $R$. Even if $\xi$ is irrelevant for the threshold stabilization mechanism in the Jordan frame, it may change the parameters needed for inflation or the maximum value of $h$ that is reached during the process.
Therefore, in order to see whether the coupling of the Higgs to $R$ changes the picture for inflation in the SMS, we have to check whether the presence of $\xi$ allows an inflationary background capable of reproducing the measured primordial spectra, enough e-folds, and such that $h\ll \Lambda_I$\,. 

If we assume that $\xi\sim \mathcal{O}(1)$ (positive or negative, it does not matter) during inflation, the last three equations show that the inflationary background will not change significantly with respect to the $\xi=0$ case whenever the inflationary dynamics  is dominated by the field $S$. This is because in this case one has $\dot h\ll \dot S$, as well as $h\ll M_P$, so that the $\xi$-dependent terms in \eqref{friedxi} become suppressed. The factor  $M_P^2-\xi h^2$ in the modified Friedmann equation \eqref{friedxi} is an effective Planck mass which has to stay positive. Then, it is clear that a non-zero $\xi$ cannot fix inflation for trajectories close to the bottom of the $h$-valley when the potential becomes negative. However, trajectories that start far enough from the bottom of the valley, in a region with $V>0$, might be saved from falling into the instability region due to the positive effective mass contribution of the curvature, as it is apparent from  \eqref{eq:heq}. We will not study this any further as we focus on  inflation along the bottom of the $h$-valley. In this case, in  order to have a relevant change in the inflationary dynamics the non-minimal coupling has to be considerably larger than $\mathcal{O}(1)$, so that the $\xi$ terms in the equations of motion become important. Notice that the maximum possible value of $h$ during inflation in the case $\xi=0$ is approximately $h^2\sim -2m_H^2/\lambda\sim \lambda_{SH} v_S^2/\tilde\lambda\sim-6\lambda_{SH} m_S^2/(\lambda_S\tilde\lambda)$, which can be expected to be much smaller than $M_P^2$ for reasonable values of the quartic couplings. A detailed analysis of inflation in the SMS for $\xi\gg 1$ is beyond the scope of this 
work.

\subsubsection{Effect of $\xi$ in a generic inflationary background}

In this section we study the possible effect of the coupling \eq{coup} on the stability from a different perspective. Assuming that $R$ is positive and constant, $\xi$ modifies the relative importance of the effective Higgs mass with respect to the quartic coupling. Therefore the coupling \eq{coup} can be seen effectively in this approximation (and in the Jordan frame) as a contribution to the potential that may in principle affect the threshold stabilization mechanism in any inflationary background. We recall that although $\xi$ is generically produced radiatively,  the $\overline{\text{MS}}$ running of the quartic couplings is not affected by it, so we will only be concerned with tree-level effects. This also means that the instability scale $\Lambda_I$ is independent of $\xi$.

Assuming an inflationary background, no matter its origin, it is easy to see that the effect of $\xi$ can be seen as a change of the scale $\Lambda_{th}$ defined in \eq{lambdath} and appearing in the scale $\Lambda$ of \eq{cd2}, which in turn sets the stability condition \eq{condstfull}. This should come as no surprise because $\Lambda_{th}$ is basically determined by the extension of the $h$-line. In the presence of a non-zero coupling $\xi$, the induced quadratic mass term for the Higgs is
\begin{align}
 m^2_{H,\xi}=\tilde m^2_H+ \delta m^2_H+3\frac{\lambda_{SH}}{\lambda_S}m^2_S,
\end{align}
where we have used the matching condition of \eqref{eq:treematching} and  $\delta m^2_H$ is given in \eq{mshift}.  Following the arguments of section~\ref{subsec:stability}, we see that we just have to replace $m^2_H$  with $m^2_{H,\xi}$ to obtain the new $\Lambda_{th}$ scale, which is
\begin{align}
 \Lambda_{th,\xi}=-2 \frac{m^2_{H,\xi}}{\lambda}\,.
\end{align}
This scale is either smaller or larger than $\Lambda_{th}$ for  positive or negative $\xi$, respectively. Notice that this result assumes that $m^2_{H,\xi}$ is negative so the equation $\partial V/\partial h=0$ has a solution, see \eq{htree}. If $\xi$ is large enough, so that the new mass squared $m^2_{H,\xi}$  becomes positive, the scale $\Lambda_{th,\xi}$ can be identified with zero, and so it will be irrelevant for the stability conditions, because the potential will not have a local minimum away from zero in the $h$ direction.

On the other hand, it is clear that the scale $\hat\Lambda_{th}$, is not affected by the coupling $\xi$ in this picture, since $\hat\Lambda_{th}$ is related to the size of the effective quadratic interaction of the singlet $S$ but not to that of the Higgs.\footnote{However, the scale $\hat\Lambda_{th}$ would be affected by a non-minimal gravitational coupling of the singlet. This would also modify the inflationary background so we do not consider this possibility.} Therefore, if $\xi>0$, the relevant scale for stability in an inflationary background,  $\Lambda_{\xi}={\rm Max}\{\Lambda_{th,\xi},\hat\Lambda_{th}\}$, can differ from the original $\Lambda$ of \eq{cd2}  only  if $\Lambda_{th}>\hat\Lambda_{th}$. In this case, $\xi>0$ lowers $\Lambda$ and enhances the stability of the potential during inflation. Given \eqref{eq:lambdasrel} and \eqref{eq:delta}, this will only happen if the coupling $\lambda_{SH}$ is large enough, i.e.\ $\lambda^2_{SH}>\lambda\lambda_S/3$. 

In the specific case of the SMS with inflation mainly driven by $S$ and with $\lambda_S\sim10^{-13}$ and $\lambda\sim 10^{-1}$, which are standard values for $\xi=0$, we can use the equations \eq{eq:delta} and \eq{eq:lambdasrel} to obtain that $\lambda_{SH}\gtrsim 10^{-7}$ is needed for $\Lambda_{th}>\hat\Lambda_{th}$. This value of $\lambda_{SH}$ is much larger than $10^{-17}$, which is approximately the maximum value for which the field $h$ remains below the instability region during inflation. Therefore, if the Higgs potential is not absolutely stable and if $\xi>0$  we see that  $\xi$ cannot enhance the threshold stabilization mechanism and protect the $h$-valley from reaching the Higgs instability region. 

In the case $\xi<0$, we have the following inequality between scales: $\Lambda_{th,\xi}>\Lambda_{th}$. If the difference between the two is sufficiently large, it will set $\Lambda_{th,\xi}$ above the instability scale $\Lambda_I$,\footnote{The instability scale is independent of $\xi$, even if we consider \eq{coup} as part of the potential. The reason is that at large field values the potential is dominated by the quartic couplings and the running of these is not affected by the direct couplings to $R$, as we already mentioned.} rendering impossible the threshold stabilization mechanism. Looking at it differently, the destabilizing effect of a negative $\xi$ in a generic inflationary background could in principle be compensated by the threshold stabilization mechanism only if $\Lambda_{th,\xi}$ remains below $\Lambda_I$. However, in the concrete case of the SMS, we have shown earlier that threshold stabilization is essentially incompatible with the singlet $S$ playing the role of the inflaton. This conclusion remains true 
with a negative $\xi$, since this coupling can only make $\Lambda_{th,\xi}\geq\Lambda$.

\subsection{\label{subsec:mtvalley}Higgs false-vacuum inflation}

If for a given Higgs mass in the SM, the top quark mass, $m_t$, is tuned with high precision to be close to its lower stability bound,\footnote{We recall that absolute stability is not preferred by the current central values of the Higgs and top masses.}  the quartic Higgs coupling can graze negative values and become positive again once the effects of the gauge couplings dominate the running of $\lambda$. As it is well known, a false (metastable) vacuum  appears in this case. In the SMS, this vacuum can extend into the $S$-direction, giving rise to a new line of minima, that we term ``top-line''.  The minima disappear for large values of $S$ due to $\lambda_{SH}>0$, which generates an effective quadratic term for $h$. For this reason, the value of $h$ along the line will not fall below the value of the local maximum located before the false vacuum at $S=0$. The line is thus approximately straight in field space, and solves \eqref{eq:lineh} for $h$ near $h_t$, where $h_t$ is the false 
vacuum appearing at $S=0$. Therefore, it is very approximately the projection of an actual valley (the ``top-valley'') on the plane $\{S,h\}$. Substituting $h$ by its value $h_{t}$ at the false vacuum, the potential along the bottom of this valley is approximately 
\begin{align} \label{eq:Vmt}
 V_{m_t}(S)=V_0+\frac{1}{2}\left(\tilde m^2_H+3\frac{\lambda_{SH}}{\lambda_S}m^2_S\right)h_t^2+\frac{1}{8}\left(\tilde\lambda+3\frac{\lambda^2_{SH}}{\lambda_S}\right)h_t^4+\frac{1}{2}{m^2_S}_{\rm eff}\,S^2+\frac{\lambda_S}{4!} S^4\,,
\end{align}
where $V_0$ is given in \eq{eq:treematching} and 
\begin{align} \label{Sefft}
{m^2_S}_{\rm eff}=m^2_S+\frac{\lambda_{SH}}{2}h^2_{t}\,.
\end{align}
In contrast to the $S$- and $h$-valleys, the top-valley is not guaranteed to be connected with the  vacuum of \eqref{veVs} along a line of decreasing potential energy. For this to happen, a first condition is that the  top-valley should slope downwards  away from $S=0$, which will occur if there is a negative effective mass for $S$ near $h=h_{t}$, i.e. ${m^2_S}_{\rm eff}<0$. Secondly, the false vacuum in the $h$-direction should disappear while the value of the potential at the bottom of the valley is still decreasing, allowing the fields to roll down to the present vacuum. The disappearance of the false vacuum in $h$ is controlled by the portal coupling since,  as explained before, the coupling acts like a mass for the field $h$ for a fixed value of $S$. If the two conditions are met, one could in principle have inflation starting along the top-valley, with the fields ending in the current Higgs vacuum. The possibility that inflation could be generated inside the top-valley was studied in \cite{Masina:
2012yd} and \cite{Notari:2014noa, Fairbairn:2014nxa}, with the latter works concluding that this was not possible for the measured value of the Higgs mass. Studies focused on the gravitational waves sourced by false-vacuum inflation were done in \cite{Masina:2011un,Masina:2014yga}, which concluded that it is possible to achieve values of the tensor-to-scalar-ratio $r\lesssim 0.2$ for values of the Higgs and top masses compatible with current measurements. 

\begin{figure}[t]
\centering
\includegraphics[scale=.9]{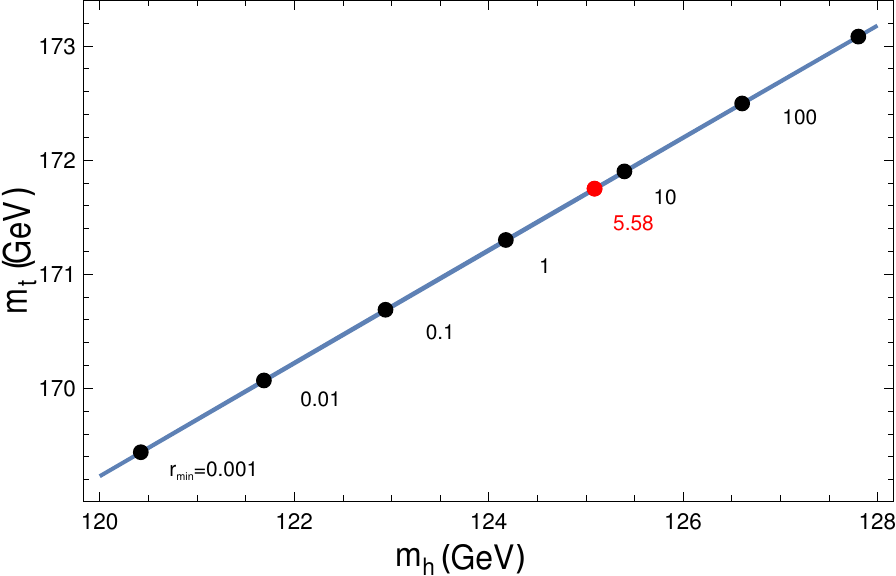}
\includegraphics[scale=.9]{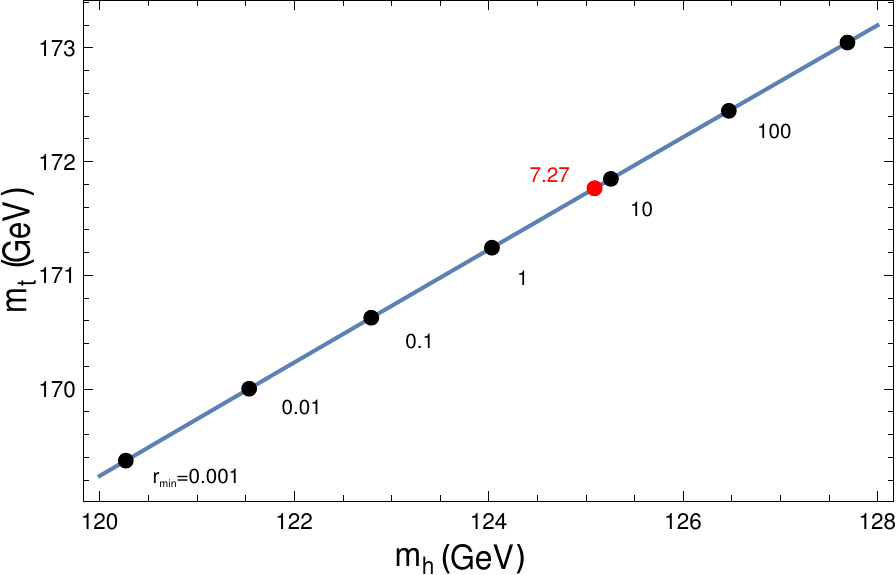}
\caption{\label{fig:mtmh}\small Values of $m_h$ and $m_t$ giving rise to a plateau in the two-loop improvement of the one-loop (left) and two-loop (right) effective potential in the SM. The black points indicate cases in which the lower bound for the tensor-to-scalar ratio, \eqref{eq:rbound}, is given by an integer power of ten. The red point corresponds to the measured value of the Higgs mass, $125.09$ GeV.
  }
\end{figure}

It is straightforward to obtain a rough estimate of the expected amount of gravitational waves that are produced in this scenario. We just have to use the expression
\begin{align}
\label{eq:rbound}
 r=\frac{2V}{3\pi^2\, A_s\, M^4_P}\,,
\end{align}
which comes from combining the slow-roll expressions of \eqref{eq:Asnsr} and \eq{eq:reps}. In \eq{eq:rbound} the only unknown  is the value of the potential, $V$, since the amplitude of scalar perturbations is well determined, and given in \eq{measuredAs}. A lower bound on the potential, and hence a lower bound on $r$, is given approximately by the energy density of the SM plateau that appears tuning the Higgs and top masses. This can be understood as follows. First, in order for the fields to escape the top valley, the false vacua in the $h$-direction have to disappear for a value of $S$ for which the energy along the top valley still decreases as a function of $S$. This has to occur before the top-line meets the $S$-line, since the latter follows the local minima of the potential in the $S$-direction. At the point in which the top-valley disappears, the false vacuum in $h$ becomes flat, corresponding to a plateau. The energy density of this plateau will be minimal when the plateau is reached just 
at the $S$-line. Along the $S$-line the potential as a function of $h$ matches the SM potential
up to higher dimensional terms. Therefore we conclude that the energy of the SM plateau corresponds to a lower bound on the potential along the top-valley, which gives a lower bound on the values of $r$ achievable inside it, thanks to \eqref{eq:rbound}. Such a connection between the values of $r$ in Higgs false-vacuum inflation and the energy of the SM plateau was already made in \cite{Masina:2011un,Masina:2014yga}.

\begin{figure}[t]\centering
  \hskip-.2cm\includegraphics[scale=.92]{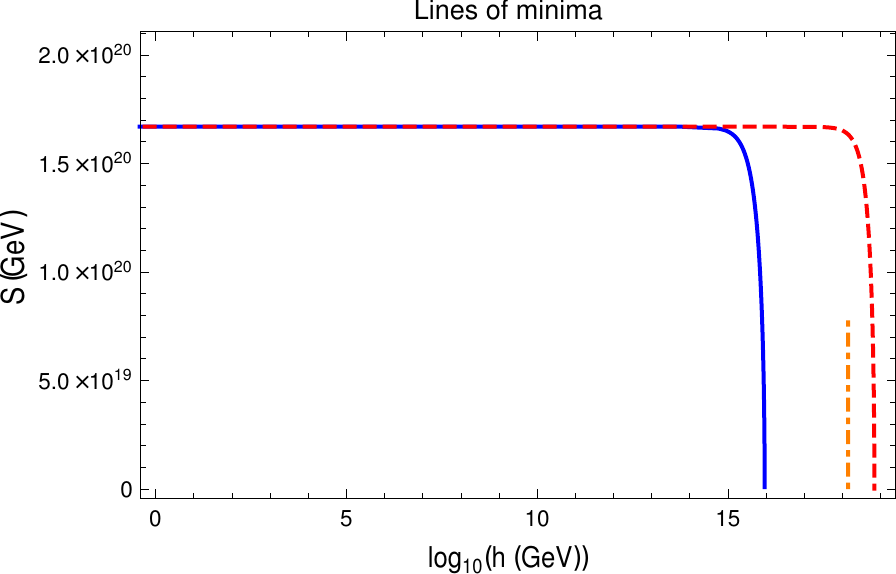}%
  \includegraphics[scale=.86]{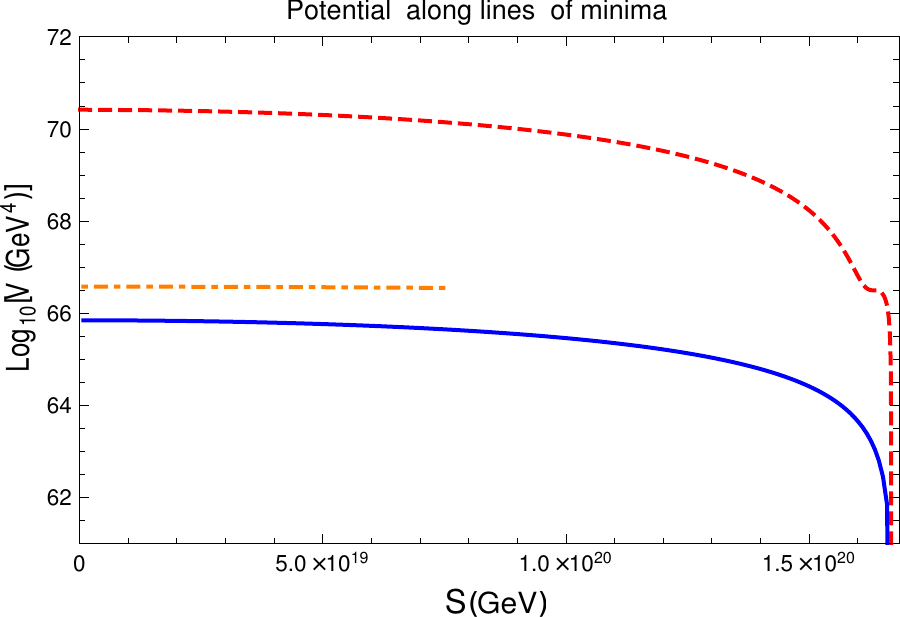}%
\caption{\label{fig:valleys1}Lines of minima and potential along them for the full one-loop potential with 2 loop RG-improvement for $m_t=171.725$ GeV,$m_h=125.09$ GeV, $\lambda_S=2.15\cdot 10^{-14},\lambda_{SH}=4.18\cdot10^{-12}$ and $m^2_S=-1\cdot10^{26} {\rm GeV}^2$. The $S$-line is shown with red dashed lines, the $h$-line with a solid blue line, and the top-line, which ends at $S=7.5\cdot10^{19}$ GeV, is depicted with a dot-dashed orange line. Near the top-line appears the quartic $\lambda$ is close to zero, which also causes the dent in the potential along the $S$-line.
}
\end{figure}

\begin{figure}[t]\centering
\includegraphics[scale=.89]{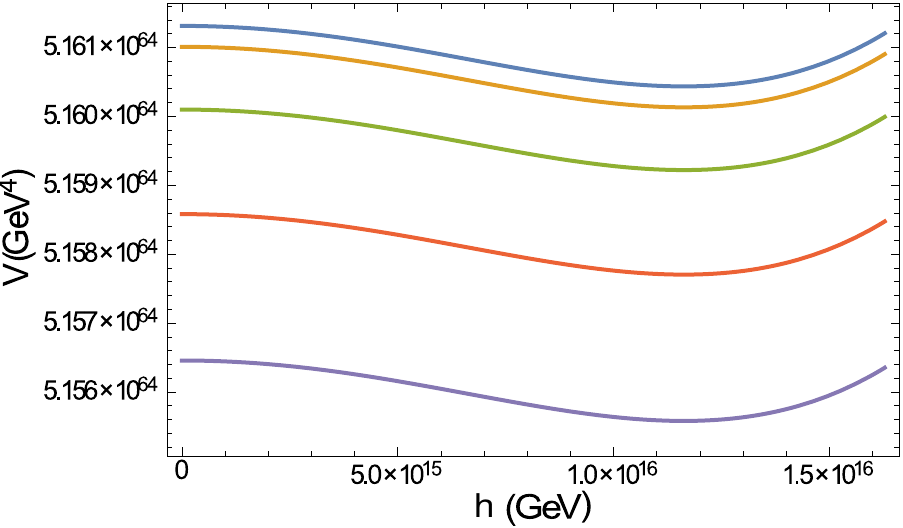}
\includegraphics[scale=.89]{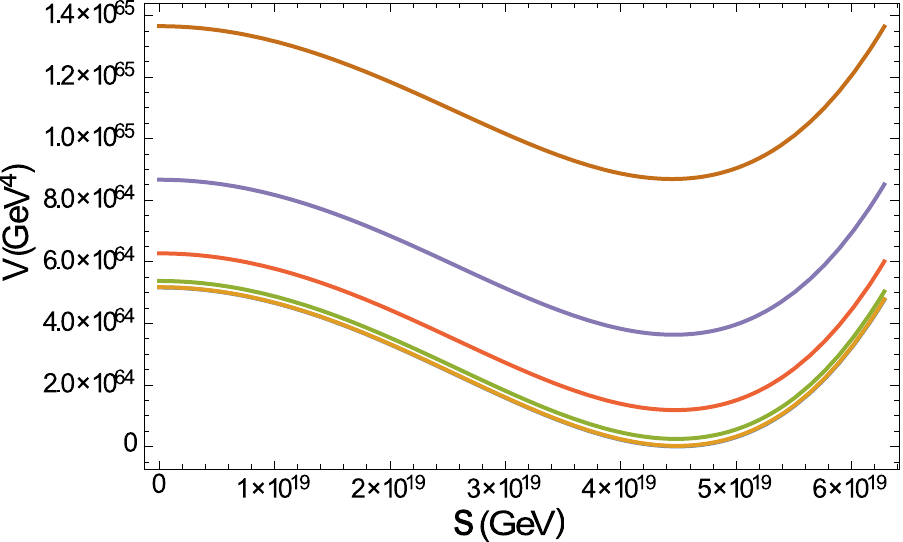}
\caption{\label{fig:valleycuts}For $m_t=171.7$ GeV, $\lambda_S=3.05\cdot10^{-13},\,\lambda_{SH}=2.66\cdot10^{-10},m^2_S=-1.02\cdot10^{-26}$: Left: Potential in the $h$-direction for $S=0,0.1 M_P,0.2 M_P, 0.3 M_P, 0.4 M_P$ (from top to bottom), illustrating the minima giving rise to the $h$-line. Right: Potential in the $S$-direction for $h=0.05 M_P,0.04 M_P,0.03 M_P, 0.02 M_P, 0.01 M_P$ (from top to bottom), illustrating the $S$-line. }
  \end{figure}

Using the two-loop improvement of the SM effective potential, and  matching couplings to experimental values as in appendix \ref{app:SM_Pars}, for $m_h=125.09$ GeV the plateau arises for $m_t=171.75$ GeV, with an energy equal to 
\begin{align}
\label{eq:Vplateau}
 V^{SM, 1-loop}_{\rm plateau}=6.22\cdot10^{66}\,{\rm GeV}^4\,.
\end{align}
The corresponding bound for $r$ is  $r>5.58$, which is far larger than the current upper bound $r\lesssim 0.11$ coming from CMB measurements. In order to estimate the robustness of this lower bound, we can check how experimental and theoretical uncertainties affect the value of the energy scale of the plateau in the SM. First, it should be pointed out that value of $V^{SM}_{\rm plateau}$ of \eqref{eq:Vplateau} is compatible with the results in \cite{Degrassi:2012ry}. We can   estimate its uncertainty by varying the Higgs mass $m_h=125.09\pm 0.21(stat.)\pm 0.11(syst.)$ and $\alpha_s(m_Z)=0.1185(6)$ within their experimental errors, as well as by varying the RG scale of the effective potential  between $\mu=1/10 h$ and $\mu=10h$. Doing this, the lowest value achieved for $V^{SM}_{\rm plateau}$ is 
\begin{align}
\label{eq:Vplateaumin}
  V^{SM,1-loop}_{\rm plateau,min}=0.95\cdot10^{66}\,{\rm GeV}^4,
\end{align}
which yields a bound of $r$ still well beyond current constraints, $r>0.86$. Moreover, we have repeated these calculations using the full two-loop effective potential of the SM (with a two-loop RG improvement), calculated with the formulae of \cite{Martin:2001vx}, and including two-loop thresholds at zero momentum in the determination of the Higgs quartic from experimental data. This yields values of $V^{SM}_{\rm plateau}$ compatible with the above within the uncertainty, and with less dispersion due to the improved two-loop scale dependence:
\begin{align}
\label{eq:Vplateau2}
 V^{SM, 2-loop}_{\rm plateau}=8.10\cdot10^{66}\,{\rm GeV}^4,\\
 V^{SM, 2-loop}_{\rm plateau,min}=2.73\cdot10^{66}\,{\rm GeV}^4.
\end{align}
The two-loop plateau is obtained for $m_t=171.763$ GeV.
Using the minimum two-loop value value of \eqref{eq:Vplateau2} yields a bound of $r\geq2.45$. It is clear then that even when taking into account theoretical and experimental uncertainties, the bounds of $r$ cannot be relaxed to anywhere near $r\simeq 0.1$.  We thus can confidently conclude that the observed cosmological parameters cannot be attained within the top-valley.\footnote{In \cite{Notari:2014noa} it was argued that the scenario could be saved with the addition of a non-minimal coupling to gravity.} In particular, values of $r\lesssim 0.2$ as were obtained in \cite{Masina:2014yga} are in conflict with our results. For completeness we show in figure~\ref{fig:mtmh} the values of $m_t$ and $m_h$ that yield a plateau in our calculations, and the corresponding values of the lower bound on the tensor-to-scalar-ratio.  We show the results obtained with both the one-loop (left) and two-loop (right) effective potentials, both improved with the two-loop RG equations; note the similarity between 
the lines that mark the values of the masses giving rise to a plateau. Finally, as already noted in \cite{Masina:2014yga}, these bounds for $r$ obtained from the limiting case of a SM plateau should also apply to very generic models of Higgs false vacuum inflation whenever the potential becomes close to  that of the SM with a plateau at the end of inflation. This is indeed generic since inflation will imply a rolling from high to low values of potential energy, such that by the end of inflation the  false vacuum in the Higgs direction should disappear, implying the appearance of a plateau. If the false vacuum appears as a consequence of the same top-Higgs interplay as in the SM, the plateau at the end of inflation will be close to the SM plateau and the bounds for $r$ derived here will apply.

The only possible caveat of the previous estimates relying on a SM calculation is that higher order effects in the matching between the SMS and the SM, and the difference in the beta functions of the two theories, might modify the value of the minimum energy inside the top valley. After all, the potential along the $S$-line only reproduces the SM accurately at scales much lower than the value of the Higgs at the plateau, $h_{plateau}\sim 2\cdot10^{18}$ GeV.
In order to discard the possibility that these effects could be significant, we have used again the two-loop RG-improvement of the one-loop effective potential  to scan the space of parameters, searching for points that are compatible with the constrains that are coming from both particle physics data and CMB measurements. Apart from the implementation of the matching between the SM and the SMS at one-loop, the use of two-loop RG equations in the SMS and a more precise matching of the SM parameters to experimental measurements, our analysis improves upon the previous ones by avoiding the assumption that the variations of $V$ along the valley are much smaller than the value of the potential at the false vacuum of $S=0$. The numerical scan was done in the following range of parameters: $\lambda_S \in\{10^{-13},10^{-20}\}$, $\lambda_{SH} \in\{10^{-6},10^{-20}\}$ and $-m^2_S \in\{10^{25},10^{30}\}\,{\rm GeV}^2$. The ranges of $\lambda_{S}$ and $m^2_S$ were chosen after performing preliminary scans using tree-
level approximations of the potential along the top-line, and choosing intervals for which $A_s$ and $n_s$ could be fit either inside the $h$-valley or the top-line. The upper value  of $\lambda_{SH}$ is motivated by requiring that the effective mass of $S$ for $h=h_v\sim 2\cdot10^{18}$ GeV remains negative.
  
Our calculations show that it is possible to fit the observed values of $A_s$ and $n_s$ inside the top-valley for a Higgs mass compatible with experimental measurements, if $|m^2_S|\sim10^{30}\,{\rm GeV}^2$. However, this can only happen if the top-valley reaches very close to the end of the $S$-line, which marks the maximum extension allowed for the top-valley if the fields are to be able to roll down towards the Higgs vacuum. 
We note that the fact that $A_s$ and $n_s$ can only be matched to their measured values near the maximum length allowed for the $h$-valley implies that one will only be able to have very few e-folds of observable inflation along the top-valley. This is similar to what was found in \cite{Fairbairn:2014nxa}; however, it is not enough to discard these scenarios since, as we have seen, one could still generate additional e-folds of inflation along the $h$-valley; but again, these scenarios are ruled out because they predict too high values of $r$. The scan confirms the bounds estimated with the SM plateau, as shown in figure~\ref{fig:mtscan}, which gives $r>5.07$, similar to the estimate of \eqref{eq:Vplateau}. The 10\% difference comes mainly from the fact that the conditions for a plateau in the SMS are different than in the SM, given the nontrivial matching relations and the different beta functions. The requirement that the fields can roll down in the $S$-direction from  the false-vacuum at $S=0$, 
$h=h_v$ gives an upper bound on $\lambda_{SH}$, so that for the range of $\lambda_S$ interesting for inflation, the tree-level threshold in the matching relation for the quartic coupling, \eqref{eq:treematching}, is very small. Given this, one-loop effects in the matching become relevant, and $\lambda$ ends up being slightly smaller than $\tilde\lambda$ at the matching scale. As a consequence of this, the values of $m_t$ needed for a plateau or a false vacuum in the SMS, as well as the energy of the latter, become smaller than in the SM, giving a slightly lower bound on $r$. For example the plateau arises for values of $ m_t$ between $171.72$ GeV and $m_t=171.73$ GeV. Still, the bound on $r$ remains an order of magnitude above current limits, so that we can confidently rule out the possibility that observable inflationary perturbations were generated inside the top-valley.

It should be noted that although a full period of successful inflation is discarded inside the top-valley, this does not rule out some inflationary period happening before inflation continues outside the valley. It was argued in \cite{Masina:2012yd,Fairbairn:2014nxa} that once the fields exit the valley, only a very small number of e-folds could be generated before the fields reached the Higgs vacuum. As we explained before, this is not true in general because of the existence of the $h$-valley, which as we saw in section~\ref{subsec:inflationtree} can support inflation. The $h$-valley was ignored in previous works because they focused on the contributions of the quartic couplings to the effective potential, and as we saw the $h$-valley arises from the interplay between quadratic and quartic interactions. We have checked that the fields in some cases can even pass from the top-valley to the $h$-valley while maintaining the slow-roll condition $\epsilon_{\mathcal H}<1$. A period of slow-roll inflation along the top-valley, followed by a period of fast roll between valleys, and a final period of slow-roll inflation along the $h$-valley might in principle be possible. Another effect ignored in earlier works takes place when $\lambda_{SH}$ is large enough so that the $h$-valley reaches values of the Higgs field close to those in which radiative false-vacua may appear. In this case the destabilizing effects of the top Yukawa can deepen and broaden the $h$-valley, to the point that this valley and the top-valley become a single valley connected to the Higgs vacuum, which may again support inflation. This situation is essentially the same as the $h$-valley inflation analyzed previously. The points for which this happens were removed from the plot of figure~\ref{fig:mtscan}.

To close this section,  we include some figures illustrating the lines of minima calculated with the RG-improved effective potential, with stabilized valleys. figure~\ref{fig:valleys1} shows the  lines in field space and the potential along them for a  value of $m_t$ for which a top-valley arises. figure~\ref{fig:valleycuts} shows the potential energy along curves parallel to the $h$ and $S$ axes. 

\section{\label{sec:Conclusions}Conclusions}
We have studied the possibility of embedding the inflationary sector into the SM through a $Z_2$ Higgs portal, in a model that we dubbed ``SMS''. This is motivated by the obvious requirement of reheating the universe at the end of inflation and by the role of the large quantum fluctuations that are induced on light fields during inflation. If as the data seem to suggest, the SM potential is metastable \cite{Degrassi:2012ry,Buttazzo:2013uya,Espinosa:2015qea}, these fluctuations can suffice to push the Higgs into the instability region while inflation is taking place, see \cite{Espinosa:2007qp,Espinosa:2015qea}. Since a heavy scalar, $S$, coupling to the Higgs through a portal (and taking a large VEV) can provide a stabilization mechanism via a tree-level threshold \cite{Lebedev:2012zw,EliasMiro:2012ay}, it is important to determine whether this scalar could drive inflation as well. We have found that successful inflation can indeed occur in this model, but the stabilization of the potential cannot be provided 
by the inflaton itself, as we explain below.

We have first studied the potential energy valley that provides an attractor trajectory for inflation, focusing in the limit of small Higgs portal coupling, $\lambda_{SH}$. We have shown that inflation can be described in a very good approximation with a single-field effective theory along this valley. The inflaton field corresponds to a combination of the heavy singlet and the Higgs, and represents the length travelled along the valley. However, in this limit inflation takes place mostly in the direction of $S$, which makes reasonable identifying this field as the inflaton. This identification is exact in the decoupling limit, i.e.\ when $\lambda_{SH}$ is sent to zero. The variation of the potential energy along the floor of the valley is described by a Mexican hat potential, which provides a good fit to current CMB data \cite{Ade:2015lrj}  if the inflation takes a large VEV of the order of 15 $M_P$. This is one of the reasons why inflation has to proceed mostly in the direction of the singlet, since the 
extension of the valley in the Higgs direction for small $\lambda_{SH}$ is much smaller than the VEV of the singlet. In addition, a small effective quartic coupling is required to fit the amplitude of primordial perturbations, which also impedes inflation from going substantially along the Higgs direction. During inflation, the field $S$ interpolates between a hilltop-like behaviour and a quadratic potential with positive curvature. In particular, we find that one may have a tensor-scalar ratio as small as $r\gtrsim0.04$ for values of $A_s$ and $n_s$ within their 95\% confidence levels. Using these results, we are able to estimate the associated mass scale of the (very heavy) singlet, which turns to be of the order of $10^{13}$ GeV, as well as its self-coupling $\lambda_S \sim 10^{-13}$. In the limit of small coupling between the SM and the singlet, $\lambda_{SH}$ cannot be determined with CMB measurements, but it may be possible to do it for larger values, thanks to the deformations that the inflationary 
valley would undergo in that case.  Assuming that the SM is indeed unstable and using a purely classical argument, we have estimated that $\lambda_{SH}$ has to be at most $10^{-17}$ if  the Higgs is to be safe during inflation. On top of this, we stress the relevance of quantum fluctuations, which are important even in the extreme case of exact decoupling limit \cite{Espinosa:2015qea}.

If the heavy singlet of the SMS plays the role of the inflaton, the tree-level threshold stabilization mechanism mentioned above does not apply. There are two reasons for this. First, the singlet mass required for successful inflation, $\sim 10^{13}$ GeV, lies above the SM instability scale $\Lambda_I$ (which is of the order of $10^{12}$ GeV) for the central values of the Higgs and top masses. And second, because even if the masses where such that the instability would be pushed beyond the mass scale of the singlet, the conditions of applicability for the mechanism are largely incompatible with the values of the couplings needed for inflation. We find that the tree-level stabilization only has a chance of being successful for a very restricted narrow band of top masses, which is most likely negligible.

Furthermore, we have shown that regardless of any consideration related to inflation, the mechanism fails in general for a sufficiently small portal coupling $\lambda_{SH}$. This is due to the appearance of a relevant scale, which was not identified in previous works. This scale can grow above the SM instability scale as $\lambda_{SH}$ decreases, eventually becoming unbounded in the decoupling limit. The actual scale that has to be compared to $\Lambda_I$, in order to determine whether the threshold effect can cure the instability, is the largest one between two competing scales, see \eq{lambdath} and \eq{lambdath2}, that have opposite behaviours under variations of $\lambda_{SH}$. The need of the new scale \eq{lambdath2} that we have identified in this work can be understood intuitively by realizing that  the value of the quartic threshold in the decoupling limit can be kept unchanged if the self-coupling of the singlet is modified accordingly. This implies that the stabilization cannot depend on the 
value of the quartic  threshold alone, since it is clear that the mechanism should not work if the SM and the singlet are completely disconnected from each other. The two competing scales, \eq{lambdath} and \eq{lambdath2}, can be identified by following the potential along the directions given by the lines of minima with respect to the Higgs and the heavy singlet. 

Coming back to inflation, the inapplicability of the threshold stabilization is worrisome given the large Higgs fluctuations sourced by inflation.\footnote{In principle, another reason of concern is that for non-zero $\lambda_{SH}$ the classical attractor trajectories may reach values of $h$ beyond the instability scale. For small $\lambda_{SH}$, we expect this to be a subdominant effect in comparison to that of the quantum fluctuations, but it can become more significant if the coupling is larger than the values we have considered here.} Stabilization, however, can be simply achieved for models with $m_t\lesssim 171.7$ GeV, which are still marginally allowed by current experimental results from CMS (less so for ATLAS and even less for CDF+D0). In this respect, it should be noted that our calculations are less precise than the state-of-the art SM results of \cite{Degrassi:2012ry}, \cite{Espinosa:2015qea}. 

Other possibilities to stabilize the potential can be imagined. One that we have considered here consists in including a second singlet stabilized at the origin, which does not change the shape of the potential energy valleys at tree-level. We have checked that whenever the potential is stabilized in this way or with an appropriate choice for $m_t$, the predictions for inflation for small $\lambda_{SH}$ including loop corrections are essentially identical to the ones obtained at tree-level. Alternatively, one could consider stabilization via higher-dimensional operators in the effective potential \cite{Datta:1996ni,Grzadkowski:2001vb,Branchina:2013jra,Eichhorn:2014qka}, which could also affect the inflationary power spectra, leading to less sharp predictions. It has been argued that their relevance for the stability is small whenever that their effects can be reliably computed \cite{Burgess:2001tj}.  

We have also studied the possible stabilizing role of a non-minimal gravitational coupling, $\xi$, of the Higgs in the SMS, showing that the picture does not change for $\xi$ of order unity. Although such a coupling can suppress quantum fluctuations of the Higgs
in an inflationary background, in general it does not alter
significantly the inflationary background itself in the SMS or the
threshold stabilization mechanism. It would be interesting to study the
interplay between the inflationary dynamics and the threshold
stabilization mechanism in the SMS for values of $|\xi|$ larger than 1.

Given that the SMS had been studied earlier in the context of Higgs false-vacuum inflation, in which an inflationary valley arises radiatively due to the effect of a tuned top quark Yukawa on the running of the the Higgs quartic coupling, we have reconsidered this scenario here. A detailed calculation of the potential, taking into account theoretical and experimental uncertainties, allows us to rule out the possibility of successful inflation in this situation, since primordial gravitational waves are overproduced, in qualitative agreement with the overall conclusions of \cite{Fairbairn:2014nxa,Notari:2014noa} and in contrast to  previous claims \cite{Masina:2011un,Masina:2014yga}. 

\section*{Acknowledgements}

GB thanks Perimeter Institute for hospitality at the very beginning of this work. Research at  Perimeter  Institute  is  supported  in  part  by  the
Government of Canada through Industry Canada, and by the Province of Ontario through the Ministry of Research and Information (MRI). GB thanks as well the Departament de F\'isica Fondamental at the Universitat de Barcelona and the CERN Theory Division for hospitality at different stages of this work. CT acknowledges support of the Spanish Government through grant FPA2011-24568 (MICINN), and thanks Rhorry Gauld and Anupam Mazumdar for useful conversations. GB thanks Brando Bellazzini,  Alberto Casas, Mikael Chala, Jos\'e Ram\'on Espinosa, Mathias Garny, Gian Giudice and Felix Kahlhoefer for valuable discussions and comments on a draft version of this work. We also thank Isabella Masina and Alessio Notari for useful exchanges.

\appendix
\section{\label{app:RGs}Two-loop RG equations}
Here we present the two-loop RG equations in the $\overline{\rm MS}$ scheme for the models obtained by adding one or two singlets to the SM. These RG equations have been used to evaluate the RG-improved effective potential in the calculations of section~\ref{subsec:inflationloop}. The beta functions and anomalous dimensions are defined as in \eqref{eq:betadef}. For couplings and fields already present in the SM, for compactness we give their beta functions/anomalous dimensions in terms of the SM results, denoted with tildes. 
The beta functions have been obtained by applying the results of references \cite{Machacek:1983tz,Machacek:1983fi,Machacek:1984zw,Luo:2002ti}. For the two-loop beta functions in the SM, see also  \cite{Luo:2002ey}.

\subsection{SM with a real singlet (SMS)} \label{eq:betassinglet} 
We consider new scalar interactions as given in \eqref{eq:V0}. The beta functions are given next: 
\begin{align}
 \nonumber\beta_{g_i}=\tilde\beta_{g_i},
  \end{align}
\begin{align}
  \nonumber\beta_{y_i}=\tilde\beta_{ y_i}+\frac{1}{4(16\pi^2)^2}\lambda^2_{SH}y_i,
   \end{align}
\begin{align}
  \nonumber\beta_{\lambda}=&\tilde\beta_{\lambda}+\frac{\lambda^2_{SH}}{16\pi^2}+\frac{1}{(16\pi^2)^2}\left(-4 \lambda_{SH}^3-5 \lambda  \lambda_{SH}^2\right),
   \end{align}
\begin{align}
 \nonumber \gamma_{H}=\tilde\gamma_{H}+\frac{\lambda^2_{SH}}{4(16\pi^2)^2},
  \end{align}
\begin{align} \nonumber
 \beta_{m^2_H}=\tilde\beta_{m^2_H}+\frac{m^2_S \lambda_{SH}}{16\pi^2}+\frac{1}{(16\pi^2)^2} \lambda_{SH}^2 \left(-\frac{m^2_H}{2}-2m^2_S\right),
 \end{align}
\begin{align}
 \nonumber  \beta_{\lambda_S}=&\frac{1}{16\pi^2}\left(3 \lambda _S^2+12 \lambda_{SH}^2\right)+\frac{1}{(16\pi^2)^2}\left[ \lambda_{SH}^2 \left(-72 y_b^2+\frac{72 g_1^2}{5}+72 g_2^2-20 \lambda _S-72 y_t^2-24 y_{\tau }^2\right)\right.\\
  \nonumber &\left.-\frac{17}{3}  \lambda _S^3-48 \lambda_{SH}^3\right],
 \end{align}
\begin{align}
  \nonumber \beta_{\lambda_{SH}}=&\frac{1}{16\pi^2}\left[ \lambda_{SH} \left(6 y_b^2-\frac{1}{10} 9 g_1^2-\frac{9 g_2^2}{2}+6 \lambda +\lambda _S+6 y_t^2+2 y_{\tau }^2\right)+4 \lambda_{SH}^2\right]+\frac{1}{(16\pi^2)^2}\left[ \lambda_{SH} \left(y_t^2 \bigg(\right.\right.\\  
 \nonumber &\left.-21 y_b^2+\frac{17 g_1^2}{4}+\frac{45 g_2^2}{4}+40 g_3^2-36 \lambda \right)+\frac{5}{4} g_1^2 y_b^2+\frac{45}{4} g_2^2 y_b^2+40 g_3^2 y_b^2-36 \lambda  y_b^2-\frac{27 y_b^4}{2}\\
 \nonumber &+\frac{36 g_1^2 \lambda }{5}+36 g_2^2 \lambda +\frac{15}{4} g_1^2 y_{\tau }^2+\frac{15}{4} g_2^2 y_{\tau }^2+\frac{1671 g_1^4}{400}+\frac{9}{8} g_2^2 g_1^2-\frac{145 g_2^4}{16}-15 \lambda ^2-\frac{5 \lambda _S^2}{6}-\frac{27 y_t^4}{2}\\
 \nonumber &\left.\left.-12 \lambda  y_{\tau }^2-\frac{9 y_{\tau }^4}{2}\right)+ \lambda_{SH}^2 \left(-12 y_b^2+\frac{3 g_1^2}{5}+3 g_2^2-36 \lambda -6 \lambda _S-12 y_t^2-4 y_{\tau }^2\right)-\frac{1}{2} 21 \lambda_{SH}^3\right],
  \end{align}
\begin{align}
 \nonumber  \gamma_{S}=\frac{1}{(16\pi^2)^2}\left[\frac{\lambda _S^2}{12}+ \lambda_{SH}^2\right],
  \end{align}
\begin{align}
 \nonumber  \beta_{m^2_S}=&\frac{1}{16\pi^2}\left[4m^2_H \lambda_{SH}+m^2_S \lambda _S\right]+\frac{1}{(16\pi^2)^2}\left[ \lambda_{SH} \left(-24 y_b^2m^2_H+\frac{24}{5} g_1^2m^2_H+24 g_2^2m^2_H\right.\right.\\
 \nonumber  &\left.\left.-24m^2_H y_t^2-8m^2_H y_{\tau }^2\right)+ \lambda_{SH}^2 \left(-8m^2_H-2m^2_S\right)-\frac{5}{6} m^2_S \lambda _S^2\right].
\end{align}

\subsection{SM with two real singlets (SMSS)} \label{eq:betas2}
The scalar interactions in this case are as in \eqref{eq:V0}. The beta functions follow:
\begin{align} 
\nonumber
\beta_{g_i}=\tilde\beta_{g_i},
\end{align}
\begin{align}
 \nonumber\beta_{y_i}=\tilde\beta_{ y_i}+\frac{1}{4(16\pi^2)^2}(\lambda^2_{SH}+\overline\lambda^2_{SH})y_i,
\end{align}
\begin{align}
  \nonumber\beta_{\lambda}=\tilde\beta_{\lambda}+\frac{1}{16\pi^2}\left(\bar\lambda_{SH}^2+ \lambda_{SH}^2\right)+\frac{1}{(16\pi^2)^2}\left[-4 \bar\lambda_{SH}^3+\lambda  \left(-5 \bar\lambda_{SH}^2-5 \lambda_{SH}^2\right)-4 \lambda_{SH}^3\right],
  \end{align}
\begin{align}
 \nonumber\gamma_{H}=\tilde\gamma_{H}+\frac{1}{(16\pi^2)^2}\left(\frac{\bar\lambda_{SH}^2}{4}+\frac{ \lambda_{SH}^2}{4}\right),
\end{align}
\begin{align}
 \nonumber \beta_{m^2_H}=\tilde\beta_{m^2_H}+\frac{1}{16\pi^2}\left(\bar{m}^2_S \bar\lambda_{SH}+m^2_S \lambda_{SH}\right)-\frac{1}{(16\pi^2)^2} \left[\bar\lambda_{SH}^2 \left(2 \bar{m}^2_S+\frac{m^2_H}{2}\right)+ \lambda_{SH}^2 \left(\frac{m^2_H}{2}+2m^2_S\right)\right],
  \end{align}
\begin{align}
 \nonumber \beta_{\lambda_S}=&\frac{1}{16\pi^2}\left[3 \lambda _{S \bar{S}}^2+3 \lambda _S^2+12 \lambda_{SH}^2\right]+\frac{1}{(16\pi^2)^2}\left[-5 \lambda _S \lambda _{S \bar{S}}^2-12 \lambda _{S \bar{S}}^3+ \lambda_{SH}^2 \left(-72 y_b^2+\frac{72 g_1^2}{5}+72 g_2^2\right.\right.\\
 \nonumber &\left.\left.-20 \lambda _S-72 y_t^2-24 y_{\tau }^2\right)-\frac{1}{3} 17 \lambda _S^3-48 \lambda_{SH}^3\right],
\end{align}
\begin{align}
 \nonumber \beta_{\lambda_{SH}}=&\frac{1}{16\pi^2}\left[\lambda _{S \bar{S}} \bar\lambda_{SH}+ \lambda_{SH} \left(6 y_b^2-\frac{1}{10} 9 g_1^2-\frac{9 g_2^2}{2}+6 \lambda +\lambda _S+6 y_t^2+2 y_{\tau }^2\right)+4 \lambda_{SH}^2\right]\\
 \nonumber &+\frac{1}{(16\pi^2)^2}\left[ \lambda_{SH} \left(-\frac{1}{2} \lambda _{S \bar{S}}^2-4 \lambda _{S \bar{S}} \bar\lambda_{SH}-\frac{\bar\lambda_{SH}^2}{2}+y_b^2 \left(\frac{5 g_1^2}{4}+\frac{45 g_2^2}{4}+40 g_3^2-36 \lambda -21 y_t^2\right)\right.\right.\\
 \nonumber &-\frac{27 y_b^4}{2}+g_1^2 \left(\frac{9 g_2^2}{8}+\frac{36 \lambda }{5}\right)+36 g_2^2 \lambda +\left(\frac{17 g_1^2}{4}+\frac{45 g_2^2}{4}+40 g_3^2-36 \lambda \right) y_t^2+\left(\frac{15 g_1^2}{4}+\frac{15 g_2^2}{4}\right.\\
 \nonumber &\left.\left.-12 \lambda \right) y_{\tau }^2+\frac{1671 g_1^4}{400}-\frac{145 g_2^4}{16}-15 \lambda ^2-\frac{5 \lambda _S^2}{6}-\frac{27 y_t^4}{2}-\frac{9 y_{\tau }^4}{2}\right)-2 \lambda _{S \bar{S}} \bar\lambda_{SH}^2-2 \lambda _{S \bar{S}}^2 \bar\lambda_{SH}\\
 \nonumber &\left.+ \lambda_{SH}^2 \left(-12 y_b^2+\frac{3 g_1^2}{5}+3 g_2^2-36 \lambda -6 \lambda _S-12 y_t^2-4 y_{\tau }^2\right)-\frac{1}{2} 21 \lambda_{SH}^3\right],
 \end{align}
\begin{align}
 \nonumber \gamma_{S}=\frac{1}{(16\pi^2)^2}\left[\frac{1}{4} \lambda _{S \bar{S}}^2+\frac{\lambda _S^2}{12}+ \lambda_{SH}^2\right],
 \end{align}
\begin{align}
  \nonumber\beta_{m^2_S} &=\frac{1}{16\pi^2}\left[\bar{m}^2_S \lambda _{S \bar{S}}+4m^2_H \lambda_{SH}+m^2_S \lambda _S\right]+\frac{1}{(16\pi^2)^2}\left[-\frac{1}{2}m^2_S \lambda _{S \bar{S}}^2-2 \bar{m}^2_S \lambda _{S \bar{S}}^2+ \lambda_{SH} \left(-24 y_b^2m^2_H\right.\right.\\
 \nonumber &\left.\left.+\frac{24}{5} g_1^2m^2_H+24 g_2^2m^2_H-24m^2_H y_t^2-8m^2_H y_{\tau }^2\right)+ \lambda_{SH}^2 \left(-8m^2_H-2m^2_S\right)-\frac{5}{6} m^2_S \lambda _S^2\right].
\end{align}
The expressions for $\beta_{\bar \lambda_S},  \beta_{\bar\lambda_{SH}},   \gamma_{\bar S},  \beta_{\bar m^2_S}$ can be obtained from the formulae above for the ``unbarred'' couplings/fields by making the substitutions $\{\lambda_S,\bar\lambda_S,\lambda_{SH},\bar\lambda_{SH}\}\leftrightarrow\{\bar\lambda_S,\lambda_S,\bar\lambda_{SH},\lambda_{SH}\}$. 


\section{\label{app:SM_Pars}Matching the SM couplings to experimental measurements}

    Throughout this paper, we work in the $\overline{MS}$ scheme and fix all the SM parameters from experimental measurements except for the top mass, which we vary so as to satisfy stability constraints or to generate a false vacuum in the Higgs direction. Nevertheless, the values that we consider are compatible with the latest results: $m_t=172.38\pm0.10 (stat.)\pm0.65 (syst.)$ GeV by CMS \cite{CMS:2014hta} and $m_t=172.99\pm0.48(stat.)\pm0.78(syst.)$ GeV by ATLAS \cite{Aad:2015nba}. Also, we note that the models of section~\ref{subsec:inflationloop} in which stability is ensured by an additional scalar can perfectly accommodate masses equal to the central values of $m_t$ arising from other experiments. In the following we briefly comment on the determination of the most important parameters affecting the effective potential, which are the gauge couplings, $\tilde m^2_H$ and $\tilde\lambda$\footnote{Recall that we define SM parameters with tildes, to distinguish them from those of the high-energy models with 
additional fields}.
    
    First, the values of the $\overline{MS}$ gauge couplings are derived from the results in the Particle 
     reviews~\cite{Agashe:2014kda}, $\alpha_s(m_Z)=0.1885(6)$, $\alpha(m_Z)^{-1}=127.916\pm0.015$, $\sin^2\theta_w(m_Z)=0.231 26(5)$. For the determination of the top Yukawa from the chosen physical top mass we include one-loop electroweak threshold corrections taken from \cite{Hempfling:1994ar}, and three-loop strong corrections from \cite{Chetyrkin:1999qi,Melnikov:2000qh}. Our determination of $y_t(m_t)$ as a function of the physical masses $m_t$ and $m_h$ coincides with the results of the numerical formulae given in \cite{Degrassi:2012ry} within a relative precision of 0.6\%.
    
    The Higgs mass is fixed by the latest  combined ATLAS and CMS results, $m_h=125.09\pm 0.21(stat.)\pm 0.11(syst.)$ GeV \cite{Aad:2015zhl}, and we take the PDG value of the Fermi
    constant $G_F= 1.1663787(6)\cdot10^{-5}\,\,{\rm GeV}^{-2}$\footnote{The Fermi constant is extracted from experimental measurements of the muon lifetime by matching them to the predictions in the QED-improved Fermi theory, and is a physical parameter independent of the RG scale (see \cite{PhysRevD.22.971} or the review~\cite{PhysRevD.86.010001})}. We
    impose the renormalization condition that the Higgs VEV corresponds to a minimum of the full effective potential\footnote{Another condition found in the literature is to set $v^2=\frac{1}{\sqrt{2}G_F}$ at a given scale, see e.g. \cite{Degrassi:2012ry}}, i.e.
\begin{align}
\label{eq:Higgsvev}
 \left.\frac{dV}{dh}\right|_{h=v}=0.
\end{align}
With this convention, one may derive a relation between the Fermi constant and the Higgs VEV by matching results in the QED improved Fermi theory and the SM, setting to zero the contributions from tadpole diagrams. Doing this in the one-loop formulae of \cite{Hempfling:1994ar}, which are based on  \cite{PhysRevD.22.971}, one gets 
\begin{align}
\label{eq:vGF} v^2=&\frac{1}{\sqrt{2}G_F}\left[1+\frac{\Pi_{WW}(0)}{m^2_W}+E\right],
\end{align}
\begin{align}
\Pi_{WW}(0)=& \Pi_{WW}^{bos}(0)+ \Pi_{WW}^{fer}(0),
 \end{align}
 \begin{align}
\Pi_{WW}^{bos}(0)=&\frac{\alpha w}{4\pi S^2_w}\left[\left(-2+\frac{1}{c^2_w}\right)\Delta(w)+\left(2+\frac
  {1}{c^2_w}-\frac{17}{4S^2_w}\right)\log c^2_w-\frac{3}{4}\frac{h}{w-h}\log\frac{w}{h}-\frac{17}{4}+\frac{7}{8c^2_w}-\frac{h}{8w}\right],
\end{align}
\begin{align}
 \Pi_{WW}^{fer}(0)=&\frac{\alpha}{8\pi S^2_w}\sum_{U,D}N(U,D)\left[-m^2_U\left(\Delta(m^2_U)+\frac{1}{2}\right)-m^2_D\left(\Delta(m^2_D)+\frac{1}{2}\right)+\frac{m^2_U m^2_D}{m^2_U-m^2_D}\log\frac{m^2_U}{m^2_D}\right],
\end{align}
\begin{align} \label{eq:vGF2}
 E=\frac{\alpha}{4\pi S^2_w}\left[4\Delta(z)+\left(\frac{7}{2S^2_w}-6\right)\log c^2_w+6\right],
\end{align}
where $S^2_w$ and $c^2_w$ denote the sine and cosine squared of the Weinberg angle, $U$ and $D$ denote up and down fermions, and we used the definitions $\Delta(x)\equiv m^2_x$, $w\equiv m^2_W$, $z\equiv m^2_Z$ and  $h\equiv m^2_h$. 

A final condition comes from requiring that the theory reproduces the measured Higgs mass $m_h=125.09$ GeV. The mass is associated with the zero of the 1PI momentum-space two-point function, and the corresponding equation can be written as
\begin{align}
\label{eq:Higgsmass}
  \frac{d^2}{dh^2}V(h)+\Delta \Pi(m^2_h)=m^2_h,
\end{align}
where the derivatives of the effective potential capture the zero momentum contribution to the 1PI 2-point function $\Pi$, while
\begin{align}
 \Delta \Pi(m^2_h)= \Pi(m^2_h)- \Pi(0)
\end{align}
implements the necessary finite-momentum correction and can be obtained  from the general results of \cite{Martin:2003it}.

The Higgs couplings $\tilde m^2_H$ and $\tilde\lambda$ can be then determined from $G_F$ and the physical mass $m_h$ by using \eqref{eq:Higgsvev} with the value of $v$ determined from \eqref{eq:vGF}--\eqref{eq:vGF2}, together with condition \eqref{eq:Higgsmass}. Even when using a different renormalization condition for the Higgs VEV than in \cite{Degrassi:2012ry}, we find that our determination of $\lambda(\mu=m_t)$ as a function of $m_t$ and $m_h$ agrees with the numerical formulae of that reference within a relative precision of 0.2\%.

\section{\label{sec:SMmatching}Matching the extended model to the SM}

In this section we elaborate on the matching between the SMS and the SM introduced in section~\ref{sec:SMS}, taking care to formulate it precisely in terms of the RG-improved effective potentials introduced in section~\ref{subsec:RGimproved}. In the notation used in that section, the matching equations~\eqref{eq:ssol0} and \eqref{eq:match0} correspond to
\begin{align}
 \label{eq:ssol}&\left.\frac{d\hat V}{dS}\right|_{S=S_{min}(h)}=0\,,\\
\label{eq:match} & \hat V^{SM}(\tilde h,\tilde t)= \left.\hat V(h,S_{min}( h), t)\right|_{ h=\rho_Z \tilde h}+O(h^6/|m^2_S|),\quad \rho_Z=\frac{Z_H[\mu_0]}{Z_H[\tilde \mu_0]}.
\end{align}
Note that we allow for different reference scales, and hence different classical fields $\tilde h$ and $h$, at both sides of the threshold; the reference scales are denoted as $\tilde\mu_0$ for the SM and $\mu_0$ for the SMS [see \eqref{eq:betadef}, \eqref{eq:t}]. The factor $\rho_Z$ in \eqref{eq:match} simply accounts for the fact that the potentials should give the same value when evaluated on fields corresponding to the same reference scale. Finally, we also allow for different values of the rescaling parameter $t$ in the high and low energy models. This is motivated by the presence of a (new) scale $m^2_S$ in the high energy theory and will be discussed in more detail in section~\ref{subsec:inflationloop}.

Once the SM potential is known, the matching \eqref{eq:match} allows to determine $V_0$, $m^2_H$ and $\lambda$ in the high energy model in terms of the SM values. We choose to proceed by matching the derivatives of order zero, one and two of the potentials in \eqref{eq:match} evaluated at a value of $h$ equal to a reference matching scale.  The latter has to be chosen low enough that the effects of the singlet do decouple. We take a matching scale $M_S$ given by $M_S^2=10^{-4}\times |m^2_S|$, which is a hundred times smaller than the expected mass of the singlet excitations around the valley.  This method of performing the matching improves upon the standard tree-level matching by considering one-loop effects, as well as not relying on a polynomial expansion of the potential around the origin. This allows a more reliable matching of the shape of the potential for the intermediate field values around the matching scale.

We have checked that varying the value of $M_S$ within a reasonably wide range has very little impact on the final results of the computations. For example using $M_S^2=10^{-2}\times |m^2_S|$ typically results in changes in the predictions for the parameters of the primordial spectrum of 2\% or less, as long as $m_t$ is not near the region triggering instabilities of the effective potential.\footnote{We remind the reader that the spectral index and the amplitude of primordial perturbations are currently determined with a precision of $\sim 0.5\%$ and $\sim 3\%$, respectively. Our theoretical predictions cannot match this level of precision, a factor that we have taken into the account in the calculations of section~\ref{subsec:inflationloop} by considering theoretical sources of errors in the computation for the cosmological parameters.}

Given all this, the matching equations that we employ are the following, using the notation $\tilde u\equiv \tilde h^2,u\equiv h^2$ (for neutral Higgs in the SM and SMS, respectively).
\begin{align}
 \nonumber \hat V^{SM}(\tilde h,\tilde t)\big|_{\tilde h=M_S}=& \hat V(h,S_{min}(h),t)\big|_{h=\rho_Z M_S},\\
 \label{eq:matchconds}\frac{d}{d\tilde u} \hat V^{SM}(\tilde h,\tilde t)\big|_{h=M_S}=&\rho^2_Z\frac{d}{d u} \hat V(h,S_{min}(h),t))\big|_{h=\rho_Z M_S},\\
 \nonumber \frac{d^2}{d\tilde u^2} \hat V^{SM}(\tilde h,\tilde t)\big|_{h=M_S}=&\rho^4_Z\frac{d^2}{d u^2} \hat V(h,S_{min}(h),t)\big|_{h=\rho_ZM_S}\,,
\end{align}
where, for convenience, we differentiate with respect to $u$ and $\tilde u$ rather than $h$ and $\tilde h$. One may obtain analytic formulae for $s_{min}(h)$ by solving \eqref{eq:ssol} perturbatively in a loop expansion. This notation makes explicit the fact that the renormalization scale can be chosen differently at each side of the threshold, see the related discussion in section~\ref{subsec:mtvalley}. Writing
\begin{align}
\label{eq:ssolloop}
 s_{min}(h)= S^{(0)}_{min}(h)+\frac{1}{16\pi^2} S^{(1)}_{min}(h)
\end{align}
then \eqref{eq:ssol} turns into
\begin{equation}
 \left.\frac{d\hat V^{(0)}}{ds}\right|_{S=S^{(0)}_{min}}=0,\quad \left.\frac{d^2\hat V^{(0)}}{dS^2}\right|_{S=S^{(0)}_{min}}S^{(1)}_{min}+\left.\frac{d \hat V^{(1)}}{ds}\right|_{S=S^{(0)}_{min}}=0.
 \label{eq:ssolloopeq}
\end{equation}
Since the one-loop expressions for $S^{(1)}_{min}$ are too lengthy to be displayed here, we just recall the tree-level result of \eq{sline} and also \eq{eq:treematching}, \eq{eq:treematching2}, noting that the full expressions with the two-loop RG improvement were used in the numerical calculations.

\bibliographystyle{hunsrt}

\end{document}